\documentclass[aip, jcp, reprint, longbibliography]{revtex4-2}

\usepackage{graphicx}
\usepackage{epsfig}
\usepackage[version = 4]{mhchem}

\usepackage{amsmath}
\usepackage{amsfonts}
\usepackage{amssymb}
\usepackage{bm}
\usepackage{mathtools}

\usepackage{lipsum}

\usepackage{tikz}
\usepackage{circuitikz}
\usetikzlibrary{arrows, shapes}
\usepackage{pgfplots}
\pgfplotsset{compat=1.14}
\usepackage{kbordermatrix}

\definecolor{webgreen}{rgb}{0,.5,0}
\definecolor{webbrown}{rgb}{.6,0,0}
\definecolor{grigio}{rgb}{.85,.85,.85} 
\definecolor{RoyalBlue}{rgb}{0.0, 0.14, 0.4}
\definecolor{skyblue1}{rgb}{0.45,0.62,0.81}
\definecolor{skyblue2}{rgb}{0.2,0.39,0.64}
\definecolor{skyblue3}{rgb}{0.13,0.29,0.53}
\definecolor{scarlet1}{rgb}{0.93,0.16,0.16}
\definecolor{scarlet2}{rgb}{0.8,0,0}
\definecolor{scarlet3}{rgb}{0.64,0,0}

\definecolor{g}{gray}{0.50}

\usepackage[colorlinks=true, linkcolor=red, citecolor=blue, urlcolor=red]{hyperref}

\begin{document}

\author{Shesha Gopal Marehalli Srinivas}
\affiliation{Complex Systems and Statistical Mechanics, Department of Physics and Materials Science,
University of Luxembourg,
30 Avenue des Hauts-Fourneaux, L-4362 Esch-sur-Alzette, Luxembourg}
\author{Massimiliano Esposito}
\affiliation{Complex Systems and Statistical Mechanics, Department of Physics and Materials Science,
University of Luxembourg,
30 Avenue des Hauts-Fourneaux, L-4362 Esch-sur-Alzette, Luxembourg}
\author{Nahuel Freitas}
\affiliation{Complex Systems and Statistical Mechanics, Department of Physics and Materials Science,
University of Luxembourg,
30 Avenue des Hauts-Fourneaux, L-4362 Esch-sur-Alzette, Luxembourg}
\affiliation{Departamento de Física, FCEyN, UBA, Pabellón 1, Ciudad Universitaria, 1428 Buenos Aires, Argentina}

\title{Hypergraph-Based Models of Random Chemical Reaction Networks: Conservation Laws, Connectivity, and Percolation}
\date{\today}

\begin{abstract}
Random graph models have been instrumental in characterizing complex networks, but chemical reaction networks (CRNs) are better represented as hypergraphs. Traditional models of random CRNs often reduce CRNs to bipartite graphs—representing species and reactions as distinct nodes—or simpler derived graphs, which can obscure the relationship between the statistical properties of these representations and the physical characteristics of the CRN.
We introduce a straightforward model for generating random CRNs that preserves their hypergraph structure as well as atomic composition, enabling the direct study of chemically relevant features. Notably, our approach distinguishes two notions of connectivity that are equivalent in graphs but differ fundamentally in hypergraphs. These notions exhibit percolation-like phase transitions, which we analyze in detail. The first type of connectivity has relevance to steady-state synthesis and transduction, determining the effective reactions an open CRN can perform at steady state. The second type is suitable to identify which species can be produced from a given initial set of species in a closed CRN. Our findings highlight the importance of hypergraph-based modeling for uncovering the complex behaviors of CRNs.
\end{abstract}

\maketitle

\section{Introduction}
Chemical reaction networks (CRNs) are defined by a set of chemical species and the reactions in which they are involved. 
In many instances (e.g. metabolism, biogeochemistry, atmospheric chemistry, astrobiology), the space of possible chemical compounds is vast and complex, even when restrictions are applied to the composition, structure, and size of molecules~\cite{Dobson2004Dec, reymondChemicalSpaceProject2015}. 
The space of possible chemical reactions is typically even larger~\cite{Stocker2020Oct, Murray2013Jan, grzybowski2009wired}. 
Thus, from biology to engineering, it is crucial to characterize the structural properties of large complex CRNs, and their statistical features.

As for the study of complex networks with random graphs~\cite{barabasicomplex2002, newmanreview2003, newman2018}, or complex quantum systems with random matrix theory~\cite{guhrRMT1998}, establishing a meaningful notion of random CRNs (under given constraints) would be highly suitable. 
It would provide a useful `null hypothesis' against which actual networks can be compared. 
The power of this approach lies in the fact that random models filter out overly specific details and highlight generic characteristics - thus revealing classes of typicality.

Attempts have been made to apply these ideas to complex CRNs by resorting to graph-based representations of CRNs. 
It has been argued, for instance, that the graphs of metabolic networks and the network of organic chemistry are scale-free~\cite{Jeong2000Oct, grzybowski2009wired}, but the biochemical relevance of such results has been questioned~\cite{tanaka2005, palssontext08, smith2021}. 
The major issue is that CRNs are hypergraphs and not graphs~\cite{Klamt2009}. 
In graphs, edges connect two nodes, whereas in hypergraphs, edges can connect multiple nodes.
Since reactions can transform multiple species into multiple products, CRNs are naturally represented as hypergraphs.
Projections into bipartite graphs can be achieved by assigning nodes to every species as well as to every reaction and placing edges from a given reaction node to every species node involved in that reaction.  
Since, in contrast with hypergraphs, many tools are available to study the statistical properties of graphs, most studies resort to such representations~\cite{Jeong2000Oct, grzybowski2009wired, smith2021, holme09, sandefur2013}.
Unfortunately, these mappings blur the relationship between the statistical properties of these representations and the physical characteristics of the CRN and can lead to misleading conclusions~\cite{Klamt2009, montanez10, Fagerberg2013Sep}. The usual graph theoretic quantities, such as degree distribution or clustering, have little chemical meaning in this context.
Developing tools to study the statistical properties of hypergraphs rather than graph representations of CRNs is thus essential to characterize chemically meaningful features of complex CRNs. 

Abstract mathematical models of random hypergraphs are starting to be studied~\cite{barthelemy22, ghoshal09, bianconiT2024}. However, the generation of random CRNs poses specific challenges related to the chemical consistency of the network~\cite{muller2022}. Indeed, since chemical reactions conserve atomic compositions, the resulting random CRNs must display correspondig conservation laws. 
Random hypergraph models of CRNs have been proposed in the literature. 
Some of these works study the probability that deficiency (a structural property) arises in random CRNs~\cite{anderson21, anderson22}, but without enforcing any conservation laws.
Others works define random CRNs that do preserve conservation laws but focus on dynamical features rather than structural ones~\cite{bigan2013,nicolaou23}.

Our contributions in this article are threefold. 
First, we propose a simple model of random CRNs that, by construction, enforces the conservation of atomic compositions. Starting from a representation solely in terms of the atomic composition of molecules, we assign reactions with a probability $p$ from the set of possible reactions allowed by atomic conservation laws.    
Second, we identify two different notions of connectivity, the first characterizing the existence of effective reactions between a set of chosen species that are exchanged in an open CRN, the second identifying which species can be produced from a given initial set of chosen species in a closed CRN.
Both notions are equivalent for graphs, but crucially different for CRNs.
Third, we characterize the statistical properties of these notions of connectivity and show that they each manifest different percolation-like phase transitions wherein the CRNs transition from low connectivity to high connectivity.
These findings have relevance in determining the likelihood of different types of synthetic pathways in complex CRNs.  

The paper is structured as follows.
Sec.~\ref{Sec:CRNS} provides the background.
We introduce CRNs and their structure in Subsec.~\ref{Sec:CRN_topintro}. 
We discuss issues with graph projections of CRNs in Subsec.~\ref{Sec:graphs}. 
We introduce our two notions of connectivity in Subsec.~\ref{Sec:synth_intro}.
In Sec.~\ref{sec:method}, we describe our method for generating chemically consistent random CRNs. 
In Sec.~\ref{Sec:result_main}, we present our main results on the various statistical features of our model of random CRNs and show that they display  percolation-like transitions. 
In Sec.~\ref{discussion}, we discuss our findings and compare our method for generating random CRNs with other methods that have been used in the literature. 
In Sec.~\ref{Sec:disc}, we conclude with a discussion on the broader implications and future scope of our results.

\section{Introduction to chemical reaction networks}
\label{Sec:CRNS}

\subsection{Structural aspects of CRNs}\label{Sec:CRN_topintro}
We consider CRNs wherein the chemical species, labeled $\sigma \in \mathcal{Z}$, are interconverted via elementary~\cite{Svehla1993}, reversible, and mass-balanced chemical reactions $\rho \in \mathcal{R}$ of the form, 
\begin{equation}
      \boldsymbol{\nu}_{+\rho} \cdot \boldsymbol{\sigma} \xrightleftharpoons[-\rho]{+\rho}  \boldsymbol{\nu}_{-\rho} \cdot \boldsymbol{\sigma}\,.
      \label{eq:chemeq}
\end{equation}
Here, $\boldsymbol{\sigma} = (\dots,\sigma,\dots)^{\intercal}$ and $\boldsymbol{\nu}_{+ \rho} = (\dots,\nu_{\sigma,+ \rho},\dots)^\intercal$ (resp. $\boldsymbol{\nu}_{-\rho}= (\dots,\nu_{\sigma,- \rho},\dots)^\intercal$) is the vector collecting the stoichiometric coefficients of the forward $+\rho$ (resp. backward $-\rho$) reaction. 
The structure (also called topology) of CRNs is encoded in the stoichiometric matrix $\nabla$ whose columns, $\nabla_{\rho} = \boldsymbol{\nu}_{-\rho}-\boldsymbol{\nu}_{+\rho}$, capture the net change in every species due to the reaction $\rho$.

The linearly independent vectors that span the left null space of the stoichiometric matrix, $\boldsymbol{\ell}^{\lambda}\cdot\nabla = 0$, are called conservation laws:
they identify parts of (or entire) molecules, called moieties, that are preserved by the reactions. 
Since we consider CRNs with all mass-balanced chemical reactions, there is always a conservation law, called the mass conservation law and denoted $\boldsymbol{\ell}^{m}$, which involves all the species:
$\boldsymbol{\ell}^{m} = (\dots,\ell^{m}_{\sigma},\dots)^{\intercal}$ with $\ell^{m}_{\sigma}\geq1$. 
%
We emphasize that every physically consistent CRN has at least one mass conservation law~\cite{muller2022}.

The linearly independent vectors that span the right null space of the stoichiometric matrix, $\nabla\cdot\boldsymbol{c} = 0$, are called (stoichiometric) cycles:
they identify series of reactions that leave all the chemical species unchanged upon completion.
Denoting the set of conservation laws as $\mathcal{L} = \{\boldsymbol{\ell}^\lambda\}$ and the set of cycles as $\mathcal{C} = \{\boldsymbol{c}\}$, the rank-nullity theorem~\cite{meyer2000} applied to the stoichiometric matrix $\nabla$ and its transpose $\nabla^{\intercal}$ gives the relation:
\begin{equation}\label{Eq:rank_nullity}
   \text{rk}(\nabla) = |\mathcal{Z}| - |\mathcal{L}| = |\mathcal{R}| - |\mathcal{C}|,
\end{equation}
where $\text{rk}(\cdot)$ denotes the rank of the matrix, and $|\cdot|$ denotes the cardinality of the set. We will make extensive use of this relation in Sec.~\ref{Sec:conserv_laws}.

The structural considerations described so far are appropriate for \textit{closed} CRNs, where no species are exchanged with the surroundings. For closed CRNs, conservation laws define dynamical invariants ~\cite{Rao2016, avanzini2020}, and both cycles and conservation laws are essential for thermodynamic descriptions~\cite{rao2018a, shimada2024}. For open CRNs where some chemical species are exchanged with the surroundings,  we partition the set of species $\mathcal{Z}$ into distinct subsets of internal species ($X$) and external species ($Y$), where the external species are exchanged with the surroundings. We refer to the selection of a set of $Y$ species as chemostatting the corresponding species.
We emphasize that this work focuses exclusively on the structural aspects of CRNs, without considering kinetics. The open/closed distinction used here represents the physical settings we use to describe the two notions of connectivity.

The structural aspects of $\textit{open}$ CRNs are derived using the (sub)stoichiometric matrix with rows that correspond only to the $X$ species and denoted by $\nabla^{X}$. 
In open CRNs, the set of conservation laws $\mathcal{L}$ splits into the sets of broken (resp. unbroken) conservation law, denoted $\mathcal{L}_{u}$ (resp. $\mathcal{L}_{b}$).
The unbroken conservation laws $\boldsymbol{\ell}^{\lambda_{u}}$ satisfy $\boldsymbol{\ell}^{\lambda_u}\cdot\nabla^{X} = 0$, i.e., they are conserved moieties of the internal species and are unaffected by the exchange.
Broken conservation laws, denoted by $\boldsymbol{\ell}^{\lambda_{b}}$, on the other hand, involve the chemostatted species and are affected by the exchange with the environment.

A key structural feature of open CRNs are \textit{emergent cycles}: effective interconversions of the external species that leave the internal species unaffected. 
Thus, the emergent cycles, denoted by $\boldsymbol{c}_{\epsilon}$, satisfy
\begin{equation}\label{Eq:emer_cyc_defn}
    \nabla^{X} \cdot \boldsymbol{c}_{\epsilon} = \boldsymbol{0} \quad \text{and} \quad \nabla^{Y} \cdot \boldsymbol{c}_{\epsilon} \neq \boldsymbol{0},
\end{equation}
where $\nabla^{Y}$ is the substoichiometric matrix with rows corresponding to the $Y$ species.
Denoting the number of emergent cycles as $|\epsilon|_{Y}$, from Eq.~\eqref{Eq:emer_cyc_defn}, we infer that
\begin{equation}\label{Eq:emer_cyc_num}
    |\epsilon|_{Y} = \text{dim}(\text{ker}(\nabla^{X})) - |\mathcal{C}|\geq 0\,.
\end{equation}
Given an emergent cycle $\boldsymbol{c}_{\epsilon}$, the stoichiometric coefficients of the effective reaction are obtained as the entries of the vector $\nabla^{Y}\cdot\boldsymbol{c}_{\epsilon}$, up to multiplication by a constant.
By collecting the negative (resp. positive) elements of $\nabla^{Y} \cdot \boldsymbol{c}_{\epsilon}$ into $\boldsymbol{\nu}^{e}_{+}$ (resp. $\boldsymbol{\nu}^{e}_{-}$), we can write the effective reaction as
\begin{equation}\label{Eq:net_reactions}
  \boldsymbol{\nu}^{e}_{+}\cdot\boldsymbol{\sigma}_{y}
   \xrightleftharpoons[]{}  \boldsymbol{\nu}^{e}_{+}\cdot\boldsymbol{\sigma}_{y}\,,
\end{equation}
where \( \boldsymbol{\sigma}_{y} = \left( \sigma \mid \sigma \in Y \right)^{\intercal} \) is the vector of external species.
We emphasize that i) emergent cycles interconvert the sets of substrates and products and ii) the number of emergent cycles depends on the choice of the external species. 

We can link the number of broken conservation laws and the number of emergent cycles by the relation~\cite{polettini2014}
\begin{equation}\label{Eq:emer_broken}
    |\mathcal{L}_{b}| + |\epsilon|_{Y} = |Y|\,,
\end{equation}
where $|Y|$ denotes the number of external species.
Eq.~\eqref{Eq:emer_broken} provides a fundamental insight: chemostatting a species either breaks a conservation law or creates an emergent cycle.
The dynamical and thermodynamical importance of emergent cycles and broken conservation laws have been thoroughly explored in the literature, and we refer readers to Refs.~\onlinecite{polettini2014, Rao2016, avanzini2020, avanzinicircuit2023} for more details.

\subsection{Graph Projections of CRNs}\label{Sec:graphs}
Hypergraphs are a generalization of graphs wherein (hyper)edges can connect more than two nodes. 
Since chemical reactions as considered in Eq.~\eqref{eq:chemeq} generically involve multiple chemical species, they are appropriately represented by hyperedges. By mapping chemical species onto nodes and reactions onto hyperedges, CRNs can be represented as hypergraphs\cite{Klamt2009, dal_cengio23}. 
Since we work with reversible reactions, the corresponding hyperedges are arbitrarily oriented. Reversing these orientations will not change any of the results.

Several graph representations of CRNs have been considered in the literature.
We briefly describe the most important representations below and direct readers to Refs.~\onlinecite{muller2022, sandefur2013} for a comprehensive review of the same.

\subsubsection{S, R and SR Graphs}\label{Sec:Bipartite}
Any CRN can be mapped into a bipartite graph by mapping the species and reactions onto two distinct types of nodes. In this representation, edges connect the species nodes to a reaction node if the species participate in the corresponding reaction.
These graphs are also called SR-graphs (species-reaction graphs) and can be used to extract information about the dynamics of the CRN\cite{shinar2013, craciun2006, feinberg_2019}. 
The SR graph can be further projected into the S graph (substrate graph) and R graph (reaction graph), respectively. 
In the S graph~\cite{kaltenbach2020}, the nodes are chemical species and two species $\sigma, \sigma'$ are connected by an edge if there is a reaction in which
$\sigma$ is the reactant and $\sigma'$ is the product (or vice versa).
In the R graph, chemical reactions are nodes and are connected by an edge if there is a chemical species involved in both reactions.
Although the S and R graphs can be unambiguously derived from the SR graph, the opposite may not be true~\cite{Fagerberg2013Sep}, implying an information loss associated with moving from the bipartite graph description to the S graph or the R graph.

\subsubsection{Graph of Complexes}\label{Sec:Complex_graph}
Another way to define a graph from a CRN is to consider the vectors $\boldsymbol{\nu}_{\pm \rho}$ (called complexes) of Eq.~\eqref{eq:chemeq} as nodes. Every reaction $\rho$ is then an edge connecting the node $\boldsymbol{\nu}_{+\rho}$ to the node $\boldsymbol{\nu}_{-\rho}$ (or vice versa).  
This description has been used in chemical reaction network theory to define a topological quantity, called deficiency, that is linked to the asymptotic dynamics and thermodynamics of CRNs~\cite{feinberg_2019, andersonkurtz2015, polettini2015, srinivas2023a}. The deficiency of the CRN, denoted by $\delta$,  is defined as the difference between the number of stoichiometric cycles and the number of cycles in the graph of complexes, i.e, 
\begin{equation}\label{Eq:def_defn}
    \delta = \text{dim}\left(\text{Ker}~\nabla\right) - \text{dim}\left(\text{Ker} ~\partial\right)\, \geq 0,
\end{equation}
where $\partial$ is the (oriented) incidence matrix of the graph of complexes.

\subsubsection{Example}\label{subs:counter_eg}
\begin{figure}
    \centering
    \includegraphics[width = .5\textwidth]{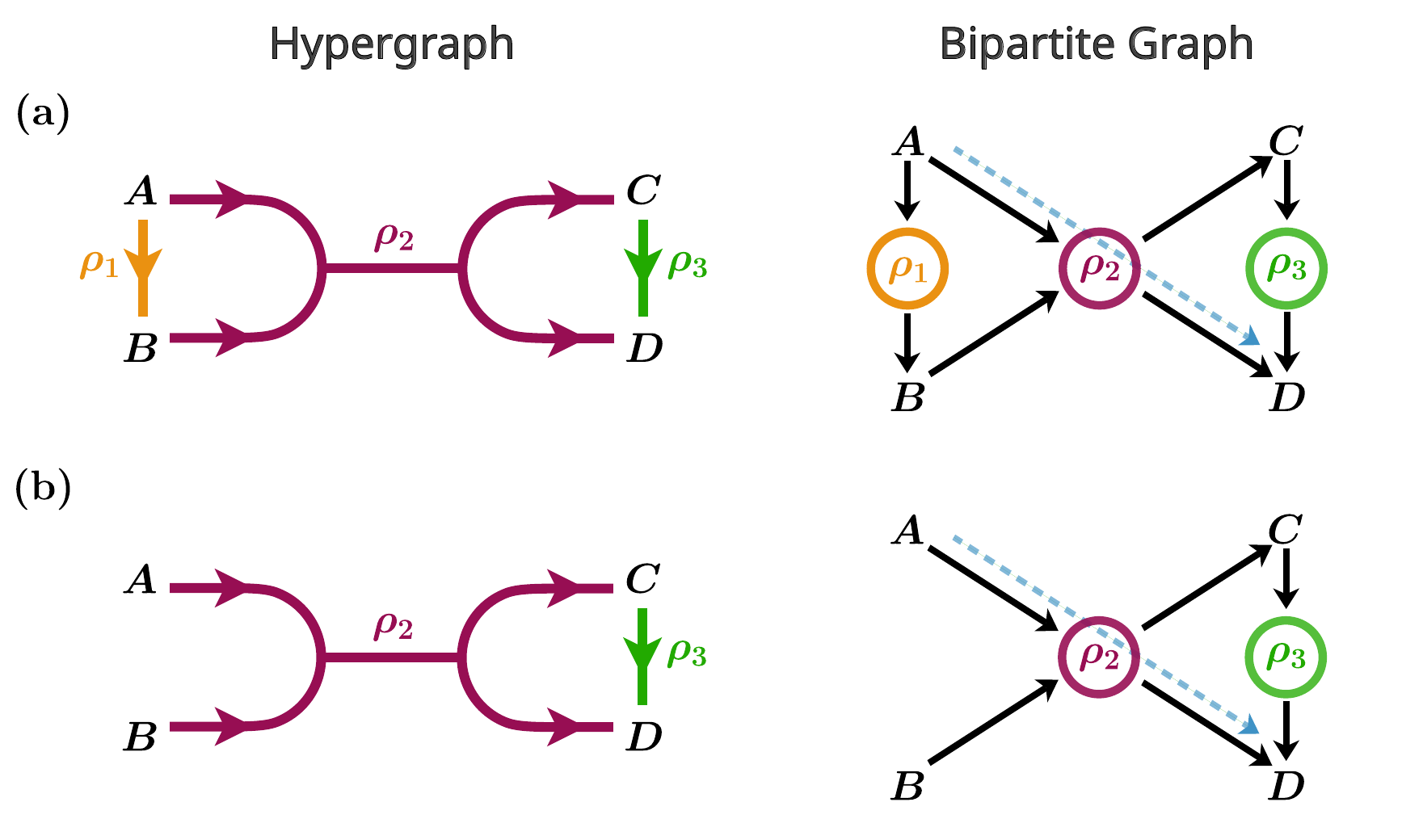}
    \caption{a) CRN in Eq.~\eqref{Eq:CRN_grapheg} represented as a hypergraph and a bipartite (SR) graph, with reaction nodes colored. b) Same representation of the CRN when reaction $\rho_1$ is removed. The blue dashed arrow depicts the path from $\text{A}$ to $\text{D}$ despite the fact that the CRN is unable to achieve it without $\rho_1$.}
    \label{fig:Hg_eg}
\end{figure}

We consider the following example, inspired by Ref.~\onlinecite{Klamt2009}, to illustrate some problems with the graphical descriptions discussed above.
Consider the following CRN:
\begin{align}\label{Eq:CRN_grapheg}
    \text{A} &\xrightleftharpoons[-1]{+1}  \text{B} \,,\notag\\
    \text{A} + \text{B} &\xrightleftharpoons[-2]{+2}   \text{C} + \text{D}\,, \notag\\
    \text{C} &\xrightleftharpoons[-3]{+3}  \text{D}\,.
\end{align} 
with the stoichiometric matrix,
\begin{equation}\label{Eq:stoc_grapheg}
      \nabla = \kbordermatrix{
       & {\color{gray} \rho_{1}} & {\color{gray} \rho_{2}} & {\color{gray} \rho_3}\\
     {\color{gray} \text{A}} &   -1 & -1 & 0 \\
     {\color{gray} \text{B}} &  1 & -1 & 0\\
      {\color{gray} \text{C}} &  0 & 1 & -1\\
     {\color{gray} \text{D}} &  0 & 1 & 1
    }\,.
\end{equation}
The hypergraph and graphs associated with Eq.~\eqref{Eq:CRN_grapheg} are shown in Fig.~\ref{fig:Hg_eg}a.

Suppose that we wish to answer the following question: Can the species $\text{D}$ be synthesized starting from the species $\text{A}$? 
By considering the set $Y = \{\text{A}, \text{D}\}$ as external species and carrying out the procedure outlined below Eq.~\eqref{Eq:net_reactions}, we find the emergent cycle $\boldsymbol{c}_{\epsilon} = (1, 1, 1)^{\intercal}$ corresponding to the net reaction $2\text{A} \xrightleftharpoons[]{} 2 \text{D}$.
This shows that starting from $\text{A}$, we can use the first reaction to create species $\text{B}$ and then perform the second reaction to create the species $\text{C}$, which then creates species $\text{D}$ via the third reaction.
If we now remove the first reaction, this is not possible anymore. Indeed, repeating the same procedure, the emergent cycle described above disappears. 
However, the SR graph still has a path starting from the node $\text{A}$ ending at the node $\text{D}$, as shown in Fig.~\ref{fig:Hg_eg}b. This illustrates that fundamental graph notions, such as connectivity in the SR graph (or in the S and R graphs as well), might not be chemically meaningful. 

\subsection{Notions of connectivity}\label{Sec:synth_intro}
From a chemical perspective, a natural question is whether a possibly large network of elementary reactions can, based solely on its topology (i.e., independent of any kinetic or thermodynamic considerations), interconvert certain chemical species into others.
More precisely, given a set of substrates and a set of desired products, we ask whether there exists a sequence of reactions that effectively consumes the substrates to produce the products.
In this subsection, we show that this question can be formulated in two distinct ways.

\emph{Chemostatting and emergent cycles.}
The first scenario characterizes the capacity of an open CRN to sustain steady-state synthesis.
Given a set of externally maintained (chemostatted) species, we ask whether at least one emergent cycle exists.
If so, the CRN can sustain the steady-state synthesis of the chemostatted products from the chemostatted substrates associated with each emergent cycle, without generating any additional waste products.

\emph{Reachability from a single species.}
The second scenario characterizes the capacity of a closed CRN, initialized with an infinite amount of a single species, to synthesize other species as it relaxes to equilibrium. 
In other words, the question is: which species can be produced from a given initial species by iterating the network's reactions until no new species are produced?
If we start with species $\sigma$, we refer to the set of species that is obtained by the previous procedure as the \textit{forward reachable set} of $\sigma$, denoted $\mathcal{F}(\sigma)$. If $\sigma' \in \mathcal{F}(\sigma)$, we say that $\sigma'$ can be reached from $\sigma$. 
Note that forward reachability is generically not symmetric, i.e., $\sigma' \in \mathcal{F}(\sigma)$ does not imply
$\sigma \in \mathcal{F}(\sigma')$.
A second key point is that, unlike in the emergent cycle scenario where all species are assumed to be available and every reaction can occur, here we must account for the availability of substrates. If a substrate for a reaction is absent, it cannot proceed, even if the reverse reaction is allowed.
As a result, even though all reactions are reversible, the choice of the initial species introduces a directionality.
Forward reachability can be viewed as a chemically grounded analogue of the notion of the out-component in a directed graph. In a directed graph, the out-component of a node is the set of nodes reachable by following directed edges \cite{newman2018}.
We note that the forward reachable set given two or more initial species can be defined analogously, but we do not consider that notion in this work.

Importantly for CRNs that can be represented as undirected graphs (i.e. CRNs solely made of unimolecular reactions $A \xrightleftharpoons[]{} B$), the two notions of connectivity introduced above are equivalent. 
Indeed, in this case, forward reachability is symmetric and any terminal species involved in an emergent cycle belongs to the same reachable set.
This shows that the two notions of connectivity introduced above can be considered non-equivalent generalizations of connectivity in undirected graphs.

As an illustration, we can go back to the CRN in Eq.~\eqref{Eq:CRN_grapheg} \textit{without} the first reaction.
In this case, chemostatting the species $Y = \{\text{A}, \text{D}\}$ does \textit{not} produce an emergent cycle.
However, the forward reachable set of the species $\text{D}$ is given by $\mathcal{F}(\text{D}) = \{\text{D}, \text{C},\text{A},\text{B}\}$. 
If we start with a large amount of $\text{D}$, then the species $\text{C}$ can be formed via the third reaction and together, they can react to form $\text{A}$ and $\text{B}$.
In turn, the forward reachable set of the species $\text{A}$ is $\mathcal{F}(\text{A}) = \{\text{A}\}$ because no reaction can solely run on the species $A$. 
These results illustrate that forward reachability and emergent cycle are two different notions of connectivity and that forward reachability is asymmetric.
It may seem that bidirectional reachability between two species $\sigma$ and $\sigma{'}$ (i.e. if $\sigma'$ can be reached from $\sigma$ and vice-versa) would imply the existence of an emergent cycle if one would  chemostat them. However, this is not generically true as shown by counterexamples in Appendix~\ref{sec:link_reach_emer}.

We end this section by discussing another common notion of connectivity in graphs: that of connected components. 
Mathematically, the number of connected components of a graph is equal to the dimension of the left nullspace of the (oriented) incidence matrix. Therefore, a natural analogue of connected components in CRNs is given by the conservation laws, which generate the left nullspace of the stoichiometric matrix~{(see Sec.~\ref{Sec:CRN_topintro})}.

\section{Generation of random CRNs}
\label{sec:method}
In this section, we explain our algorithm for generating random CRNs. We begin by explaining the problem to be overcome.
Let $\nabla = \{\nabla^\sigma_\rho\}_{\sigma=1,\cdots,|\mathcal{Z}|, \rho=1,\cdots,|\mathcal{R}|}$ be the $|\mathcal{Z}| \times |\mathcal{R}|$ stoichiometric matrix of a generic system of $|\mathcal{Z}|$ chemical species and $|\mathcal{R}|$ reactions. 
Since we work with closed CRNs, each column of $\nabla$ must have at least one entry of each sign. 
If we restrict ourselves to reactions where all species appear with a stoichiometric coefficient less than or equal to $\kappa$, then the components of $\nabla$ are integer numbers between $-\kappa$ and $\kappa$.
If the order of the reaction is restricted to be less than or equal to $K$, i.e, $\sum_{\sigma} \nu_{\sigma, \pm \rho} \leq K$, then the sum of positive entries in each column $\nabla$ must be less than or equal to $K$, and the same must be true for $-\nabla$.
The entries of $\nabla$ could be filled with random integers in a way such that those conditions are verified. However, for consistency, the resulting CRN must also have at least a mass conservation law.
This condition is not trivial to satisfy in a random assignment of elements to $\nabla$, since it introduces a constraint between different columns or reactions (in contrast to all the previous conditions, that are internal to each column). For example, the reactions $A+B \xrightleftharpoons[]{} C$ and $A+B \xrightleftharpoons[]{} 2C$ cannot be part of the same system. 
Furthermore, if the $|\mathcal{Z}|$ molecular species are formed by different types of atoms, then one might have additional conservation laws beyond that of the total mass, since chemical reactions cannot alter atomic identities.

We propose the following solution to the problem. We start by fixing the number $N_A$ of different atomic types forming the possible molecules, and we consider that the maximum number of atoms of type $a=1,\cdots,N_A$ involved in any molecular species is $Q_a$. In addition, we assume that a molecular species is fully determined by the number of atoms of each type that make it (the chemical formula). In other words, we represent a chemical species by its molecular formula, completely disregarding its internal structure. For example, the molecules pentene and cyclopentane would be the same for us, since both have 5 carbons and 10 hydrogen atoms. Then, at maximum, we can have 
\begin{equation}
|\mathcal{Z}|^{\text{max}} = \prod_{a=1}^{N_A} (Q_a+1) -1
\label{eq:NS_max}
\end{equation}
molecular species. Each one of them is identified by the tuple $\sigma=(n_1, n_2,\cdots, n_{N_A})$, where $n_a$ is the number of atoms of each type. 
We then focus on the case in which the overall order of any reaction is at most two, as is typically the case for elementary reactions. 
Given that we are disregarding the internal structure of the chemical species, unimolecular reactions like $A \xrightleftharpoons[]{} B$ are meaningless, since $A$ and $B$ would be the same species due to atom conservation by reactions. 
Thus, we consider reactions of the kind $A + B \xrightleftharpoons[]{} C$ or $2A \xrightleftharpoons[]{} B$. 
Although bimolecular reactions such as $A+B \xrightleftharpoons[]{} C+D$ could also be taken into account, they can be considered as effective descriptions of the process $A+B \xrightleftharpoons[]{} X \xrightleftharpoons[]{} C+D$, and therefore, for simplicity, will not be counted separately in what follows. 
For a molecule $(n_1,\cdots,n_{N_A})$ we have the possible dissociations:
\begin{align}
    (n_1,n_2,\cdots, n_{N_A}) \xrightleftharpoons[]{} &(n_1\! -\! m_1, n_2\! -\! m_2, \cdots, n_{N_A} \! -\! m_{N_A})\nonumber\\&+ (m_1, m_2,\cdots, m_{N_A}) \label{Eq:rxn_chem}
\end{align}
for any values of $m_1,m_2,\cdots m_{N_A}$ such that  $\min\{m_a\} > 0$, $\min\{n_a-m_a\}>0$ and that ${0 \leq m_a \leq n_a}$ for all $1 \leq a \leq A$. Noting that the substitution of $n_a - m_a$ by $m_a$ in \eqref{Eq:rxn_chem} leaves the reaction unchanged, the total number of distinct dissociations of the molecule $\sigma=(n_1,\cdots,n_{N_A})$ is 
\begin{equation}
D(\sigma) = \frac{\prod_{a=1}^{N_A} (n_a + 1)}{2} - \pi(\sigma),
\label{eq:diss}
\end{equation}
where $\pi(\sigma)$ is $1/2$ if all the $n_a$ are even, or $1$ otherwise. Since any possible reaction is a dissociation of some species, the total number of possible reactions is (by convention, we define that $D(\sigma=0)=0$):
\begin{equation}
\begin{split}
    |\mathcal{R}|^\text{max} =& \sum_\sigma D(\sigma)\\ 
    =& \sum_{n_1=0}^{Q_1}\sum_{n_2=0}^{Q_2} \cdots \sum_{n_{N_A}=0}^{Q_{N_A}} D(n_1,\cdots,n_{N_A}) \\
    =& \: 2^{-(N_A+1)} \prod_{a=1}^{N_A} (Q_a+1)(Q_a+2) + \\
    & \: 2^{-1} \prod_{a=1}^{N_A} (\lfloor Q_a/2 \rfloor + 1) - |\mathcal{Z}|^{\text{max}} - 1.
    \label{eq:NR_max}
\end{split}
\end{equation}
To derive the last line in Eq.~\eqref{eq:NR_max}, observe that $\pi(\sigma)=1$, except for those $\sigma$ with even entries, for which $\pi(\sigma)=1/2$. Hence,
$\sum_{\sigma}\pi(\sigma)= 1 + |\mathcal{Z}|^{\max}-(1/2)\prod_{a=1}^{N_A}(\lfloor Q_a/2\rfloor+1)$, where the negative term corrects for the $\sigma$ with all even entries.
Note that if $Q_a \gg 1$, then we have:
\begin{equation}
    |\mathcal{R}|^\text{max} \simeq \frac{|\mathcal{Z}|^{\text{max}}(|\mathcal{Z}|^{\text{max}}+1)}{2^{N_A+1}} - |\mathcal{Z}|^{\text{max}}.
    \label{eq:NR_max_asymp}
\end{equation}
All the possible $|\mathcal{R}|^\text{max}$ reactions can be easily enumerated and form the total set of reactions we consider. Appendix \ref{ap:example_counting} illustrates Eq.~\eqref{eq:NS_max} and \eqref{eq:NR_max} for the simple case of $N_{A} = 2, Q_{1} = Q_{2} = 2$.

To create a random stoichiometric matrix $\nabla$, one possible approach is to randomly select exactly $|\mathcal{R}|$ reactions from the total set of $|\mathcal{R}|^{\text{max}}$ possible reactions. This ensures that the number of active reactions is fixed and that all subsets of size $|\mathcal{R}|$ are sampled uniformly.
Another possibility, which is the one we use in what follows, is to assign to each of the $|\mathcal{R}|^{\text{max}}$ reactions an independent probability $p$ of being active. In this case, each reaction is included independently with probability $p$, and the total number of active reactions $|\mathcal{R}|$ becomes a random variable that follows a binomial distribution with mean $p \cdot |\mathcal{R}|^{\text{max}}$. Unlike the first method, this procedure does not constrain the total number of reactions to a fixed value, but instead defines it probabilistically.

The proposed model is similar to the Erdös-Rényi (ER) model of random graphs {(see Appendix~\ref{Sec:ER_intro} for a brief description of the ER model)}, but extended to hypergraphs describing CRNs.
Note that the scaling relation between the total number of possible reactions and the total number of chemical species in our model is similar to the ER random graph~(in the sense that $|\mathcal{R}|^\text{max}$ grows quadratically in $|\mathcal{Z}|^{\text{max}}$, see Eq. \eqref{eq:NR_max_asymp} and compare with Appendix~\ref{Sec:ER_intro}). 

The same process can be used to construct networks with reactions of an order higher than $2$, or elementary reactions of the kind $A+B \xrightleftharpoons[]{} C+D$, that were discarded before. The scaling of $|\mathcal{R}|^\text{max}$ and $|\mathcal{Z}|^{\text{max}}$ will be different and also the precise counting of $|\mathcal{R}|^\text{max}$ will be more involved. 
Finally, we note that by appropriately defining the way reactions are selected, CRNs models analogous to well-studied random graph models (Watts-Strogatz, Barabasi-Albert) can also be implemented~\cite{newman2018, barabasicomplex2002,dorogovtsevcomplexnet2022}.

\section{Statistical properties of Random CRNs}\label{Sec:result_main}
In this section, we consider ensembles of randomly generated CRNs and present our main findings of their statistical properties.
We first study the statistics of the number of species and the number of reactions they are involved in. These results will then be used to study the statistics of their number of conservation laws, cycles, and emergent cycles, as well as the statistics of the deficiency, and the size of the largest forward reachable set.
As for the ER random graph (see Appendix~\ref{Sec:ER_intro}), we scale the probability of choosing a reaction $p$ as
\begin{equation}\label{Eq:prob_rescaled}
    p = \gamma/|\mathcal{Z}|^{\text{max}}\,.
\end{equation}
This ensures that the CRN we generate is \textit{sparse}, i.e. the average number of reactions in which a given species participates will be fixed by $\gamma$, even as we scale the number of reactions and species. 
We also focus on the \textit{extensive limit}, when $|\mathcal{Z}|^{\text{max}} \to \infty$.
Many of the results in this section are semi-analytic, especially for $N_{A}>1$. 
{Closed-form expressions for the case $N_{A}=1$, together with additional remarks, are collected in Appendix \ref{Sec:one_atom}.}

\subsection{Selected reactions and species}\label{Sec:reac}
Given a set of reactions $\mathcal{R}$, $\mathcal{Z} \subseteq {\mathcal{Z}^{\text{max}}}$ is the set of species that are involved in at least one of the reactions in $\mathcal{R}$. 
From hereon, we restrict our focus to $\sigma \in \mathcal{Z}$ when generating a random stoichiometric matrix $\nabla$.
Since the set $\mathcal{R}$ is constructed by independently choosing possible reactions with probability $p$, the sizes $|\mathcal{R}|$ and $|\mathcal{Z}|$ of both sets will be stochastic quantities. 
The probability of a particular value of $|\mathcal{R}|$ is given by the binomial distribution:
\begin{equation}
    \mathbb{P}(|\mathcal{R}|) = {|\mathcal{R}|^\text{max} \choose |\mathcal{R}|} p^{|\mathcal{R}|} (1-p)^{|\mathcal{R}|^\text{max}-|\mathcal{R}|}\,,
\end{equation}
with the average number of reactions given by,
\begin{equation}\label{Eq:Avg_rxn}
    \langle|\mathcal{R}|\rangle = p |\mathcal{R}|^\text{max} \,.
\end{equation}

The probability of a particular value of $|\mathcal{Z}|$ is highly non-trivial, since the events $\sigma \in \mathcal{Z}$
and $\sigma' \in \mathcal{Z}$ are not independent. 
However the probability for a particular species $\sigma$ to not be in $\mathcal{Z}$ is:
\begin{equation}\label{Eq:rxn_total_def}
    P(\sigma \notin \mathcal{Z}) = (1-p)^{I(\sigma)},
\end{equation}
where $I(\sigma)$ is the number of possible reactions involving $\sigma$. This number can be written as 
\begin{equation}\label{Eq:tot_rxn_I}
    I(\sigma)=D(\sigma)+A(\sigma)\,,
\end{equation}
where $D(\sigma)$ is the number of possible dissociations given in Eq. \eqref{eq:diss}, 
and
\begin{equation}\label{Eq:assoc_rxn_def}
    A(\sigma) = \prod_{a=1}^{N_A} (Q_a - n_a + 1) - 1
\end{equation}
is the number of possible associations of the form:
\begin{equation}\label{Eq:association_rxn}
\begin{split}
    (n_1,n_2,\cdots, &n_{N_A}) \: + \: (m_1, m_2,\cdots, m_{N_A}) \\ 
    &\xrightleftharpoons[]{} (n_1+m_1, n_2+m_2, \cdots, n_{N_A} + m_{N_A}) 
\end{split}
\end{equation}
for any values of $m_1,m_2,\cdots m_{N_A}$ such that  $\min\{m_a\} > 0$ and ${0 \leq m_a \leq Q_a - n_a}$ for all $1 \leq a \leq A$. 

\subsection{Number and Degree distributions} \label{Sec:degree_gen}

Given a random set of reactions $\mathcal{R}$, the degree $d(\sigma)$ of a given species $\sigma=(n_1, \cdot, n_{N_A})$ is defined as the number of reactions in $\mathcal{R}$ in which it participates. The probability for the species $\sigma$ to have degree $k$ is given by the binomial distribution:
\begin{equation}
    p_{\sigma}(k) = {I(\sigma) \choose k} p^k (1-p)^{I(\sigma)-k}.
    \label{eq:degree_prob}
\end{equation}

Note that in the Erdös-Rényi model each node in a random graph also has a binomial distribution for its degree, but this distribution is the same for all nodes~\cite{newman2018}. This is not the case here, since Eq. \eqref{eq:degree_prob} depends explicitly on the species $\sigma$. 

\begin{figure}
    \centering
    \includegraphics[scale=.50]{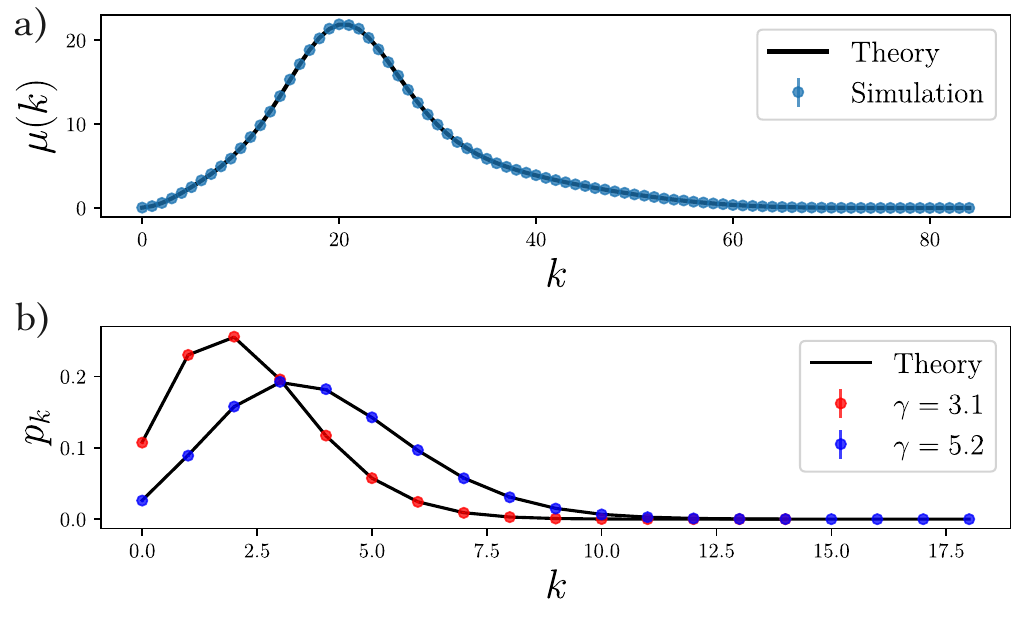}
    \caption{a) Average number of species with degree $k$ as a function of $k$ for the parameters:  $N_A = 2$, $Q_{1} = 27, Q_{2} = 15$, $p = 0.13$ averaged over $5000$ reaction sets compared with the theoretical curve (black line) computed using Eq.~\eqref{Eq:Mean_deg}. The standard error of the mean for each point is indicated by the errorbar which is not visible in the plot. b) Probability for a randomly chosen species to have degree $k$ as a function of $k$ for the parameters: $N_{A} = 1$, $Q = 500$ averaged over $5000$ reaction sets. Two different datasets with $\gamma = 3.1$ and $\gamma = 5.2$ are plotted against the theoretical curves (black line) computed using Eqs.~\eqref{Eq:deg_distri_full} and \eqref{Eq:deg_distri_1D_full}. For both curves, the standard error of the mean is plotted on the data and is not visible on the plot.}
    \label{fig:Deg_dist_eg}
\end{figure}

We define the degree distribution as the probability that a randomly chosen species has a degree $k$. The analogous quantity in the context of random graphs offers useful information about the graph \cite{newman2018}.
We first compute a quantity similar to the degree distribution: the number distribution, which is the \textit{number} of species with degree $k$. 
For this computation, we assume that species $\sigma$ having degree $k$ and species $\sigma'$ having degree $k$ are independent events. 
This approximation can be expected to hold in the limit of a large number of reactions as there can only be at most two different reactions involving the species $\sigma$ and $\sigma'$ (see Eq.~\eqref{Eq:rxn_chem}). 
Thus, the probability that $n_{k}$ species have degree $k$ is given by,
\begin{equation}\label{Eq:deg_distribution}
  P(n_{k}) = \sum_{\substack{A \subset \mathcal{Z}^{\text{max}} \\ |A| = n_{k}}} \prod_{\sigma \in A } p_{\sigma}(k)\prod_{\sigma' \notin A} (1-p_{\sigma'}(k))\,,
\end{equation}
where the sum runs over all subsets  $A$ of  $\mathcal{Z}$ with cardinality $n_k$. We note that the resulting probability density in Eq.~\eqref{Eq:deg_distribution} is a Poisson-Binomial distribution~\cite{tangPBD2023} with mean
\begin{equation}\label{Eq:Mean_deg}
    \mu(k) = \sum_{\sigma} p_{\sigma}(k)\,,
\end{equation}
and variance 
\begin{equation}
    \text{Var}(k) = \sum_{\sigma} p_{\sigma}(k)(1-p_{\sigma}(k))\,.
\end{equation}
It is seen from Fig.~\ref{fig:Deg_dist_eg}a) that Eq.~\eqref{Eq:Mean_deg} captures the mean very well.
In the scaling regime of Eq.~\eqref{Eq:prob_rescaled},  we can expect that the probability of a randomly chosen species to have degree $k$, i.e, the degree distribution denoted $p_{k}$, to be equal to (in the limit of large 
$|\mathcal{Z}|^{\text{max}}$),
\begin{equation}\label{Eq:deg_from_num}
    p_{k} = \lim_{|\mathcal{Z}|^{\text{max}} \to \infty}\frac{\mu(k)}{|\mathcal{Z}|^{\text{max}}}\,.
\end{equation}
We show in Appendix~\ref{Sec:sp_proof} that Eq.~\eqref{Eq:deg_from_num} can be expressed as
\begin{align}\label{Eq:deg_distri_full}
    p_{k} 
    &= { \frac{\gamma^{k}}{k!}}\int_{0}^{1} \! dx_1 \cdots \int_{0}^{1} \! dx_{N_A} \left( \frac{1}{2} \prod_{k = 1}^{N_A}x_{k} + \prod_{j = 1}^{N_A}(1 - x_{j}) \right)^{k} \\
    &\quad \times e^{-\gamma \left( \frac{1}{2} \prod_{k = 1}^{N_A}x_{k} + \prod_{j = 1}^{N_A}(1 - x_{j}) \right)}\,, \notag
\end{align}
which is valid for any number of atomic types.  Fig.~\ref{fig:Deg_dist_eg}b) plots the degree distribution obtained from simulation for two different values of $\gamma$ against Eq.~\eqref{Eq:deg_distri_full} (also see Eq.~\eqref{Eq:deg_distri_1D_full} for a closed form expression for the case of $N_{A} = 1$). We see that Eq.~\eqref{Eq:deg_distri_full} captures the behavior of the degree distribution very well.

\begin{figure}
    \centering
    \includegraphics[scale=0.5]{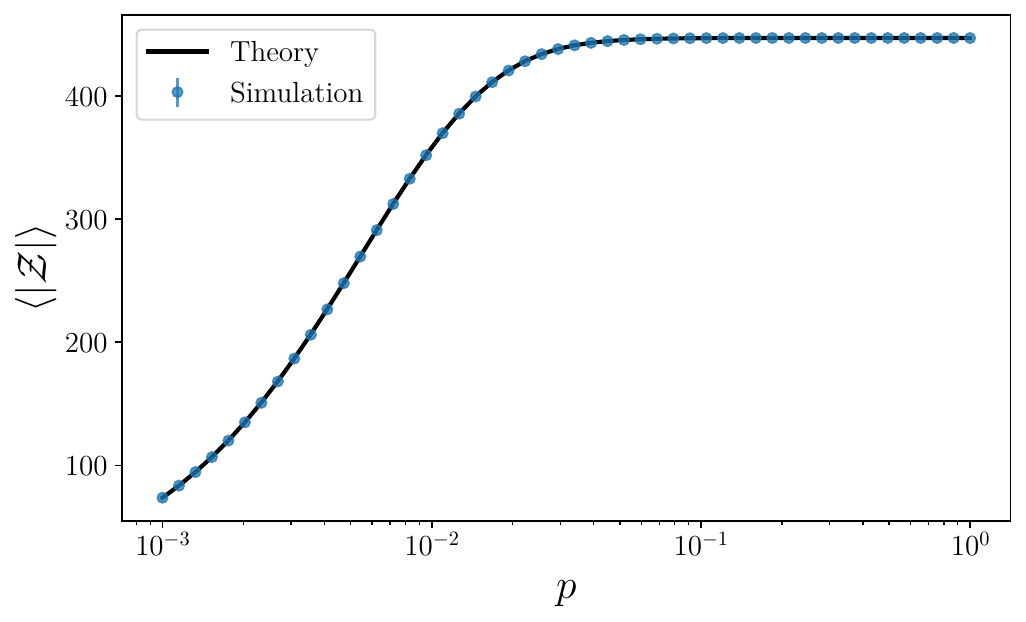}
    \caption{Average number of species versus $p$ averaged over $10000$ reaction sets for the parameters: $N_A = 2$, $Q_{1} = 27, Q_{2} = 15$, $p = 0.13$. The theoretical curve (black line) is calculated using Eq.~\eqref{Eq:exp_species}. The standard error for each point is indicated by the errorbar which is not visible in the plot.}
    \label{fig:sp_dist}
\end{figure}

The average number of species participating in at least one reaction (denoted $\langle|\mathcal{Z}|\rangle$) can be computed from the average of the number distribution $\mu(k)$, as follows:
\begin{align}\label{Eq:exp_species}
    \langle|\mathcal{Z}|\rangle &= |\mathcal{Z}|^{\text{max}} - \mu(0)\,,\notag\\
    &= \prod_{a=1}^{N_A} (Q_a+1) - 1 - \sum_{\sigma} (1-p)^{I(\sigma)}\,,
\end{align}
where we used Eqs.~\eqref{eq:NS_max} and \eqref{Eq:Mean_deg} (also see Eq.~\eqref{Eq:Avg_sp_1D} for a closed form expression for the case of $N_{A} = 1$). 
At very small probabilities, using the binomial approximation $(1-x)^{n} \approx 1 - nx$ along with Eq.~\eqref{Eq:exp_species}, we find that $\langle|\mathcal{Z}|\rangle$ scales linearly in $p$ with,
\begin{equation}\label{Eq:exp_sp_small}
  \langle|\mathcal{Z}|\rangle \simeq |\mathcal{Z}|^{\text{max}} - \sum_{\sigma} (1 - {I(\sigma)} p) = p \sum_{\sigma} I(\sigma)\,,  
\end{equation}
whereas in the limit $p \to 1$, the expected number of species approaches the limiting value $|\mathcal{Z}|$.
We compare Eq.~\eqref{Eq:exp_species} to simulations in Fig.~\ref{fig:sp_dist}.


\subsection{Connectivity I:  Emergent Cycles and Conservation Laws}\label{Sec:conserv_laws}

We now analyze connectivity in terms of emergent cycles, as defined in Subs.~\ref{Sec:synth_intro}. Since our random CRNs may include any possible stoichiometrically allowed reaction, we focus on the probability that randomly selected substrates can interconvert to form randomly chosen products, as $p$ and $Q_a$ are varied. We work in the regime described by Eq.~\eqref{Eq:prob_rescaled} and in the extensive limit.

Numerically, to estimate the probability of generating an emergent cycle, we first generate ensembles of random CRNs for given values of $\gamma$ and $\{Q_{1}, Q_{2},\dots Q_{N_{A}}\}$. We then randomly select a subset of species in $\mathcal{Z}$ and check if chemostatting them forms an emergent cycle.
Importantly, in a CRN with $N_A$ atoms, at least $N_A + 1$ chemostatted species are typically needed to obtain a non-negligible chance of forming an emergent cycle. Since reactions conserve atomic identities, synthesizing a species with $N_A$ atoms generally requires $N_A$ distinct precursors. Using fewer precursors imposes strict linear constraints on atomic balances in the products, which are unlikely to be satisfied by randomly chosen products, greatly reducing the chance of forming an emergent cycle linking them.
We model the outcome as a Bernoulli random variable $n_i$, which is one if an emergent cycle is formed and zero otherwise: i.e.,
\begin{equation}
    n_{i} = \begin{cases}
        1~~\text{with prob.}~~\mathbb{P}_{\epsilon}(\gamma, \{Q_{1},...Q_{A}\})\\
        0~~\text{with prob.}~~1-\mathbb{P}_{\epsilon}(\gamma, \{Q_{1},...Q_{A}\})\,,
    \end{cases}
\end{equation}
and study how the probability $\mathbb{P}_{\epsilon}(\gamma, \{Q_{1},...Q_{A}\})$
changes with $\gamma$ in the extensive limit.  Typically, we observe that $\mathbb{P}_{\epsilon}(\gamma, \{Q_{1},...Q_{A}\})$ is zero below some threshold and nonzero above it, resembling a percolation transition. Below the threshold, a randomly chosen subset of species is unlikely to be connected via an emergent cycle, while above it, the probability becomes finite.

We begin with the simplest case of $N_A = 1$. For fixed values of $\gamma$ and $Q$, we generate ensembles of random CRNs. In each CRN, we randomly select and chemostat two species and then check whether an emergent cycle is formed (note that at low values of $\gamma$, whenever $\mathcal{Z}$ involves less than two species, the answer is automatically taken to be zero). The probability $\mathbb{P}_{\epsilon}(\gamma, Q)$ is estimated as the mean of the observed $n_{i}$.

\begin{figure}
    \centering
    \includegraphics[scale=.5]{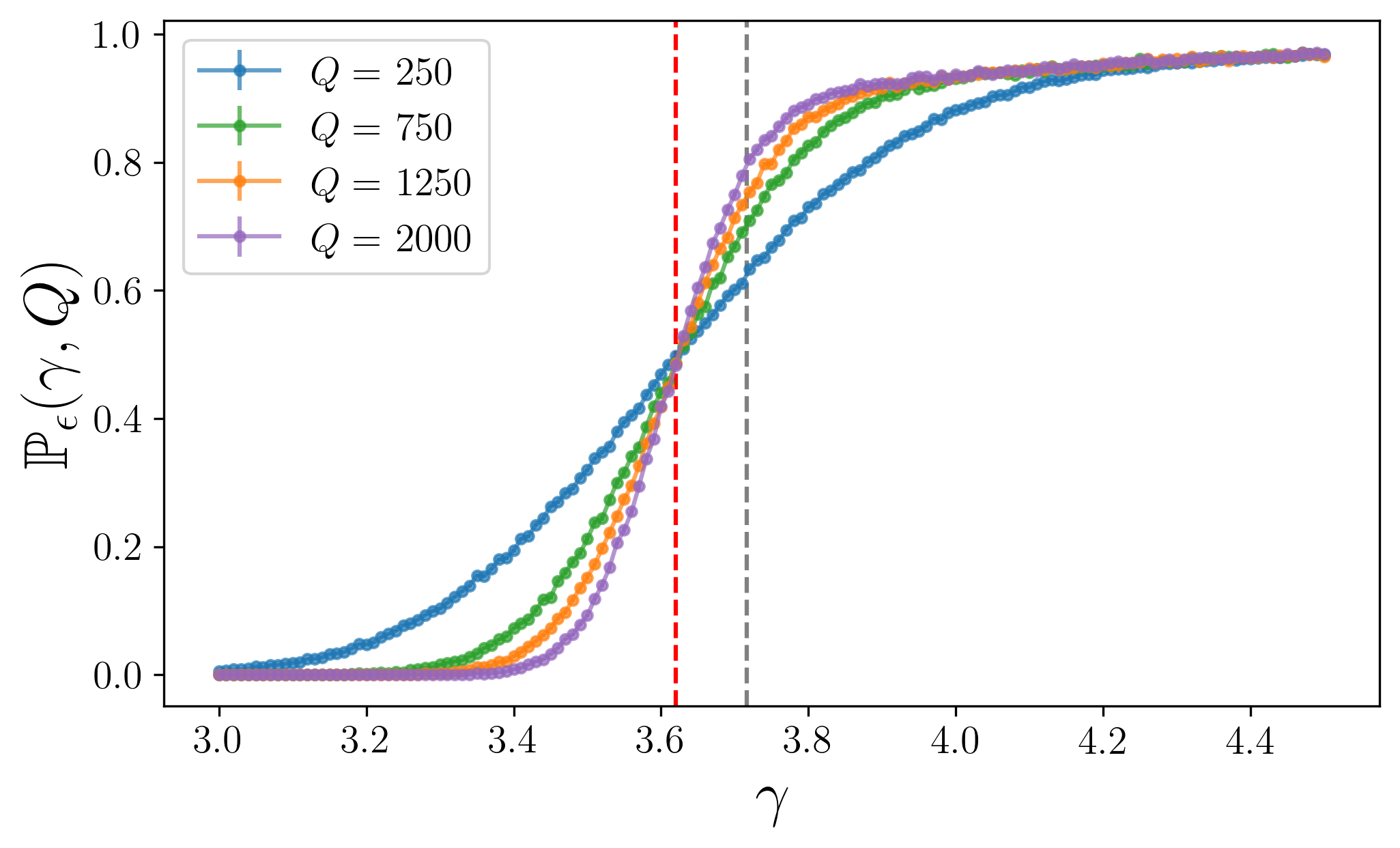}
    \caption{Estimate of the probability of having an emergent cycle when two random species involved in the CRN are chemostatted plotted as a function of the rescaled probability $\gamma$. The parameter is $N_{A} = 1$ . The grey dashed line is the values of $\gamma^{*}$ obtained numerically following Eq.~\eqref{Eq:claw_hyp2_1atom_univ}. The red dashed line is the accurate threshold $\gamma_{c}$ obtained from numerical interpolation of the crossing points in the data. For each value of $\gamma$, if $Q = 100,250$, we generated 25000 reaction sets, if $Q = 500$, we generated 20000 reaction sets and 10000 reaction sets for the rest for averaging.  The sample error of the mean is shown by errorbars which are not visible on the plot.}
    \label{fig:emer_cyc1}
\end{figure}

\begin{figure}
    \centering
    \includegraphics[scale=.5]{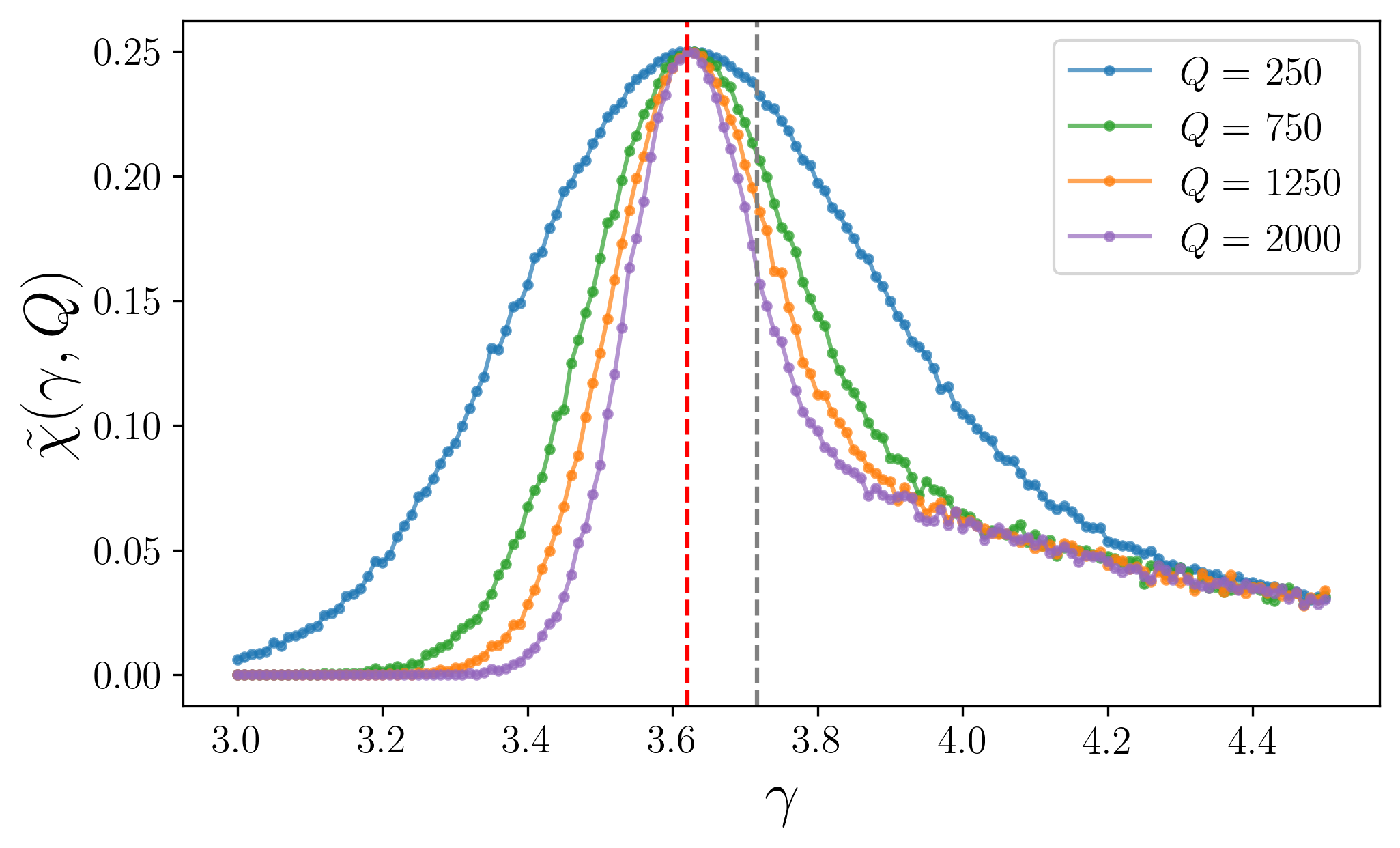}
    \caption{Variance of the probability of having an emergent cycle when two random species involved in the CRN are chemostatted plotted as a function of the rescaled probability $\gamma$. The parameter is $N_{A} = 1$. The grey dashed line is the values of $\gamma^{*}$ obtained numerically following Eq.~\eqref{Eq:claw_hyp2_1atom_univ}. For each value of $\gamma$, if $Q = 100,250$, we generated 25000 reaction sets, if $Q = 500$, we generated 20000 reaction sets and 10000 reaction sets for the rest for averaging.}
    \label{fig:emer_cyc1_sus}
\end{figure}

The result of our numerical simulations is plotted in Fig.~\ref{fig:emer_cyc1}. We observe the following:
First, for all $Q$, the probability is close to zero until some $\gamma_{c}$, after which it rapidly increases to one. 
This indicates that the low $\gamma$ regime is highly disconnected, as any random pair of species is unlikely to be interconvertible. On the other hand, in the high $\gamma$ regime, almost any species can be synthesized from another. 
Secondly, in Fig.~\ref{fig:emer_cyc1}, the curves for different $Q$ intersect each other.
This suggests that in the limit of $Q \to \infty$, the probability of an emergent cycle approaches the limiting form:
\begin{equation}\label{Eq:First_order_form}
   \lim_{Q \to \infty} \mathbb{P}_{\epsilon}(\gamma, Q) = \begin{cases}
       0~\text{if}~\gamma < \gamma_{c}\\
       1~\text{if}~\gamma >\gamma_{c}\,.
   \end{cases} 
\end{equation}
This is similar to a discontinuous (first-order) transition at the threshold $\gamma_{c}$. The threshold $\gamma_{c}$ can be estimated by extrapolating the crossing point of the curves for consecutive $Q$ in Fig.~\ref{fig:emer_cyc1}. 
Carrying out this extrapolation, we find that the percolation threshold is $\gamma_{c} = 3.62 \pm 0.02$ (see Appendix~\ref{Subsec:data_emer1D} for full details of the methodology).
{To further confirm the discontinuity, we analyze the variance of $n_i$, denoted $\tilde{\chi}(\gamma, Q)$. Since $n_i$ is a Bernoulli variable, its variance is given by
\begin{equation}\label{Eq:first_order_sus}
    \tilde{\chi} = \mathbb{P}_{\epsilon}\left(1-\mathbb{P}_{\epsilon}\right) \xrightarrow[Q \to \infty]{}  \Theta(\gamma-\gamma_{c})\Theta(\gamma_{c}-\gamma)\,,
\end{equation}
where we used Eq.~\eqref{Eq:First_order_form} and $\Theta(.)$ denotes the Heaviside function with $\Theta(0) = 1/2$. Fig.~\ref{fig:emer_cyc1_sus} confirms that the variance away from $\gamma_{c}$ decreases with increasing $Q$, consistent with this expectation.}

\begin{figure}
    \centering
    \includegraphics[scale=.5]{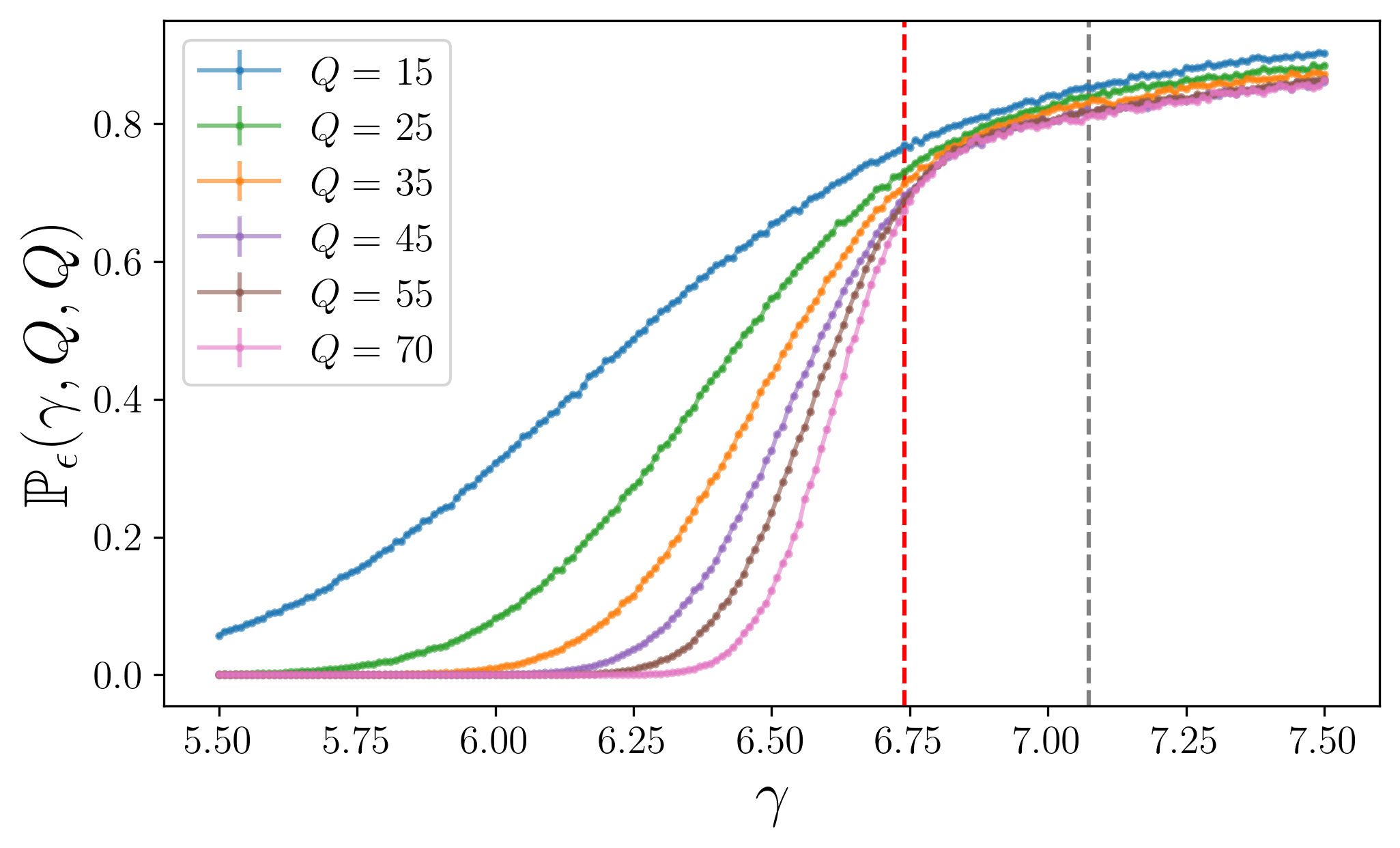}
    \caption{Estimate of the probability of having an emergent cycle when three random species involved in the CRN are chemostatted plotted as a function of the rescaled probability $\gamma$. The parameters are $N_{A} = 2$ and $Q_{1} = Q_{2} = Q$. The grey dashed line is the values of $\gamma^{*}$ obtained numerically following Eq.~\eqref{Eq:claw_hyp2_2atom_final}. The red line is the accurate threshold $\gamma_{c}$ obtained via a scaling analysis. For each value of $\gamma$, for $Q = 15$ till $Q = 45$ , we generated 32000 reaction sets for averaging. For all other $Q$, we generated 16000 reaction sets for averaging. The sample error of the mean is shown by errorbars which are not visible on the plot.}
    \label{fig:emer_cyc2}
\end{figure}

\begin{figure}
    \centering
    \includegraphics[scale=.5]{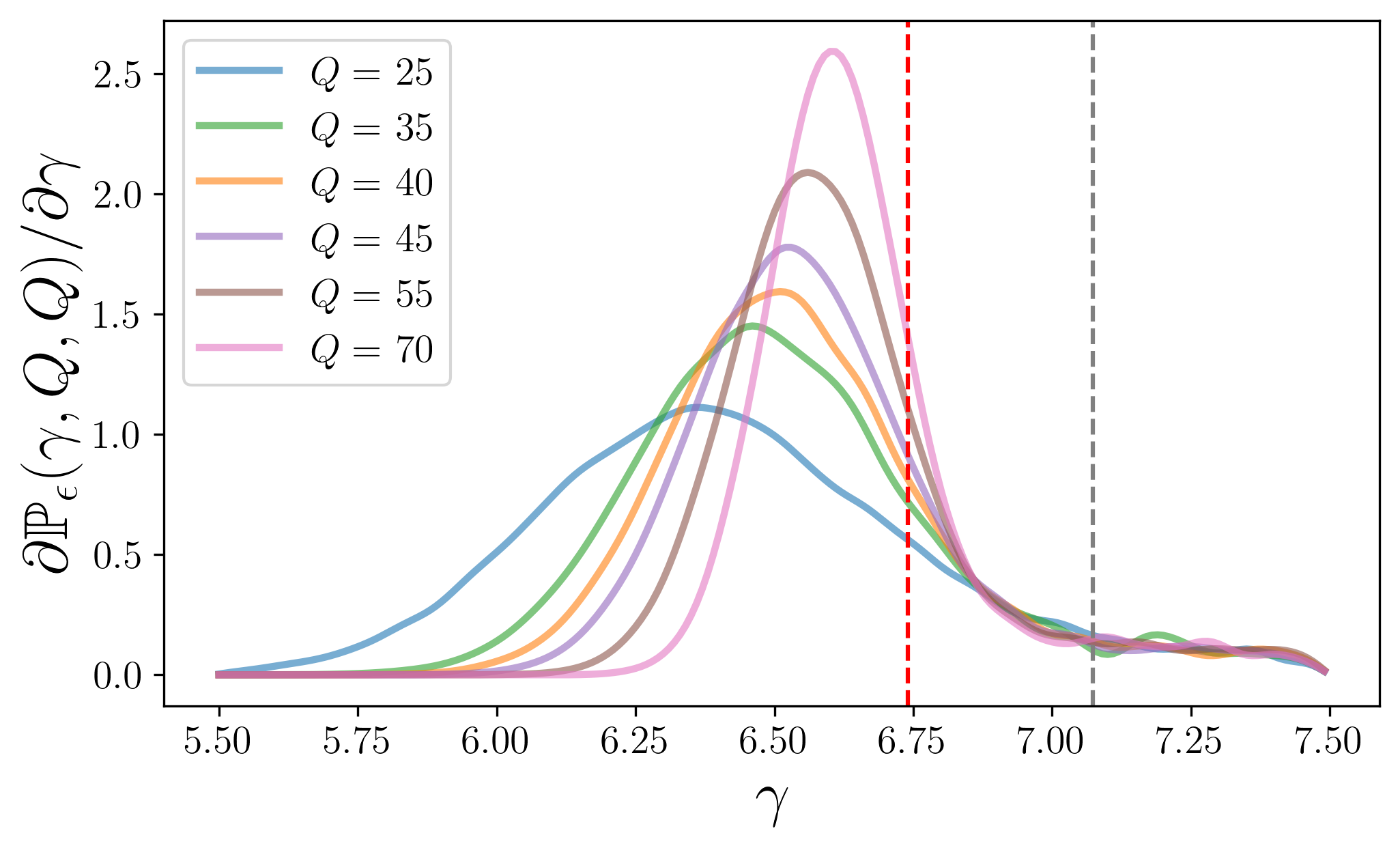}
    \caption{Derivative of the probability in Fig.~\ref{fig:emer_cyc2} of having an emergent cycle when three random species involved in the CRN are chemostatted wrt $\gamma$ plotted as a function of the rescaled probability $\gamma$. The parameters are $N_{A} = 2$ and $Q_{1} = Q_{2} = Q$. The red line is the values of $\gamma_{c}$ obtained numerically. The grey dashed line is the values of $\gamma^{*}$ obtained numerically following Eq.~\eqref{Eq:claw_hyp2_2atom_final}. The data in Fig.~\ref{fig:emer_cyc2} was smoothened via a Gaussian filter before taking a numerical derivative.}
    \label{fig:emer_cyc2_der}
\end{figure}

We now discuss the case of two atoms. For $N_{A} = 2$, we plot in Fig.~\ref{fig:emer_cyc2} the probability of obtaining an emergent cycle when chemostatting three species as a function of the rescaled probability $\gamma$, for different values of $Q_{1} = Q_{2} = Q$. As in the case of the single atom, at low $\gamma$, the probability is nearly zero and increases after some critical probability $\gamma_{c}$. However, in contrast to Fig.~\ref{fig:emer_cyc1}, the curves in Fig.~\ref{fig:emer_cyc2} do not intersect.
In fact, the data in Fig.~\ref{fig:emer_cyc2} indicates that for $N_{A} = 2$, the probability $\mathbb{P}_{\epsilon}(\gamma, Q_{1}, Q_{2})$ undergoes a \textit{continous transition}, i.e, the probability is zero below $\gamma_{c}$ and increases continuously afterward. Near the threshold, we expect the probability to follow a power law: 
\begin{equation}\label{Eq:power_law}
   \lim_{Q \to \infty} \mathbb{P}_{\epsilon}(\gamma, Q, Q) \propto |\gamma - \gamma_{c}|^{b} \,.
\end{equation}
Furthermore, in analogy with standard finite-size scaling near continuous transitions, we expect deviations from the limiting behavior to follow a universal form~\cite{staufferbook}. Specifically, curves for different $|\mathcal{Z}^{\max}|$ should collapse onto a master curve when suitably rescaled:
\begin{equation}\label{Eq:data_collapse}
    \mathbb{P}_{\epsilon}(\gamma, Q, Q)(|\mathcal{Z}|^{\max})^{\frac{b}{a}} = f((\gamma - \gamma_{c})(|\mathcal{Z}|^{\max})^{\frac{1}{a}})\,.
\end{equation}
Using Eq.~\eqref{Eq:data_collapse}, following the method outlined in Refs.~\onlinecite{houdayerhartmann2004, melchert2009}, we extract the critical exponents: $\gamma_{c} = 6.74 \pm 0.01$, $a = 2.219 \pm 0.131$ and $b= 0.095 \pm 0.005$. The resulting data collapse is shown in Appendix~\ref{Subsec:data_emer2D}, along with details of the methodology (see Figs.~\ref{fig:emer_cyc2_collapse}–\ref{fig:emer_cyc2_der_pl2}). To further support this picture, we analyze the derivative $\partial \mathbb{P}_\epsilon(\gamma, Q, Q)/\partial \gamma$.  If the probability $\mathbb{P}_{\epsilon}(\gamma, Q, Q)$ follows a power law with an exponent $b < 1$, then the derivative must diverge at $\gamma_{c}$. Fig.~\ref{fig:emer_cyc2_der} shows the derivative (smoothed with a Gaussian filter due to numerical noise), which becomes progressively sharper with increasing $Q$, consistent with this expectation. Let $\gamma_r$ denote the location of the peak in the derivative for a given $Q$. From Eq.~\eqref{Eq:power_law} and Eq.~\eqref{Eq:data_collapse}, the peak height can be expected to scale as
\begin{equation}\label{Eq:Prob_2D_der_scaling}
    \frac{\partial \mathbb{P}_\epsilon(\gamma, Q, Q)}{\partial \gamma}\Bigg{|}_{\gamma = \gamma_r} \propto (|\mathcal{Z}|^{\text{max}})^{\frac{1-b}{a}}\,.
\end{equation}
As shown in Appendix~\ref{Subsec:data_emer2D}, Eq.~\eqref{Eq:Prob_2D_der_scaling} fits the data well, with an exponent: $0.419 \pm 0.004$, close to the predicted value $0.408 \pm 0.024$ obtained from the scaling exponents. Further, in the limit $|\mathcal{Z}^{\max}| \to \infty$, we expect $\gamma_r \to \gamma_c$. We also analyze how the peak location $\gamma_r$ approaches $\gamma_c$ as $|\mathcal{Z}^{\max}|$ increases  in Appendix~\ref{Subsec:data_emer2D}. The data are consistent with the scaling: $|\gamma_{r} - \gamma_{c}| \propto |\mathcal{Z}^\text{max}|^{\frac{-1}{a}}$ with a fitted exponent $1/a = 0.499 \pm 0.011$, close to the expected value $0.451 \pm 0.003$. Note that the exponents $a$ and $b$ cannot be directly compared with known universality classes in percolation theory, as our order parameter  $\mathbb{P}_{\epsilon}(\gamma, \{Q_{1},...Q_{A}\})$ differs from the standard description of percolation in terms of the size of the largest connected component~\cite{staufferbook} and does not have a simple geometric interpretation.

While the transitions for $N_A = 1$ and $N_A = 2$ differ as the former is discontinuous while the latter is continuous, in both cases the system undergoes a qualitative change from a disconnected to a connected regime. We therefore refer to both as percolation-like transitions.

\begin{figure}
    \centering
    \includegraphics[scale=.48]{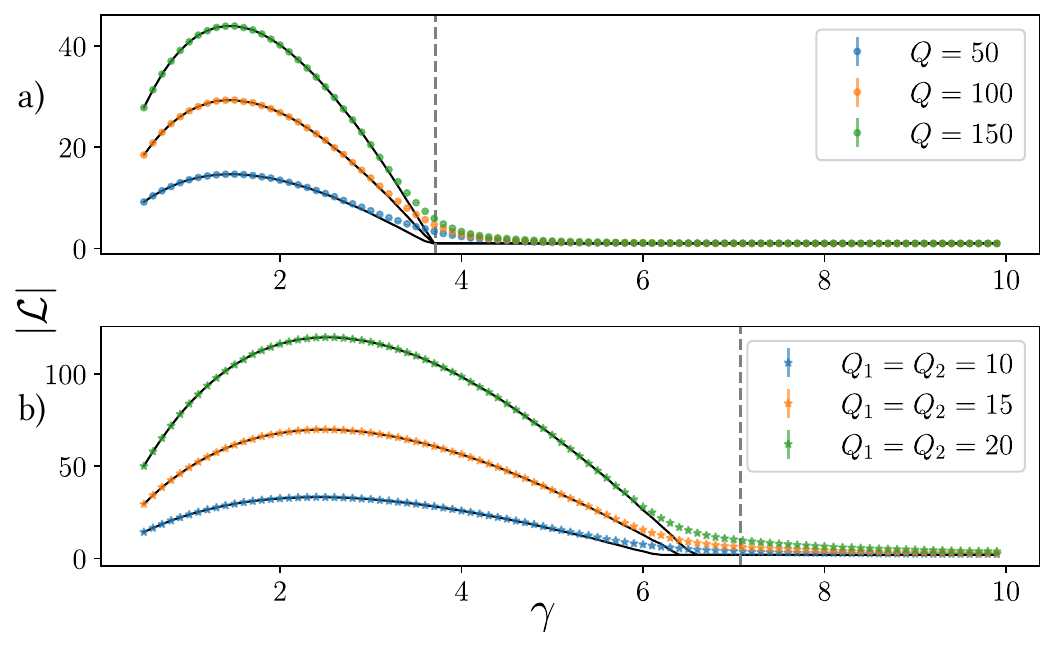}
    \caption{a) Average number of conservation laws as a function of the probability to select a reaction $\gamma$ compared with the theoretical result of Eq.~\eqref{Eq:claw_hypothesisb} (solid black line).  The parameters are $N_A=1$. The gray dashed line indicates the value of $\gamma^{*}$ obtained from Eq.~\eqref{Eq:claw_hyp2_1atom_univ}. b)  Average number of conservation laws as a function of the probability to select a reaction $\gamma$ compared with the theoretical result of Eq.~\eqref{Eq:claw_hypothesisb} (solid black line).  The parameters are $N_A=2$ and $Q_1=Q_2=Q$. The gray dashed line indicates the value of $\gamma^{*}$ obtained from Eq.~\eqref{Eq:claw_hyp2_2atom_final}. For both subplots, for each value of $Q$ and $p$ we generated $\sim 10000$ reaction sets for averaging. Errorbars indicate sample error of the mean and are not visible in the plot.}
    \label{fig:claw_hyp1}
\end{figure}

We now turn to an analytical perspective based on the conservation laws. Indeed, as Eq.~\eqref{Eq:emer_broken} showed, the number of emergent cycles is linked to the number of (broken) conservation laws. Clearly, if there are multiple conservation laws, creating an emergent cycle is difficult and vice-versa. 
We compute the average number of conservation laws as a function of the rescaled probability $\gamma$, as shown in Fig.~\ref{fig:claw_hyp1}.
We see that the average number of conservation laws initially increases and reaches a maximum value before rapidly falling to the value $N_A$. This nonmonotonic behavior is observed for both different values of $Q_1 \cdots Q_{N_A}$ and different values of $N_{A}$.

Intuitively, for low values of $p$, a very low fraction of the possible reactions are selected. With each reaction added, we can consider that two or three new species are added to the network, since the probability of a given species being selected twice scales as $p^2$ for low $p$ (see Eq. \ref{eq:degree_prob}). Thus, in the limit $p\to 0$ the generated CRN will have more species than reactions, and these reactions will be independent (i.e., they will not share any species). Thus, each added reaction will typically add a conservation law.
In this regime, chemostatting different species thus typically breaks different conservation laws.

For high values of $p$ almost all possible reactions, and consequently almost all species, will be included in the network. Thus, it is almost certain that no group of atoms will be conserved by all the reactions, except the individual atoms themselves. Therefore, the number of independent conservation laws must tend to $N_A$  (see Appendix~\ref{ap:proof_basis}). 
Consequently, chemostatting $N_{A} + 1$ species will result in an emergent cycle with high probability.

In the following, we show how the previous arguments can be used to analytically estimate the percolation threshold. 

\subsubsection{Estimating the number of conservation laws}\label{Sec:rank_arguments}
A simple bound on the number of average conservation laws is obtained from the rank of the stoichiometric matrix.
Indeed, using Eq.~\eqref{Eq:rank_nullity} we note that the rank of the stoichiometric matrix is always upper bounded by the minimum of the number of selected species and reactions, i.e,
\begin{equation}\label{Eq:rank_bound}
    \langle \text{rk}(\nabla)\rangle \leq \text{min}\left[\left<|\mathcal{R}|\right>, \langle|\mathcal{Z}|\rangle\right]\,.
\end{equation}
This implies that the average number of conservation laws satisfies $\langle |\mathcal{L}| \rangle \geq \langle|\mathcal{Z}|\rangle - \text{min}\left[\left<|\mathcal{R}|\right>, \langle|\mathcal{Z}|\rangle\right] $ (see Eq. \eqref{Eq:rank_nullity}). This bound can be broken down further:
On one hand, at small $p$, 
while both $\left<|\mathcal{R}|\right>$ and $\langle|\mathcal{Z}|\rangle$ scale linearly with $p$ (see Eqs.~\eqref{Eq:exp_sp_small} and \eqref{Eq:Avg_rxn}), since $\sum_{\sigma} I(\sigma) > |\mathcal{R}|^{\text{max}}$, the rank must be upper bounded by the number of reactions. 
On the other hand, at large $p$, $\langle|\mathcal{Z}|\rangle$ saturates (see Fig.~\ref{fig:sp_dist}) while the average number of reactions keeps increasing, implying that the rank must then be upper bounded by the number of species. 
Thus, we obtain a lower bound on the number of conservation laws
\begin{equation}\label{Eq:claw_hypothesisa}
    \langle |\mathcal{L}| \rangle \geq \begin{cases}
         \langle|\mathcal{Z}|\rangle - \left<|\mathcal{R}|\right>  ~~ & p \leq p_{1} \\
        0    ~~ &p> p_{1}\,.
     \end{cases} 
\end{equation}
where $p_{1}$ is a crossover probability. Actually, the previous bound can be tightened by just noting that for large $p$ the number of conservation laws is always larger than the number of atomic types $N_A$. Therefore, 
\begin{equation}\label{Eq:claw_hypothesisb}
    \langle |\mathcal{L}| \rangle \geq \begin{cases}
         \langle|\mathcal{Z}|\rangle - \left<|\mathcal{R}|\right>  ~~ & p \leq p^* \\
        N_A    ~~ &p> p^*\,.
     \end{cases} 
\end{equation}
The crossover probability $p^*$ is obtained by equating the different cases:
\begin{equation}\label{Eq:cross_prob}
     \langle|\mathcal{Z}|\rangle(p^*) - \langle|\mathcal{R}|\rangle(p^*) = N_A.
\end{equation}
Fig.~\ref{fig:claw_hyp1} shows that the above expression captures the nonmonotonicity in the number of conservation laws very well. 

\begin{figure}
    \centering
    \includegraphics[scale=.5]{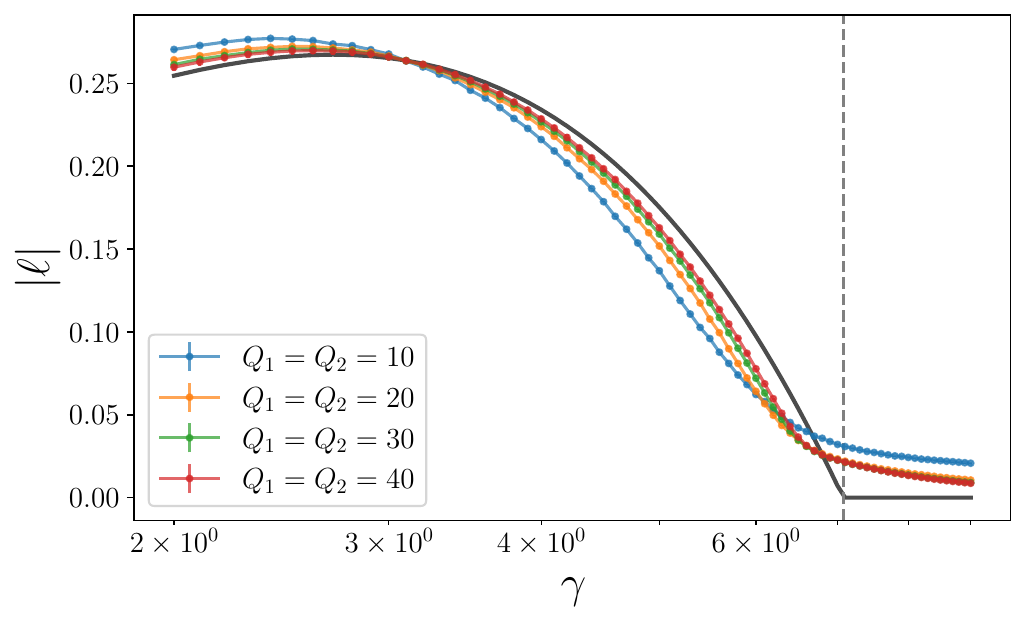}
    \caption{Scaled average number of conservation laws $\langle |{\ell}| \rangle$ for $N_{A} = 2$ plotted as a function of $\gamma$ and compared with the theoretical result in Eq.~\eqref{Eq:claw_hyp2_2atom_final} (black line). The grey dashed line is the location of $\gamma^{*}$ in Eq.~\eqref{Eq:claw_hyp2_2atom_final}, obtained numerically. For each data point, we generated $10000$ reaction sets for averaging. The sample error of the mean is shown by errorbars which are not visible in the plot.}
    \label{fig:ell_2D}
\end{figure}

To analytically obtain $p^{*}$, we work in the extensive limit using Eq.~\eqref{Eq:prob_rescaled}. In first place we derive a universal form for the behavior of the rescaled average number of conservation laws, defined as:
\begin{equation}\label{Eq:Scale_avg_defn}
    \langle|{\ell}|\rangle = \langle |\mathcal{L}| \rangle /|\mathcal{Z}|^{\text{max}}\,.
\end{equation}
From Eq.~\eqref{Eq:claw_hypothesisb}, we have the following expression for the scaled number of conservation laws,
\begin{equation}\label{Eq:claw_hyp2}
\langle |\ell| \rangle \geq \begin{cases}
         \langle|z|\rangle - \left<|r|\right>  ~~ & \gamma \leq \gamma^{*} \\
         N_{A}/ |\mathcal{Z}|^{\text{max}}   ~~ &\gamma > \gamma^{*}\,,
     \end{cases} 
\end{equation} 
where we have used lowercase letters to denote rescaled quantities. We now evaluate each quantity in Eq.~\eqref{Eq:claw_hyp2}. Using Eqs.~\eqref{eq:NR_max_asymp} and \eqref{Eq:Avg_rxn}, we can write, 
\begin{equation}\label{Eq:scaled_rxn_limit}
    \left<|r|\right> =  \frac{|\mathcal{R}|^\text{max}}{(|\mathcal{Z}|^{\text{max}})^2} \gamma \ \xrightarrow[|\mathcal{Z}|^{\text{max}} \to \infty]{}  \frac{\gamma}{2^{N_{A}+1}}\,.
\end{equation}
Similarly, using Eq.~\eqref{Eq:exp_species}, the rescaled average number of species is given by
\begin{equation}\label{Eq:scaled_sp_gen}
    \left<|z|\right> = 1 - \frac{\mu(0)}{|\mathcal{Z}|^{\text{max}}}\,.
\end{equation}
{
By using Eq.~\eqref{Eq:deg_distri_full} for the special case of $k = 0$, we have}
\begin{equation}\label{Eq:mu0_final}
\begin{split}
    \frac{\mu(0)}{|\mathcal{Z}|^{\text{max}}} &= \int_{0}^{1} \! dx_1.. \! \int_{0}^{1} \! dx_{N_A} \: e^{-\gamma\Bigl(\frac{1}{2}\prod_{k = 1}^{N_A}x_{k} + \prod_{j = 1}^{N_{A}} (1 - x_{j})\Bigr)} \\
    &\quad + \mathcal{O}~\!\Bigl(\frac{1}{Q_{a}}\Bigr).
\end{split}
\end{equation}
In the limit of all the $Q_{i} \to \infty$, substituting Eqs.~\eqref{Eq:mu0_final}, ~\eqref{Eq:scaled_sp_gen} and ~\eqref{Eq:scaled_rxn_limit} in Eq.~\eqref{Eq:claw_hyp2}, we get a bound on $\langle |\ell| \rangle$ of the form:
\begin{equation}\label{Eq:claw_hyp_gen}
\langle |\ell| \rangle \geq \begin{cases}
         1 - \frac{\mu(0)}{|\mathcal{Z}|^{\text{max}}} - \frac{\gamma}{2^{N_{A}+1}}  ~~ & \gamma \leq \gamma^{*} \\
         0   ~~ &\gamma > \gamma^{*}\,.
    \end{cases} 
\end{equation} 
Note that the critical value $\gamma^{*}$ and the form of the universal curve in Eq.~\eqref{Eq:claw_hyp_gen} depend only on the number of atoms.  
For the case of one atom, Eqs.~\eqref{Eq:claw_hyp_gen} and \eqref{Eq:mu0_final} reduce to
\begin{equation}\label{Eq:claw_hyp2_1atom_univ}
\langle |\ell| \rangle \geq \begin{cases}
        1 - \frac{2}{\gamma}e^{-\gamma}(e^{\gamma/2} -1) - \frac{\gamma}{4}  ~~ & \gamma \leq \gamma^{*} \\
         0   ~~ &\gamma > \gamma^{*}\,,
     \end{cases} 
\end{equation} 
where the value $\gamma^{*} \approx 3.72$ is obtained numerically by equating the two possible cases. {
Clearly $\gamma^{*}$ is an estimate of the percolation threshold and we see that the value of $\gamma^{*} = 3.72$ differs from the actual percolation threshold $\gamma_{c} = 3.62$ by only about $3\%$~(see the red line and the grey line in Fig.~\ref{fig:emer_cyc1}).} 
\begin{figure}
    \centering
    \includegraphics[scale=.5]{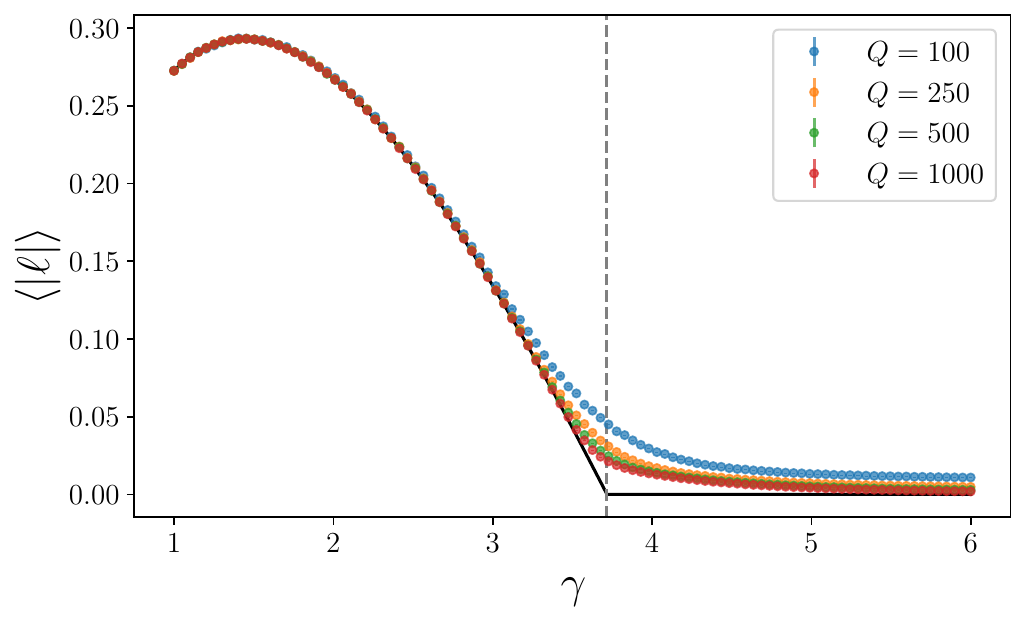}
    \caption{Scaled average number of conservation laws $\langle |{\ell}| \rangle$ for $N_{A} = 1$ plotted as a function of $\gamma$ and compared with the theoretical result in Eq.~\eqref{Eq:claw_hyp2_1atom_univ} (black line). The grey dashed line is the location of $\gamma^{*}$ in Eq.~\eqref{Eq:claw_hyp2_1atom_univ}, obtained numerically. For each data point, we generated $10000$ reaction sets for averaging. The sample error of the mean is shown by errorbars which are not visible in the plot.}
    \label{fig:ell_larged}
\end{figure}
We compare Eq.~\eqref{Eq:claw_hyp2_1atom_univ} with simulations in Fig.~\ref{fig:ell_larged}. We see that a very good agreement is obtained for large $Q$ and $\gamma < \gamma^*$. For the case of two atoms, Eqs.~\eqref{Eq:claw_hyp_gen} and \eqref{Eq:mu0_final} reduce to
\begin{equation}\label{Eq:claw_hyp2_2atom_final}
\langle |\ell| \rangle \geq \begin{cases}             \begin{multlined}[t]
        1 - \frac{\gamma}{8}
        - \frac{e^{-\gamma/3}}{\gamma}\bigg[\frac{4\text{Ei}(\gamma/3)}{3}\\ -\frac{2\text{Ei}(-2\gamma/3)}{3} - \frac{2\text{Ei}(-\gamma/6)}{3}\bigg]\end{multlined}   & \gamma \leq \gamma^{*} \\ \\ 
         0  &\gamma > \gamma^{*}\,,
     \end{cases} 
\end{equation}
where $\text{Ei}(.)$ refers to the exponential integral function and the critical value $\gamma^{*} \approx 7.07$ is obtained numerically. 
Once again, the difference between the actual threshold for $N_{A} = 2$, which is $\gamma_{c} = 6.74$ (see redline in Fig.~\ref{fig:emer_cyc2}) and $\gamma^{*} = 7.07$ (grey line in Fig.~\ref{fig:emer_cyc2}) is about $5\%$. We compare Eq.~\eqref{Eq:claw_hyp2_2atom_final} with simulations in Fig.~\ref{fig:ell_2D} for $Q_{1} = Q_{2} = Q$. We see that as $Q$ increases, the curves approach Eq.~\eqref{Eq:claw_hyp2_2atom_final}. However, since the $Q$ we use are quite small, the corrections to Eq.~\eqref{Eq:claw_hyp2_2atom_final} of order $\mathcal{O}(1/Q)$ ensure that the convergence to the regime indicated by the universal curve is quite slow.

\begin{figure}
    \centering
    \includegraphics[scale=.50]{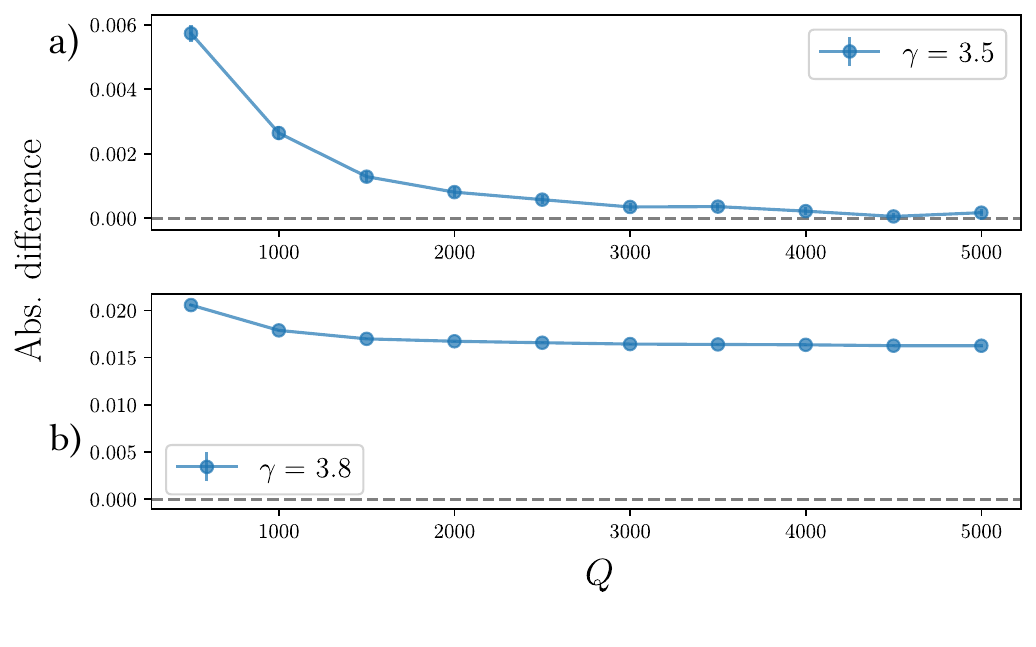}
    \caption{Comparing the absolute difference of the simulated $\langle|\ell| \rangle(Q)$ at different values of $Q$ and the theoretical prediction of the same in Eq.~\eqref{Eq:claw_hyp2_1atom_univ} at the values: a) $\gamma = 3.5$ and 
     b) $\gamma = 3.8$. For both graphs, $N_{A} = 1$. For each data point, we generated $10000$ reaction sets for averaging. The sample error of the mean is shown by errorbars.}
    \label{fig:mean_descent}
\end{figure}

It is tempting to believe that the previous results, along with Figs.~\ref{fig:ell_larged} and \ref{fig:ell_2D}, show that there is a transition in the number of conservation laws at $\gamma = \gamma^{*}$. However, this is inaccurate. We show this in the case of $N_{A} = 1$ in Fig.~\ref{fig:mean_descent}. By numerically estimating the value of $\langle|\ell| \rangle(Q)$ for different values of $Q$, we see that for values like $\gamma = 3.5$, we have convergence to the predicted value whereas for values such as $\gamma = 3.8$, we see convergence to a nonzero value, as depicted in Fig.~\ref{fig:mean_descent}.
The second argument against the presence of a phase transition in the number of scaled conservation laws comes from studying the analogous quantity for ER graphs, which is given by the scaled number of connected components. As shown in Appendix~\ref{Sec:ER_appendix}, the scaled number of connected components is a continuous function of $\gamma$, even at $\gamma_c = 1.0$, the critical value for the giant component transition. In fact, repeating the analysis of this section for ER graphs, one finds that the threshold predicted by this approach would be $\gamma^{*} = 1.49$. Thus, we conclude that, while there is an underlying percolation transition that is captured by this ansatz, it does not occur precisely at $\gamma^{*}$.

{
\subsubsection{Finding the conserved moieties} \label{Sec:roots}
While the arguments about the average rank of the stoichiometric matrix in Secs.~\ref{Sec:rank_arguments}  allow us to predict the average number of conservation laws, 
it is not clear which specific moieties are conserved in the low $p$ regime.
In this subsection, we identify the conserved moieties by introducing \textit{root} species.

For convenience, we orient all reactions to be like an association, i.e, like Eq.~\eqref{Eq:association_rxn}. 
Note that this reorientation of reactions is always possible as we are dealing with reversible reactions.
Denoting the product species of reaction $\rho$ by $\sigma_{p(\rho)}$ and the reactant species as  $\sigma_{r(\rho)},\sigma_{r'(\rho)}$,  we rewrite all reactions $\rho$ of the form of Eq.~\eqref{Eq:rxn_chem} as 
\begin{equation}\label{Eq:rxn_redefined}
    \sigma_{r(\rho)} + \sigma_{r'(\rho)} \xrightleftharpoons[]{\rho} \sigma_{p(\rho)}\,.
\end{equation}
Consider a random CRN generated with all reactions written in the form of Eq.~\eqref{Eq:rxn_redefined} with the corresponding stoichiometric matrix $\nabla$. 
Define the \textit{root} species (labeled $\hat{\sigma}$) which are reactants in every reaction that they are in, i.e, any reaction involving roots would be of the form,
\begin{align}
   \hat{\sigma}_{r(\rho)} + \hat{\sigma}_{r'(\rho)}&\xrightleftharpoons[]{\rho} \sigma_{p(\rho)}\,,\label{Eq:reac1_root} \\
   \hat{\sigma}_{r(\rho)} + \sigma_{r'(\rho)}     &\xrightleftharpoons[]{\rho} \sigma_{p(\rho)}\,.\label{Eq:reac2_root}
\end{align}
where we denoted arbitrary non-root species with the labels $\sigma$. 
Operationally, given a stoichiometric matrix $\nabla$ with reactions oriented as in Eq.~\eqref{Eq:rxn_redefined}, the root species have rows with only nonpositive entries. We direct the readers to Appendix~\ref{Sec:root_apeg} for an example of a CRN and the corresponding root species.

Appendix~\ref{Sec:ap_root_proof} shows that starting from the root species we can construct every non-root species in the CRN.
Thus, they play the role of effective atoms and we can expect that moieties corresponding to the roots should be conserved.
However, not all the root species are independent. 
Indeed, if we identify the root species as the chemostatted species, it is possible that some root species can be constructed from others,  i.e., that there are emergent cycles. 
Putting both these ideas together, we show (see Appendix~\ref{Sec:ap_root_proof}) that 
\begin{equation}\label{Eq:root_res}
    |\mathcal{Z}^{r}| = |\mathcal{L}| + |\epsilon|_{r} \   \,,
\end{equation}
where we denote the number of root species in the CRN as $|\mathcal{Z}^{r}|$, and $|\epsilon|_{r}$ is the number of emergent cycles when the root species are chosen as the chemostatted species. 
Eq.~\eqref{Eq:root_res} has remarkable implications:
Firstly, since it holds for any realization of a random CRN, the averages must satisfy:
\begin{equation}\label{Eq:claw_upper_bound}
    \langle |\mathcal{L}| \rangle \leq \langle |\mathcal{Z}^{r}| \rangle  \,.
\end{equation}
We thus obtain an upper bound on the average number of conservation laws ({also see Eq.~\eqref{Eqn:conserv_law_bound} for a closed form expression of Eq.~\eqref{Eq:claw_upper_bound} when $N_{A} = 1$}).
Secondly, from Eq.~\eqref{Eq:root_res}, if we can choose a subset of the root species that cannot be constructed from one another, then the number of these \textit{independent} roots must be equal to the number of conservation laws. 
{
Furthermore, in Appendix~\ref{Sec:ap_root_proof}, we show how to identify the independent roots and provide a constructive procedure to build a basis of the conservation laws, given the set of independent roots}, thus completing the identification of the independent roots with conserved moieties.
This tells us the origin of the multiple conservation laws at low $p$. 
In this regime, it is more probable that many species with low atom numbers are involved in reactions that construct species with higher atom numbers, as opposed to being involved in reactions that dissociate them.
They would thus be roots and create conservation laws. 
However, as we increase $p$, either the newly added reactions are dissociations of the root species or they start creating emergent cycles between the roots, reducing the number of conservation laws.
}

\subsection{Connectivity II: Forward reachability}\label{Sec:reach}
We now discuss the notion of forward reachability as introduced in Sec.~\ref{Sec:synth_intro}.
We reiterate that the notion of reachability is similar to the idea of an out-component (i.e, the set of nodes which can be reached starting from a given node) in directed networks, where every edge has a directionality associated with it. 
Analogous to studies of directed networks~\cite{newmanRandomGraphsArbitrary2001, dorogovtsevdirected2001}, we investigate reachability by studying the behavior of the largest (rescaled) forward reachable set, denoted $|{f}|$ as we change $\gamma$, i.e, 
\begin{equation}\label{Eq:largest_reach_set}
    |f|(\gamma, Q) = \frac{\text{max}_{\sigma}|\mathcal{F}(\sigma)|}{|\mathcal{Z}|^{\text{max}}}\,.
\end{equation}

\begin{figure}
    \centering
    \includegraphics[scale=.50]{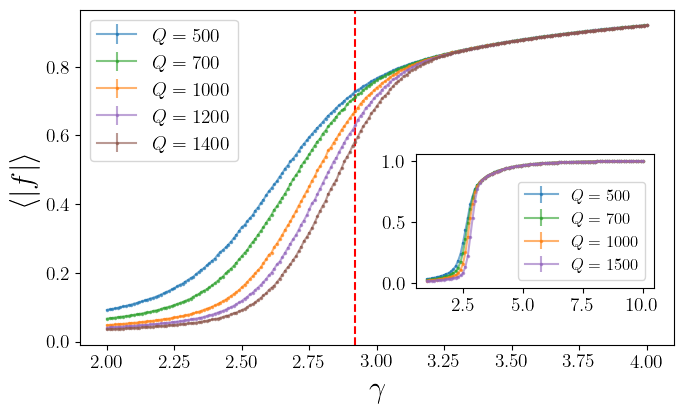}
    \caption{Average size of the rescaled largest reachable set, $\langle|f|\rangle$ as a function of $\gamma$ for  the parameters $N_A=1$ and $Q$. For each value of $Q$ and $\gamma$ we generated $39000$ reaction sets for averaging. Errorbars indicate sample error and are not visible in the plot. The red line is the estimated percolation threshold, $\gamma_{c}$. The inset shows the variation of $\langle|f|\rangle$ over a larger range of $\gamma$. The data for the inset was averaged over $2800$ reaction sets.}
    \label{fig:reach_op_1}
\end{figure}

We first consider the case of $N_{A} = 1$. For each value of $\gamma$ and $Q$, we first generate an ensemble of random CRNs. Then, for every realization,  we compute the reachable sets of each species and record the largest reachable set (see Appendix~\ref{Subsec:data_reach1D} for a description of how to find the reachable set of a species). The resulting average reachability is plotted in Fig.~\ref{fig:reach_op_1} with the inset showing the same quantity $\langle |f| \rangle$ but over a larger range of $\gamma$.
As we increase the number of species in the system, from Fig.~\ref{fig:reach_op_1}, we see that a transition-like scenario emerges: There is a threshold $\gamma$ below which $|f|$ is almost zero and above which $|f|$ increases to a value $\mathcal{O}(1)$.

\begin{figure}
    \centering
    \includegraphics[scale=.50]{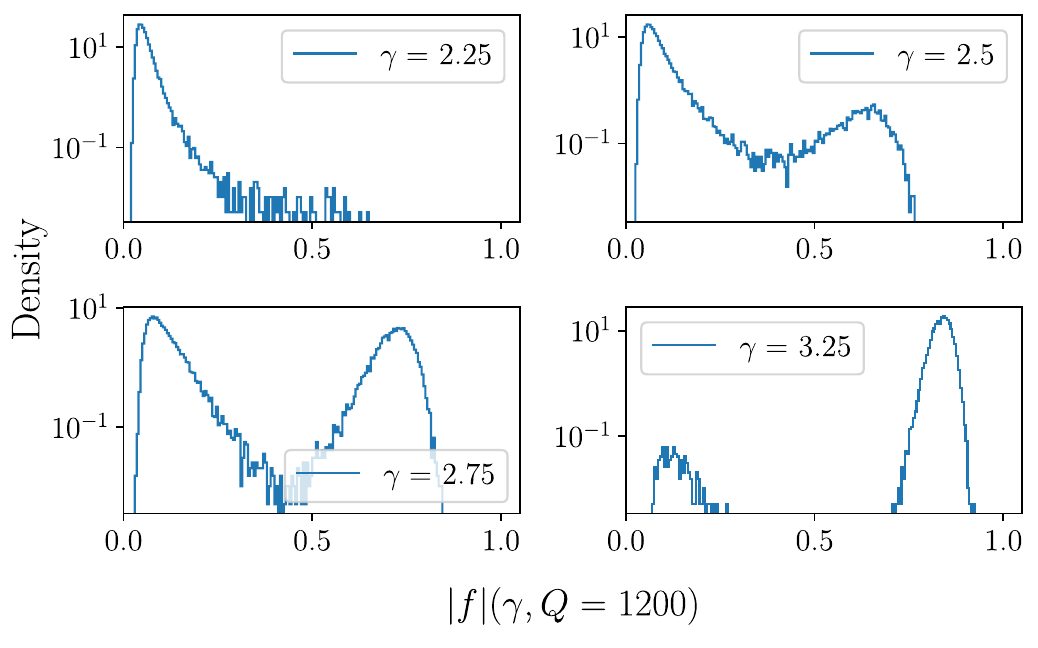}
    \caption{Histogram of $|f|$ for four fixed values of $\gamma$ and with  the parameters $N_A=1$ and $Q = 1000$. For each value of $Q$ and $\gamma$ we generated $39000$ samples and the data was binned in $201$ bins between zero and one and resulting normalized histogram is plotted.}
    \label{fig:reach_hsit1}
\end{figure}

To understand the nature of this transition, we examine the distribution of $|f|$ across realizations. In Fig.~\ref{fig:reach_hsit1}, we plot histograms of $|f|$ for $Q = 1200$ at four different values of $\gamma$. Initially, for low $\gamma$, the histogram is single-peaked with a mean value close to zero. As we increase $\gamma$,  a second peak emerges at a higher value of $|f|$. As $\gamma$ increases, the second peak begins to dominate and eventually the second peak takes over and the first peak disappears. This bimodal structure indicates that independent CRN realizations fluctuate between a disconnected regime (low reachability) and a connected regime (high reachability) and is a hallmark of a discontinuous (first-order) transition\cite{binderfirstordermonte1992}.

\begin{figure}
    \centering
    \includegraphics[scale=.50]{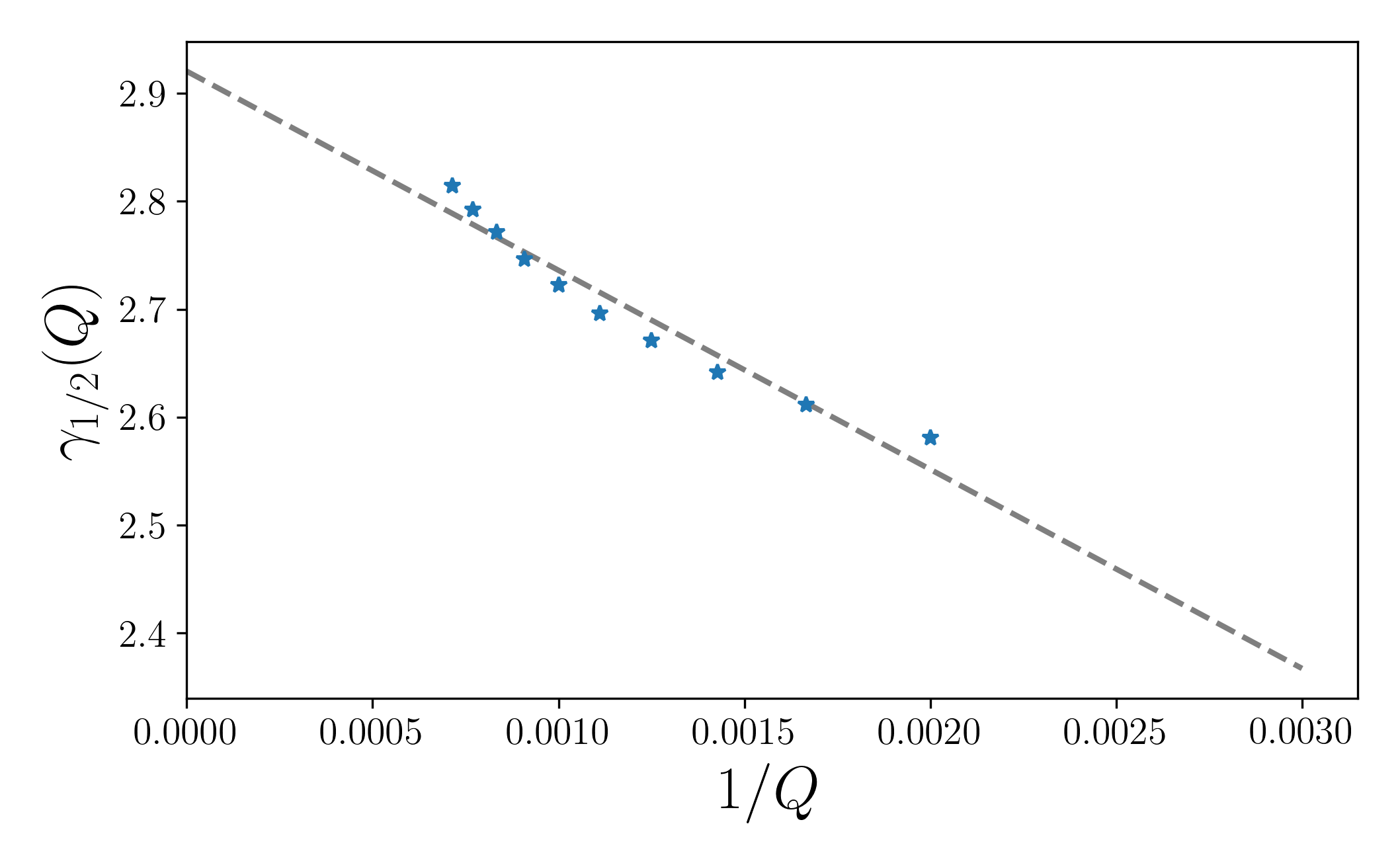}
    \caption{Equal weight probability $\gamma_{1/2}(Q)$ of the order parameter histogram (as in Fig.~\ref{fig:reach_hsit1}) plotted against $1/Q$. The parameters are $N_{A} = 1$. The grey dotted line is a linear fit with intercept $2.92$}
    \label{fig:reach_extrapol1}
\end{figure}

We estimate the critical point $\gamma_{c}$ by finding for each $Q$, the probability at which both peaks of the histogram are equally weighted (see Appendix~\ref{Subsec:data_reach1D} for the procedure), denoted $\gamma_{1/2}(Q)$, and then extrapolating to $Q \to \infty$. The result of this procedure is seen in Fig.~\ref{fig:reach_extrapol1} with the result $\gamma_{c} = 2.92 \pm 0.01$. The extrapolation is a linear fit to $1/Q$ ($r^{2} \approx 0.94$). To further support this picture, we track the locations of the two histogram peaks, denoted by $|f|_{\pm}(Q)$ and confirm that they remain separated as $Q$ increases. The resulting plot is shown in Appendix~\ref{Subsec:data_reach1D}.

\begin{figure}
    \centering
    \includegraphics[scale=.5]{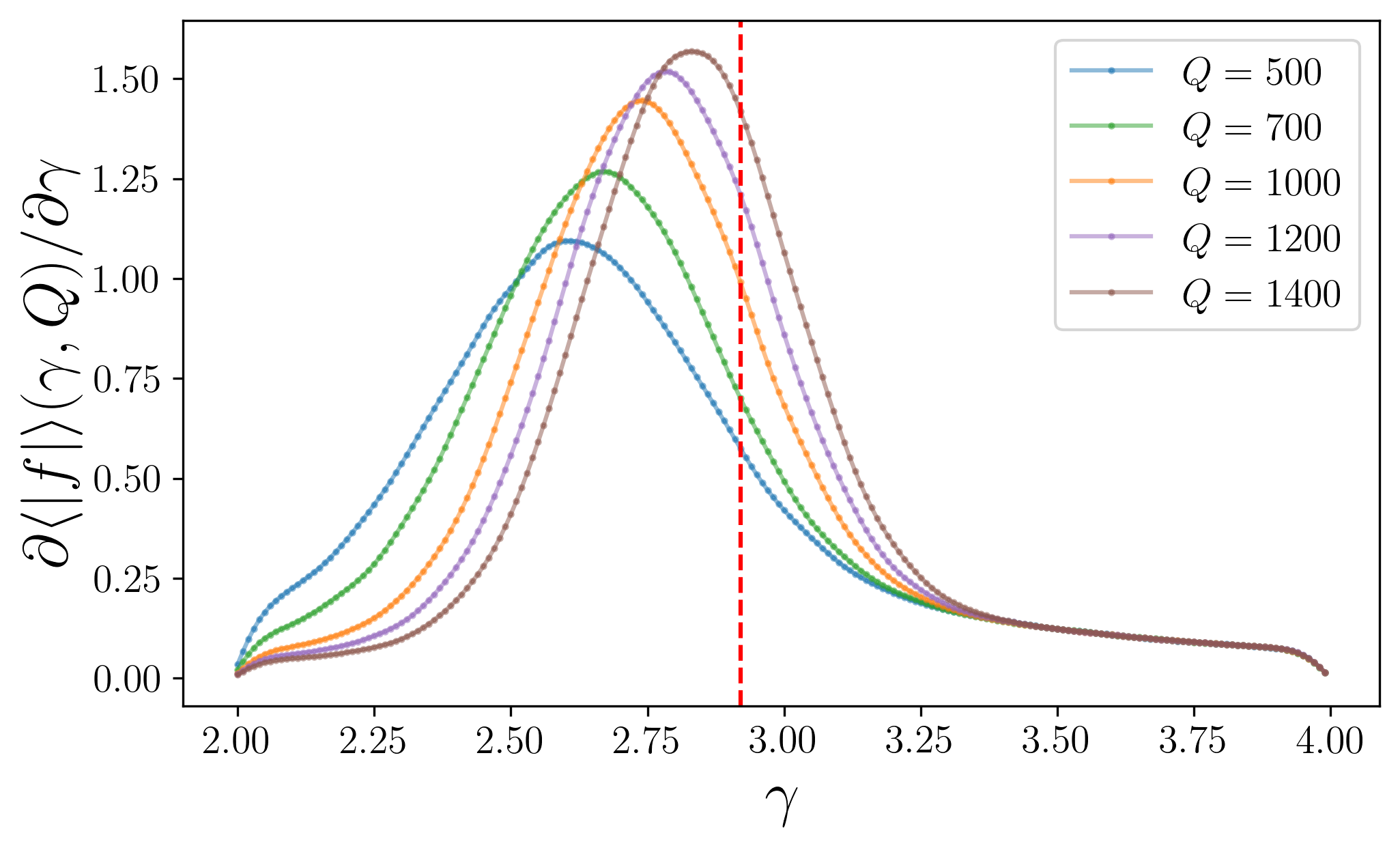}
    \caption{Derivative of the average reachability in Fig.~\ref{fig:reach_op_1} plotted as a function of the rescaled probability $\gamma$. The parameters are $N_{A} = 1$. The red line is the value of $\gamma_{c}$ obtained numerically. The data in Fig.~\ref{fig:reach_op_1} was smoothened via a Gaussian filter before taking a numerical derivative.}
    \label{fig:reach_der}
\end{figure}

In the limit of $Q \to \infty$, a consequence of the shift of the probability weight is a jump in the average reachability as $\gamma$ is varied. However, no discontinuity can be seen since the $Q$ we use for simulations are quite small. To test this, we examine the derivative of the average reachability with $\gamma$, that is, $\partial\langle |f| \rangle/\partial \gamma$. A jump in the average reachability would correspond to a delta-function singularity in its derivative. Fig.~\ref{fig:reach_der} shows the derivative (smoothed with a Gaussian filter due to numerical noise), which becomes progressively sharper with increasing $Q$, consistent with this expectation.  The location of the peak of the derivative, denoted $\gamma_r(Q)$, also extrapolates to $2.90$ as $Q \to \infty$ consistent with the previous estimate (see Appendix~\ref{Subsec:data_reach1D}). 
Further, we note that the maximum of the derivative scales as a power law in $Q$ with an exponent $0.35$, as shown in Appendix~\ref{Subsec:data_reach1D}.
We also analyze the fluctuations of the order parameter, defined as $\tilde{\chi}(\gamma, Q) = \langle |f|^{2} \rangle - \langle |f| \rangle^{2}$.  As shown in Appendix~\ref{Subsec:data_reach1D}, the fluctuations of the reachability for different $Q$ show a peak whose height scales with $Q$ and the location of which moves towards $\gamma_{c}$.

\begin{figure}
    \centering
    \includegraphics[scale=.50]{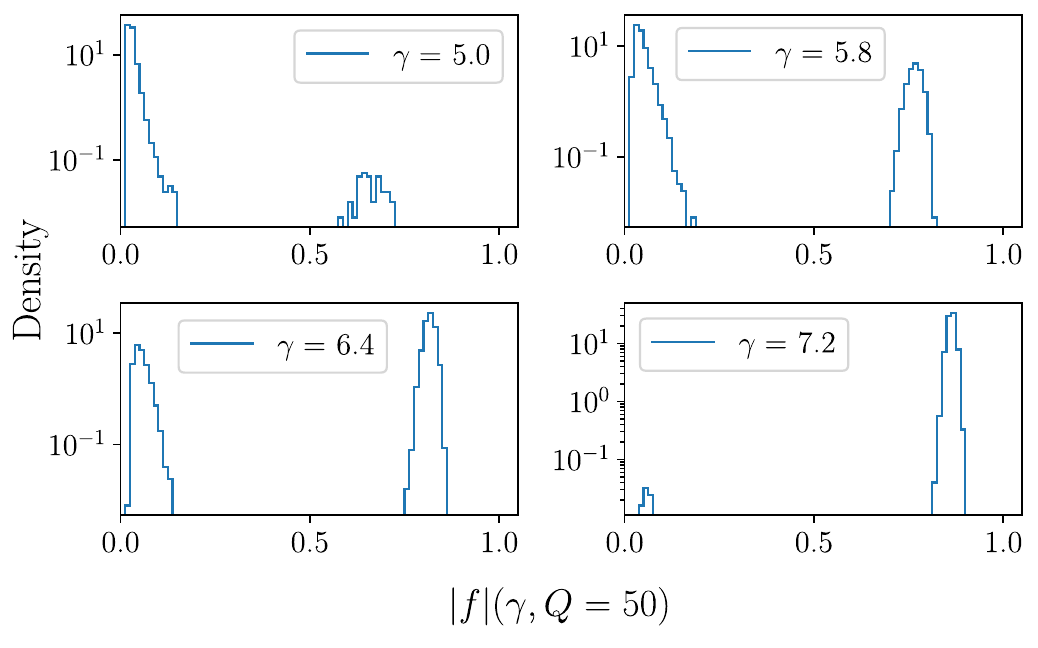}
    \caption{Histogram of $|f|$ for four fixed values of $\gamma$ and with  the parameters $N_A=2$ and $Q_{1} = Q_{2} = 50$. For each value of $Q$ and $\gamma$ we generated $10000$ samples and the data was binned in $81$ bins between zero and one and resulting normalized histogram is plotted.}
    \label{fig:reach_hsit2}
\end{figure}

Finally, we briefly consider the case of $N_{A} = 2$ and investigate the reachability from a single species for the case of $Q_{1} = Q_{2} = Q$.
We again plot the histogram of the order parameter for $Q_{1} = Q_{2} = 50$. The resulting histograms are shown in Fig.~\ref{fig:reach_hsit2}. Once again, we see the emergence of a double-peaked histogram similar to the case of $N_{A} = 1$ (see Fig.~\ref{fig:reach_hsit1}). Thus, we conclude that the reachability transition for $N_{A} = 2$ is discontinuous as in the case of $N_{A} = 1$.

\subsection{Deficiency}\label{Sec:cycles}
\begin{figure}
    \centering
    \includegraphics[scale=0.48]{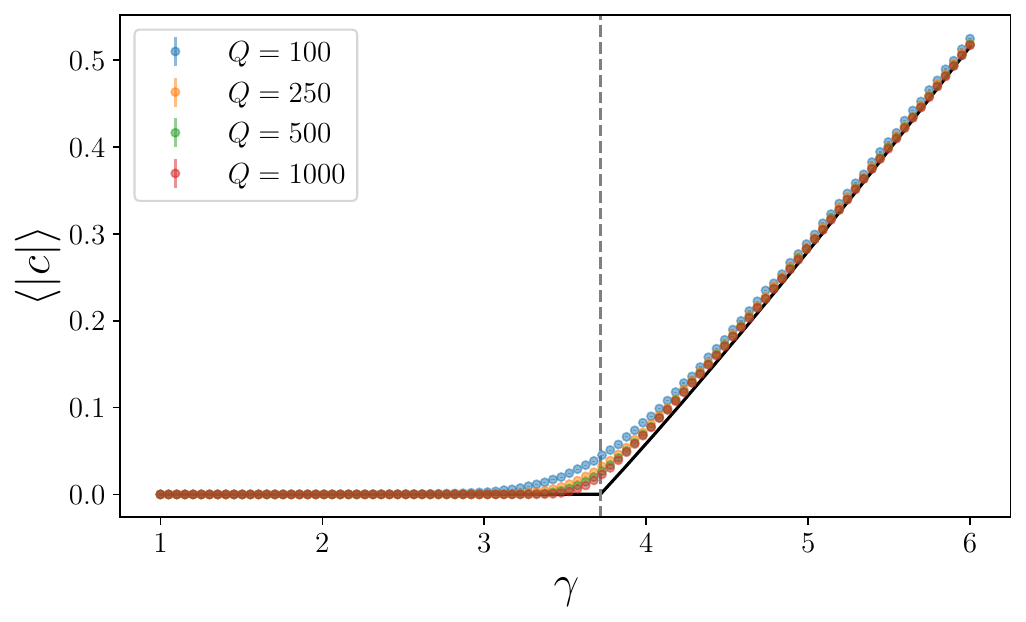}
    \caption{Scaled average number of stoichiometric cycles $\langle |{c}| \rangle$ for $N_{A} = 1$ plotted as a function of $\gamma$ and compared with the theoretical result in Eq.~\eqref{Eq:cycle_hyp_gen} (black line). The grey dashed line is the location of $\gamma^{*}$ in Eq.~\eqref{Eq:claw_hyp2_1atom_univ}, obtained numerically. For each data point, we generated $10000$ reaction sets for averaging. The sample error of the mean is shown by errorbars which are not visible in the plot.}
    \label{fig:avg_cyclesans_2D}
\end{figure}

We now turn to the deficiency, as defined in Sec.~\ref{Sec:Complex_graph}. 
Firstly, for any CRN, constructed with reactions of the form Eq.~\eqref{Eq:rxn_redefined}, we have that $\delta = |\mathcal{C}|$, i.e,  the deficiency is identical to the number of (stoichiometric) cycles.
To see this, we note that the set of complexes can be written as the union of the sets $\mathcal{Z} = \{\sigma\}$ and $\mathcal{Z}_{2} = \{(\sigma,\sigma^{'})\}$. 
The reactions $\rho$ then connect complexes in the set $\mathcal{Z}_{2}$ with complexes in  the set $\mathcal{Z}$ (see Eq.~\eqref{Eq:rxn_redefined}). 
A given complex in $\mathcal{Z}$ can be connected to multiple $\mathcal{Z}_{2}$ complexes. However, there are no connections between complexes within $\mathcal{Z}$ and $\mathcal{Z}_{2}$. Thus, we get a graph with a bipartite structure with the strong restriction that no two complexes of $\mathcal{Z}$ are connected to the same complex in $\mathcal{Z}_{2}$, removing the possibility of graph cycles.
We can thus focus on the statistical properties of the number of stoichiometric cycles. 
Using the rank-nullity theorem Eq.~\eqref{Eq:rank_nullity} along with Eq.~\eqref{Eq:claw_hypothesisb} we lower bound the average number of cycles as
\begin{equation}\label{Eq:cycle_hyp}
\langle |\mathcal{C}| \rangle \geq \begin{cases}
        0   ~~ & p \leq p^{*} \\
        \left<|\mathcal{R}|\right> + N_{A} - \left<|\mathcal{Z}|\right>    ~~ &p> p^{*}\,.
     \end{cases} 
\end{equation}
Working in the extensive limit and with the probability expressed in the form of Eq.~\eqref{Eq:prob_rescaled}, we find the universal form
\begin{equation}\label{Eq:cycle_hyp_gen}
\langle |c| \rangle \geq \begin{cases}
         0  ~~ & \gamma \leq \gamma^{*}(N_{A}) \\
         \frac{\mu(0)}{|\mathcal{Z}|^{\text{max}}} + \frac{\gamma}{2^{N_{A}+1}} - 1    ~~ &\gamma > \gamma^{*}(N_{A})\,.
    \end{cases} 
\end{equation} 
analogous to Eq.~\eqref{Eq:claw_hyp_gen}. 
Fig.~\ref{fig:avg_cyclesans_2D} shows that Eq.~\eqref{Eq:cycle_hyp_gen} matches well with simulations {(also see Eq.~\eqref{Eq:Cycle_lower_bound_1D} for a different bound in the case of $N_{A} = 1$).}

From Eq.~\eqref{Eq:cycle_hyp_gen} we note that random CRNs tend to have zero deficiency when $\gamma<\gamma^*$. However, for $\gamma>\gamma^*$, the average deficiency increases with the probability. This observation has interesting implications for random CRN dynamics: CRNs with zero deficiency always reach stable steady states~\cite{feinberg_2019}, whereas those with higher deficiency may exhibit complex dynamics like oscillations and multistability. Moreover, when $Q_{a}$ are large, the crossover probability $p^{*} = \gamma^{*}/|\mathcal{Z}|^{\text{max}}$ below which the average deficiency is zero scales inversely to the maximum possible number of species. As a result, deficiency-zero CRNs are rare, especially in the limit of large  $Q_{a}$.

\section{Discussion}\label{discussion}

\subsection{Connecting to the giant component transition}
{
We now connect the transitions in the emergent cycles and the forward reachability discussed in Secs.~\ref{Sec:conserv_laws} and \ref{Sec:reach} with the percolation transition of Erdös-Rényi random graphs (see Appendix~\ref{Sec:ER_intro}).

We start with conservation laws which, as discussed in Section \ref{Sec:synth_intro}, are the analogues of connected components in graphs.
In ER random graphs, the number of components decreases as $p$ increases, with a single giant component forming at large $p$. Similarly, in random CRNs, at small $p$, many species act as independent roots, resulting in numerous conservation laws— similar to disconnected components. As $p$ increases, the number of independent roots decreases, reducing the number of conservation laws to $N_A$, analogous to the emergence of a giant component in graphs.

We now turn to emergent cycles, which are the mathematical analogues of paths in undirected graphs.
In the ER random graph, at small $p$, since two random nodes most probably belong to different connected components, the probability of a path connecting them is very low. 
Whereas at high $p$, since the giant component corresponds to an $\mathcal{O}(N)$ number of nodes, there is a high probability that arbitrary nodes are connected. 
In our random CRNs, we see a similar behavior: at small $p$, the probability of an emergent cycle is nearly zero. At larger $p$, given an arbitrary subset, the probability of an emergent cycle associated with the subset is finite, implying the species in the subset are connected, in the sense of emergent cycles.

{
However, it is also important to emphasize the differences between percolation in CRNs and graphs. 
First, each node in a graph lies in exactly one component, so component size (number of nodes in the component) is unambiguous, and the size of the largest component serves as the standard order parameter used to study percolation~\cite{staufferbook,newman2018, dorogovtsevcomplexnet2022}. However, in CRNs, species can be part of multiple conservation laws. Furthermore, since conservation laws are a basis of the nullspace of $\nabla$, any notion of `size' for a conservation law must either use a special basis or be invariant under basis change. Secondly, paths in graphs connect only two nodes, whereas an emergent cycle involves an entire set of chemostatted species and is meaningful only as an effective reaction involving all the species. This prevents a geometric interpretation of emergent cycles as compared to paths in graphs.}

Finally,  we turn to forward reachability.
Forward reachability is the analogue of an \textit{out-component} in directed graphs. 
In studies of directed graphs, a giant out-component typically emerges beyond a critical probability, as observed in real-world networks such as the Internet. 
Our numerical results on the behavior of the largest reachable set similarly imply two regimes: At small $p$, many small reachable sets exist, akin to fragmented components below the critical threshold. Beyond the critical probability, a giant reachable set emerges, analogous to the giant out-component in directed graphs.
Based on our numerical results, such as a bimodal distribution, we concluded that the reachability transition is first-order. However, we also observe some deviations from the known theory for first-order transitions. Specifically, the maximum of the derivative, $\partial\langle |f| \rangle/\partial \gamma$ scales as a power law in $Q$ with an exponent $0.35$ rather than linearly~\cite{binderfirstordermonte1992}. Similarly, the fluctuations of $|f|$ show a growing peak that shifts towards $\gamma_{c}$ as $Q$ increases (see Appendix~\ref{Subsec:data_reach1D}). We believe that these behaviors may indicate strong finite-size effects.

\subsection{Relation with other CRN Models}
\label{Sec:rand_lit}
The complete set of reactions we defined in Sec.~\ref{sec:method} has 
been used in multiple contexts in the CRN literature. The key distinction in our approach is the focus on physical consistency, through the presence of conservation laws and the study of connectivity based on hypergraph methods.
We first discuss prior work that uses the same set of reactions, before focusing on different algorithms for generating random CRNs.

Models of polymerization have used the same network of reactions in the case of one atom.
In particular, the full set of reactions corresponds to the Smoluchowski aggregation model~\cite{Smoluchowski1916, sharko22,wattis06}.
Furthermore, Riehl~\cite{riehl2010} et al. have used the same one-atom CRN to study the properties of optimal balanced pathways, which are special vectors in the space of emergent cycles and have shown that this simple model can already reproduce generic properties of the metabolic network of E.coli.
We note that our formulation is more general as we also consider the case of multiple atoms and focus on other structural quantities such as conservation laws. 

In the case of $N_{A} > 1$, the full reaction network has also been explored under the name `cluster chemical reaction networks' (CCRNs) in Ref.~\onlinecite{gagrani2023}, where it was used to study the occurrence and geometric properties of autocatalytic cores. 
Interestingly, in their conclusion, the authors suggest future work on randomly generated CCRNs in an Erdős–Rényi framework to bound the existence and persistence of autocatalytic cycles as a function of the probability. This is a direction we aim to address using our framework, which provides tools to analyze such questions.
For $N_{A} = 2$, studies by Kauffman, Farmer, Bagley, Steel, and others~\cite{kauffman1986, bagley1990, steel2000, filisetti14} have used similar models to investigate the formation of autocatalytic sets. However, two key differences must be noted here.
First, many of these models focus on polymers where the precise order of monomers is important, a distinction that is not made in our model. Second, they assume that every reaction requires an enzyme, assigned probabilistically. Consequently, their arguments center on the emergence of a giant autocatalytic set as the probability of \textit{catalysis} varies. In contrast, our approach emphasizes structural properties of the reaction network itself, an aspect largely unexplored in these studies.
Finally, we note the work of Himeoka et al.~\cite{himeoka2024}, who analyzed the response of nonequilibrium steady states in CRNs to perturbations. Their study focuses on special subnetworks that correspond to $N_A = 2$ in our terminology. However, their approach does not include a broader discussion of the network’s structure, which is a major aspect of our work. 

We now compare our method of generating random CRNs with other algorithms used in the literature. 
Anderson et al.~\cite{anderson21, anderson22} construct random CRNs by generating random graphs in the space of complexes (see Sec.~\ref{Sec:Complex_graph}), and then study the deficiency of the associated CRN, but without taking conservation laws into account. Garcia-Chung et al.~\cite{garcia-chung2024} construct Erdos-Renyi like CRNs by randomly choosing hyperedges between nonintersecting subsets of species. However, they do not include any stoichiometric coefficients in their analysis and thus, do not explicitly conserve mass.
Fischer et al.~\cite{fischer2015} construct random open CRNs by first generating S graphs randomly with different prescriptions i.e, such as the Watts-Strogatz model or the Barabasi-Albert model.
They then rewire some edges in the S graph to obtain multimolecular reactions also. 
We note that none of the studies above have focused on constraining their reaction networks to conserve mass. 

Both Bigan et al.\cite{bigan2013} and Nicolau et al.\cite{nicolaou23} adopt methods close to ours by enforcing conservation laws. 
Bigan et al. construct all possible elementary reactions between species and select a subset randomly while ensuring mass conservation. 
Nicolau et al. assign random atomic compositions to species and generate reactions by randomly choosing reactants, products, and stoichiometric coefficients. However, unlike both approaches, we do not rely on random selection or external algorithms to enforce conservation. Instead, conservation laws are directly built into the reaction-generation process, ensuring that every reaction is chemically meaningful.
We note that the percolation transition for a random CRN model was also studied by Nicolau et al.
They projected the resulting CRN topology onto an S-graph for analysis~(see Sec.~\ref{Sec:Bipartite}) and showed that the critical value of the parameter $\langle|r|\rangle$ was approximately $0.15$.
In our analysis, the corresponding value of the same parameter is approximately $0.93$ (for $N_{A} = 1$ and considering emergent cycles). 
This difference can be attributed to the fact that adding a reaction to the CRN adds a fully connected clique of reactant nodes and product nodes to the S graph.
Furthermore, it has been shown in the literature that projections of the bipartite graph may show markedly different connectivity behavior compared to the bipartite graph~\cite{montanez10}.
It should be noted that both Nicolau et al.\cite{nicolaou23} and Bigan et al.~\cite{bigan2013} focus primarily on studying the dynamics of random CRNs, which is a direction we do not pursue in this work.
The dynamics of randomly generated catalytic CRNs, including trimolecular or higher-order reactions, have been used to model metabolism and growth in protocells~\cite{furusawa2003, kondo2011a}. These studies focus on higher-order reaction kinetics, which lie outside the scope of our framework.

\section{Conclusions and Perspectives} \label{Sec:disc}

In this work, we have proposed an algorithm for the generation of random CRNs that are physically consistent. 
We first construct a `universe' of possible reactions that respect mass balance as outlined in Sec.~\ref{sec:method} and then sample reactions randomly from it. 
This procedure allows us to statistically study the topology of CRNs under mass conservation constraints.
We focused on two physically motivated notions of connectivity in CRNs: emergent cycles, which describe steady-state synthesis pathways in open systems, and forward reachability, which describes the set of species that can be generated from a single substrate in a closed system.
Both notions exhibit percolation-like transitions as the probability of including a reaction increases.

We analyzed the probability that chemostatting an arbitrary subset of species creates an emergent cycle. 
Interestingly, when $N_{A} = 1$, the probability of an emergent cycle increased discontinuously across the transition while for the case of $N_{A} = 2$, the corresponding increase was continuous. 
We estimated the critical probability for the transitions in the emergent cycles using the number of conservation laws.
The average number of conservation laws showed marked nonmonotonic behavior at low $p$ before saturating to a constant value (the number of atoms) at high $p$. 
This directly implies that at low $p$, chemostatting species only breaks conservation laws, without creating effective interconversions. Whereas at high $p$, chemostatting an arbitrary set of species is enough to create emergent cycles. 

For reachability, we showed via numerical simulations that there is a critical probability below which (resp. above which) the largest reachable set involves a vanishing fraction of species (resp. an extensive fraction of species). Numerically, we further established that the largest reachable set undergoes a discontinuous transition for both $N_{A} = 1$ and $N_{A} = 2$.

Crucially, the percolation transitions for the emergent cycles and the forward reachability occur at different critical probabilities, that is, $3.62$ vs $2.92$ when $N_{A} = 1$. This stresses that both of these quantities capture different aspects of connectivity in CRNs.
This is a clear departure from undirected graphs, where both notions are equivalent, which again demonstrates the problems with using graph-based representations of CRNs to study topological properties. 

These transitions mark shifts in the capacity of the reaction network to take a set of substrates and create a desired product. Below the threshold, most species cannot be converted into one another.
Above the threshold, the accessible space of products is greatly enhanced and products can be created, either via emergent cycles or via reachability cascades.

Our work can be extended in various directions.
Our analysis of the transitions was limited to showing that percolation-like transitions exist in the emergent cycles and forward reachability and to estimating the respective critical probabilities.
A classification of these transitions into universality classes is an important future direction.
For the emergent cycles, a geometric perspective allowing us to define notions of (generalized) components and studying their properties is necessary. 
For reachability, we note that our quantitative analysis of the derivative of the order parameter and fluctuations revealed some discrepancies: the peak in the derivative grows as a power law in $Q$, and the susceptibility-like fluctuations also increase with $Q$. Although we believe these could be strong finite-size effects, it is important to note that similar phenomena have been observed in explosive percolation models~\cite{araujopercreview2014}, where transitions appear discontinuous at small system sizes but are actually continuous transitions that exhibit unusual finite-size properties. Thus, an important future step is to develop algorithms to calculate reachable sets in an efficient fashion, which would allow simulations at much larger $Q$, as well as analytical theories to describe the transition.

A key point in the construction of our set of reactions was to focus only on the composition of the chemical species and not their structure.
Consequently, unimolecular reactions such as isomerization reactions cannot be considered in this setup because they involve only structural changes. 
A simple solution to include unimolecular reactions is to add a new label for each species, representing possible conformations, and then add unimolecular reactions between the conformations~\cite{gagrani2023}.
Alternatively, more nuanced studies could start with string-based representations~\cite{warr2011, wigh2022} of molecules. 
In this context, we also mention the field of artificial chemistry, where the focus has been on studying abstract systems that are `chemistry-like'~\cite{dittrichreview2001}. Particularly, we note the work of Benk\"{o} et al.~\cite{benko2003, benko2003b} who developed a framework to generate CRNs based on graph-based representations of molecules, with reactions formulated as procedures that rewrite molecular graphs. This approach is particularly interesting as a starting point, as the graph-rewriting procedures inherently respect conservation laws. We direct the reader to Refs.~\onlinecite{andersen2013, andersen2016, arya2022} for further details and extensions of these ideas.

While we focused on measures of connectivity like emergent cycles and reachability, other topological features of CRN are of importance as well. 
Elementary flux modes~\cite{zanghelliniEFM2013,schusterefm2002} are nondecomposable vectors in the space of emergent cycles that are of importance for the analysis of metabolic pathways. 
Studying the statistics of properties such as the number and length of elementary flux modes, as done in Ref.~\onlinecite{riehl2010} for metabolic networks, is an important question.
Finally, the power of using randomness to understand the typical dynamics of complex systems is exemplified by the application of ideas like random matrix theory in understanding ecosystems~\cite{may1972,cui2021}. 
We believe that adding dynamics to our current formalism along the lines of work done in Refs.~\onlinecite{bigan2013, nicolaou23} and studying the response of steady states of random CRNs~\cite{himeoka2024, aslyamov2024} will be crucial steps for the future.

\begin{acknowledgments}
SGMS is grateful to Nalina Vadakkayil and Benedikt Remlein for their help with the scaling analysis, to Danilo Forastiere for pointing out a mistake in a previous draft.
SGMS was supported by the Luxembourg National Research Fund (FNR), via the research funding schemes PRIDE (Grant No. 19/14063202/ACTIVE), ME and NF by CORE project ChemComplex (Grant No. C21/MS/16356329) and by project INTER/FNRS/20/15074473 funded by F.R.S.-FNRS (Belgium) and FNR.
Simulations for the project were carried out
using the HPC facilities of the University of Luxembourg~\cite{VCPKVO_HPCCT22}
(see \texttt{\href{http://hpc.uni.lu}{hpc.uni.lu}}).
\end{acknowledgments}

\bibliographystyle{aipnum4-1}
\bibliography{references.bib}

\appendix

\section{Absence of link between emergent cycles and reachability}\label{sec:link_reach_emer}

We first note that, given a choice of chemostatted species, it is possible that there exists an emergent cycle linking them, but that no reachability relation exists between them.
An example of this is provided by the Michaelis-Menten CRN:
\begin{align}\label{Eq:CRN_MM}
    \text{E} + \text{S} &\xrightleftharpoons[-1]{+1} \text{ES} \,, \notag \\
    \text{ES} &\xrightleftharpoons[-2]{+2} \text{E} + \text{P}\,.
\end{align}
If the substrate species $\text{S}$ and the product species $\text P$ are chemostatted, the emergent cycle $\boldsymbol{c}_{\epsilon} = (1,1)^{\intercal}$ is created. However, the forward reachable sets of both $\text S$ and $\text P$ are just the species themselves (without the enzyme species $\text E$ and $\text{ES}$ no reaction can occur).

Conversely, assuming that we choose two species $\sigma$ and $\sigma'$ such that $\sigma \in \mathcal{F}(\sigma')$ and vice versa, if $Y = \{\sigma, \sigma'\}$, an emergent cycle need not be created. Consider the following CRN:
\begin{align}
      2\text{S} &\xrightleftharpoons[-1]{+1} \text{P} + \text{B}\,, \notag \\
      \text{P} &\xrightleftharpoons[-2]{+2}  \text{S} + \text{A}\,.  
\end{align}
Here, the forward reachable set of $\text{S}$ is $\mathcal{F}(\text{S}) = \left\{\text{S}, \text{P}, \text{A},\text{B}\right\}$ and is the same as the forward reachable set of $\text{P}$. However, if we chemostat $\text{S}$ and $\text{P}$, we \emph{do not} create an emergent cycle.

\section{Example}
\label{ap:example_counting}

Consider the simple case of $N_A=2$ atomic types, and $Q_1 = Q_2 = 2$. The possible molecular species is given by the set:
\begin{equation}
   \mathcal{Z}^{\text{max}} = \{ (0,1) \:\: (1,0) \:\: (0,2) \:\: (2,0) \:\: (1,1) \:\: (1,2) \:\: (2,1) \:\: (2,2)\}\,.
\end{equation}
From \eqref{eq:NS_max}, we indeed find that $|\mathcal{Z}|^\text{max} = 3 \times 3 -1 = 8$.
The possible reactions are:
\begin{equation}
\begin{split}
    &(2,0) \xrightleftharpoons[]{} (1,0) + (1,0) \\
    &\\
    &(0,2) \xrightleftharpoons[]{}(0,1) + (0,1) \\
    &\\
    &(1,1) \xrightleftharpoons[]{} (1,0) + (0,1) \\
    &\\
    &(1,2) \xrightleftharpoons[]{} (1,0) + (0,2) \\
    &(1,2) \xrightleftharpoons[]{} (1,1) + (0,1) \\
    &\\
    &(2,1) \xrightleftharpoons[]{} (2,0) + (0,1) \\
    &(2,1) \xrightleftharpoons[]{} (1,1) + (1,0) \\
    &\\
    &(2,2) \xrightleftharpoons[]{}(2,0) + (0,2) \\
    &(2,2) \xrightleftharpoons[]{}(1,0) + (1,2) \\
    &(2,2) \xrightleftharpoons[]{} (2,1) + (0,1) \\
    &(2,2) \xrightleftharpoons[]{} (1,1) + (1,1).
\end{split}
\end{equation}
Indeed, from \eqref{eq:NR_max} we find $|\mathcal{R}|^\text{max} = (3\times 4)^2/8 + 2^2/2 - 8 - 1 = 11$.

\section{Erdös-Rényi Random graphs}\label{Sec:ER_intro}
The Erdös-Rényi~(ER) random graph is the simplest and most well-studied random graph model. We briefly describe it here and direct readers to Refs.~\onlinecite{newman2018, dorogovtsevcomplexnet2022, bollobasRG2001} for comprehensive introductions.
In the ER model, a graph is generated with a given number of nodes, denoted by $|\mathcal{N}|$, where each possible pair of nodes is connected by an edge with a probability $p$. 
The total number of potential edges in such a graph is given by $|\mathcal{E}|^{\text{max}} = |\mathcal{N}|(|\mathcal{N}| - 1)/2 \sim |\mathcal{N}|^{2}/2$.
One of the most well-studied properties of the ER model is its component structure as a function of the connection probability $p$~\cite{newman2018, barabasicomplex2002, engel2004}, in the limit of large $|\mathcal{N}|$. 
As $p$ varies, the graph transitions between different connectivity regimes. 
When $p < 1/|\mathcal{N}|$, the graph is fragmented into many small components, each with a maximum size of order $\mathcal{O}(\ln |\mathcal{N}|)$. 
However, as $p$ exceeds $1/|\mathcal{N}|$, a `giant component' of size $\mathcal{O}(|\mathcal{N}|)$ suddenly emerges, containing an extensive fraction of the nodes.
This emergence of a giant component is a second-order phase transition, belonging to the universality class of mean field percolation~\cite{christensen98, staufferbook}.
The control parameter for this transition is given by $\gamma = (|\mathcal{N}| - 1)p$, while the order parameter is the fraction of nodes in the largest component.
As we vary $\gamma$, the order parameter remains zero below the critical value $\gamma_{c} = 1$,  beyond which it is nonzero.

\section{One atom}\label{Sec:one_atom}
In this section, we collect all the results for the case of $N_{A} = 1$ with the maximum possible number of atoms being $Q$. 
There are $|\mathcal{Z}|^{\text{max}} = Q$ species labeled by the index $\sigma$, i.e, the set of species is of the form $\{(1), (2),\dots,\sigma,\dots(Q)\}$. Using Eqs.~\eqref{eq:diss} and \eqref{Eq:assoc_rxn_def}, the total number of reactions a species $\sigma$ can participate in is given by the sum of the number of dissociations $D(\sigma) = \lfloor n_\sigma/2 \rfloor$ (with $\lfloor . \rfloor$ being the floor function) and $n_{\sigma}$ being the number of atoms in $\sigma$. Similarly, the number of associations $A(\sigma) = Q - n_\sigma$ and, 
\begin{equation}\label{Eq:tot_1D_sprxns}
    I(\sigma) = Q - \lceil n_\sigma/2 \rceil \,,
\end{equation}
where $\lceil.\rceil$ is the ceiling function.
The total number of possible reactions thus follows from Eq.~\eqref{eq:NR_max}, 
\begin{equation}\label{Eq:Nr_max_1d}
    |\mathcal{R}|^\text{max} = \lfloor (Q/2)^{2} \rfloor\,.
\end{equation}

\subsection{Number Distribution and Average Species}
The probability for a species $\sigma$ to have degree $k$ is given by the binomial distribution,
\begin{equation}\label{eq:deg_distri_sp1D}
    p_{\sigma}(k) = { Q - \lceil n_\sigma/2 \rceil \choose k} p^k (1-p)^{Q - \lceil n_\sigma/2 \rceil-k}\,,
\end{equation}
where we also used Eq.~\eqref{Eq:tot_1D_sprxns}. Using the independence approximation stated in Sec.~\ref{Sec:degree_gen}, the number distribution given by Eq.~\eqref{Eq:deg_distribution}, takes the form
\begin{equation}\label{eq:deg_distri_1D}
     P(n_{k}) = \sum_{\substack{A \subset \{(1), (2),\dots(Q)\} \\ |A| = n_{k}}} \prod_{\sigma \in A } p_{\sigma}(k)\prod_{\sigma' \notin A} (1 - p_{\sigma'}(k))\,,
\end{equation}
with the mean value
\begin{align}\label{Eq:mean_deg_1D}
    \mu(k) &= \sum_{\sigma} p_{\sigma}(k) \,, \notag\\
    &= \frac{1}{k!}\left(\frac{p}{1-p}\right)^{k}\sum_{n_\sigma = 1}^{Q} \frac {I(\sigma)!}{\left(I(\sigma)-k\right)!} (1-p)^{I(\sigma)}\,.
\end{align}
In the regime of Eq.~\eqref{Eq:prob_rescaled} and in the limit of $Q \to \infty$, we obtain a closed form expression for the degree distribution given in Eq.~\eqref{Eq:deg_distri_full},
\begin{equation}\label{Eq:deg_distri_1D_full}
    p_{k} = \frac{2}{\gamma k!}\left[g(k+1, \gamma) - g(k+1,\gamma/2)\right]\,,
\end{equation}
with $g(k,\gamma)$ denoting the lower incomplete gamma function~\cite{AS}.

We now turn to the expected number of species.
Using Eq.~\eqref{Eq:exp_species} with Eq.~\eqref{Eq:mean_deg_1D}, the average number of species $\langle|\mathcal{Z}|\rangle$ participating in a reaction takes the form
\begin{align}\label{Eq:Avg_sp_1D}
    \langle|\mathcal{Z}|\rangle &= Q - \sum_{\sigma} (1-p)^{Q - \lceil n_\sigma/2 \rceil}\,,\notag\\
    &\approx Q - (1-p)^{Q-1/2}\left[\frac{1-(1-p)^{-Q/2}}{1 - (1-p)^{-1/2}}\right]\,,
\end{align}
where we have replaced $\lceil n_\sigma/2 \rceil$ with $n_\sigma/2$ in the second line. 
In the limit of $p \to 0$, Eq.~\eqref{Eq:Avg_sp_1D} goes to the value $Q(Q-1/2)p$ which also vanishes. This is consistent as at very low probabilities, no reactions are chosen, implying the average number of chosen species goes to zero. Similarly, in the limit of $p \to 1$,  Eq.~\eqref{Eq:Avg_sp_1D} goes to the value $Q$, implying all species are chosen on average.

{
\subsection{Upper bound on Conservation Laws}
As argued in Sec.~\ref{Sec:roots}, the number of conservation laws is upper bounded by the number of root species (see Eqs.~\eqref{Eq:claw_upper_bound} and \eqref{Eq:root_res}). When $N_{A} = 1$, we can use this result to find a closed form upper bound on the average number of conservation laws. 
We denote the probability that a species $\sigma$ is a root as $\mathbb{P}[\sigma \in \mathcal{Z}^{r}]$. Assuming that $[\sigma \in \mathcal{Z}^{r}]$ and $[\sigma' \in \mathcal{Z}^{r}]$ are independent events, the average number of roots is the sum 
\begin{equation}\label{Eq:numroots_1D}
    \langle|\mathcal{Z}^{r}|\rangle = \sum_{\sigma} \mathbb{P}[\sigma \in \mathcal{Z}^{r}]\,.
\end{equation}
We now condition the probability that $\sigma$ is a root on the number of reactions $|\mathcal{R}|$, i.e,
\begin{multline}\label{Eq:root_prob_defn}
   \mathbb{P}[\sigma \in \mathcal{Z}^{r}]\\
    = \sum_{|\mathcal{R}| = 1}^{|\mathcal{R}|^{\text{max}}} \mathbb{P}[\sigma \in \mathcal{Z}^{r}~|~|\mathcal{R}|] {|\mathcal{R}|^{\text{max}} \choose |\mathcal{R}| } p^{|\mathcal{R}|}(1-p)^{|\mathcal{R}|^{\text{max}} - |\mathcal{R}|}\,.   
\end{multline}
The conditional probability that a species $\sigma$ is a root given a fixed number of reactions $|\mathcal{R}|$ is the fraction of possible CRNs with $|\mathcal{R}|$ reactions such that there is at least one association reaction involving $\sigma$ but no dissociation reaction of $\sigma$ i.e,
\begin{equation}\label{Eq:cond_prob_conserv}
  \mathbb{P}[\sigma \in \mathcal{Z}^{r}~|~|\mathcal{R}|]  = \frac{{|\mathcal{R}|^{\text{max}} - \lfloor n_\sigma/2 \rfloor \choose |\mathcal{R}| } - {|\mathcal{R}|^{\text{max}}- Q + \lceil n_\sigma/2 \rceil \choose |\mathcal{R}| }}{{|\mathcal{R}|^{\text{max}} \choose |\mathcal{R}|}}\,. 
\end{equation}
Substituting Eq.~\eqref{Eq:cond_prob_conserv} in Eq.~\eqref{Eq:root_prob_defn} and Eq.~\eqref{Eq:numroots_1D}, we get the result
\begin{align}\label{Eqn:conserv_law_bound}
&     \langle |\mathcal{L}| \rangle \leq \langle|\mathcal{Z}^{r}|\rangle  = \notag\\
&\sum_{\sigma} \sum_{|\mathcal{R}| = 1}^{|\mathcal{R}|^{\text{max}}} p^{|\mathcal{R}|}(1-p)^{|\mathcal{R}|^{\text{max}} - |\mathcal{R}|} \times \\
&\left\{ {|\mathcal{R}|^{\text{max}} - \lfloor n_\sigma/2 \rfloor \choose |\mathcal{R}| } - {|\mathcal{R}|^{\text{max}}- Q + \lceil n_\sigma/2 \rceil \choose |\mathcal{R}| } \right\}\notag
\end{align}

\begin{figure}
    \centering
    \includegraphics[scale=.50]{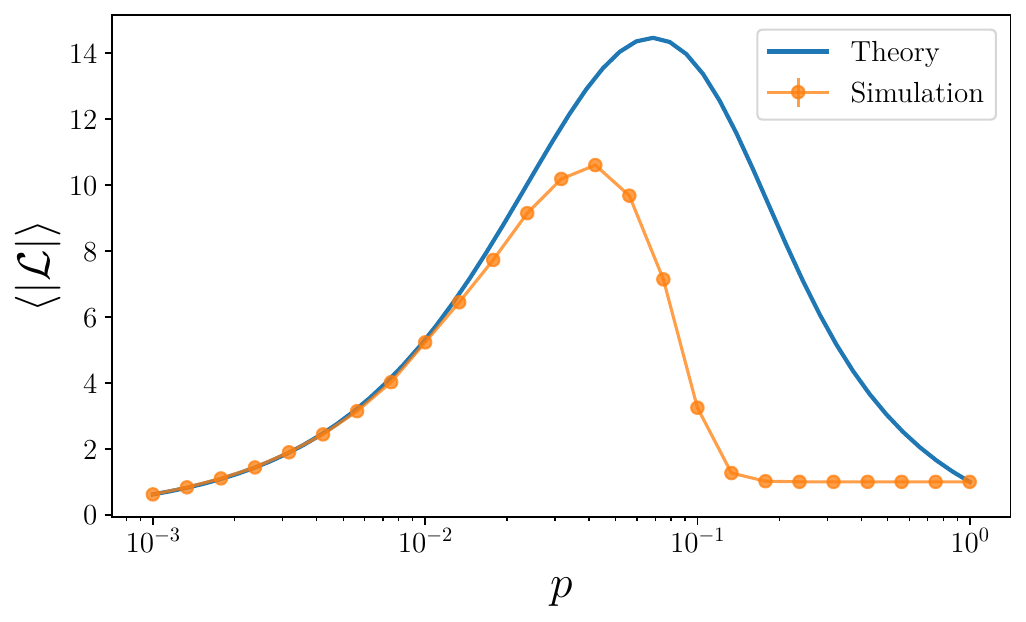}
    \caption{Average number of conservation laws as a function of the probability to select a reaction (dotted) compared with the theoretical result in Eq.~\eqref{Eqn:conserv_law_bound}. The parameters are $N_A = 1$ and $Q = 36$. For each value of $p$ we generated $5000$ reaction sets for averaging.}
    \label{fig:conserv_p_1D}
\end{figure}

Eq.~\eqref{Eqn:conserv_law_bound} captures the nonmonotonic behavior of the average number of conservation laws as seen in Fig.~\ref{fig:conserv_p_1D} and is a good approximation at low $p$ but is not a tight bound at higher $p$.
}

\subsection{Cycles}
In this subsection, we first build a basis set for cycles when $p = 1$ i.e, all reactions are considered inductively. Using this basis set, we construct a lower bound for the average number of cycles. 

We illustrate the inductive approach via an example:
When $Q< 4$, the CRN has no cycles. The first cycle emerges when $Q = 4$ and is given by 
\begin{align*}
     (2) + (2) &\xrightleftharpoons[-1]{+1}  (4) \xrightleftharpoons[-2]{+2} (3) + (1)\,,\\
    (1) + (1) &\xrightleftharpoons[-3]{+3}  (2)\,,\\
    (1) + (2) &\xrightleftharpoons[-4]{+4}  (3)\,.
\end{align*}
If we increase the number of species $Q$ to 5, any new cycle that emerges must involve the species $(5)$. There is once again one new  cycle:
\begin{align*}
     (2) + (3) &\xrightleftharpoons[-1]{+1}  (5) \xrightleftharpoons[-2]{+2} (4) + (1)\,,\\
    (1) + (3) &\xrightleftharpoons[-3]{+3}  (4)\,,\\
    (2) + (2) &\xrightleftharpoons[-4]{+4}  (4)\,.
\end{align*}
We see that both the cycles have exactly $4$ reactions. We now extend the approach to a general $Q$. Firstly, a general set of $4$ reactions which form a cycle are of the form,
\begin{align}\label{Eq:gen_cycle_structure_1D}
    (i) + (m-i) &\xrightleftharpoons[-1]{+1}  (m) \xrightleftharpoons[-2]{+2} (i') + (m-i')\,,\notag\\
    (i) + (i'-i) &\xrightleftharpoons[-3]{+3}  (i')\,,\\
    (m-i') + (i'-i) &\xrightleftharpoons[-4]{+4}  (m-i)\notag\,.
\end{align}
Finally, let the number of cycles for $Q-1$ species be $N_{c}(Q-1)$. If we now add another species, the number of cycles increases to $N_{c}(Q)$. The extra cycles must all involve reactions of species $(Q)$, which are $\lfloor Q/2 \rfloor$ in number. Fix the first reaction to be $(1) + (Q-1) \xrightleftharpoons[]{} (Q)$. Choosing any of the other $\lfloor Q/2 \rfloor - 1$ reactions gives different independent cycles. Thus, we have
\begin{equation}\label{Eq:Cycle_reccurence_relation}
    N_{c}(Q) = N_{c}(Q-1) + \lfloor Q/2 \rfloor - 1\,,
\end{equation}
with solution
\begin{equation}\label{Eq:cycle_num_1D}
    N_{c}(Q) \approx \frac{Q^{2} - 3Q}{4}\,.
\end{equation}
\begin{figure}
    \centering
    \includegraphics[scale=.40]{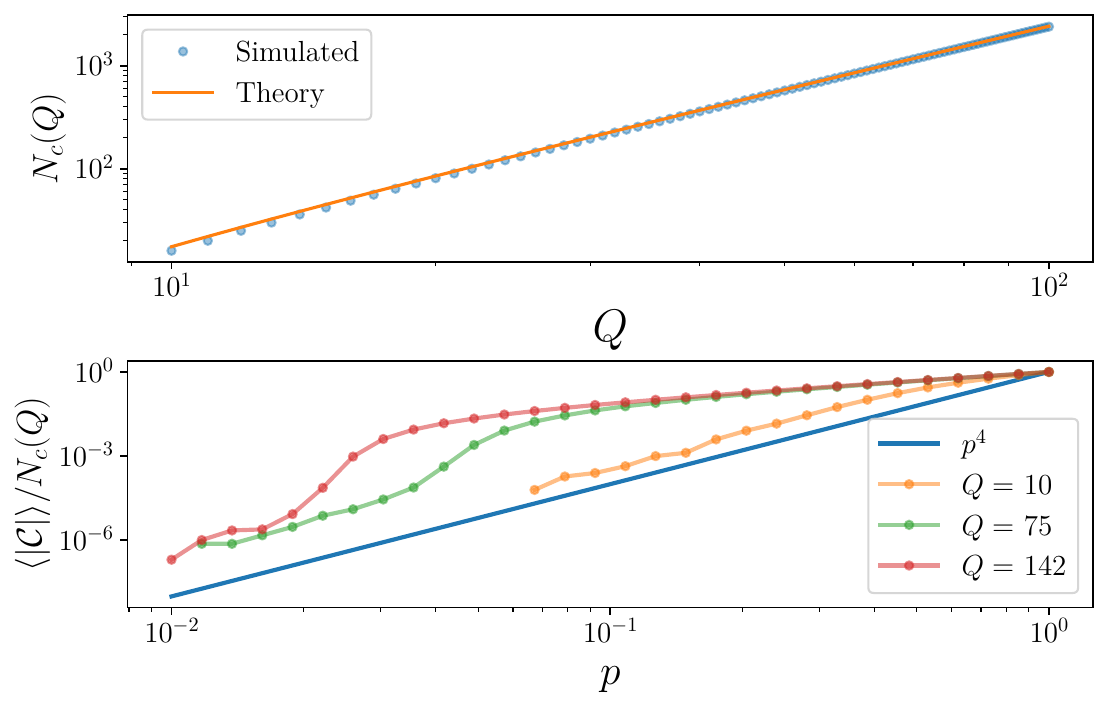}
    \caption{a) Total possible number of cycles plotted for $N_{A} = 1$ as a function of the maximum size of a molecule $Q$ (dotted) compared with the theoretical result in Eq.~\eqref{Eq:cycle_num_1D}, b) Average number of cycles scaled by the maximum possible number of cycles for $N_{A} = 1$ plotted as a function of the probability to select a reaction (dotted) compared with the theoretical result in Eq.~\eqref{Eq:Cycle_lower_bound_1D}. For each value of $p$ we generated $1000$ reaction sets for averaging.}
    \label{fig:cycle_1D}
\end{figure}
The same result was also derived differently in Ref.~\onlinecite{gagrani2023}.
We compare Eq.~\eqref{Eq:Cycle_reccurence_relation} at different $Q$ to the exact number of cycles that can be obtained numerically in Fig.~\ref{fig:cycle_1D}a). We see that at large $Q$, we get a very good agreement. 
While the construction above gives an independent basis for the kernel of $\nabla$, the different cycles in the basis share reactions. Thus, the average number of cycles at any $p$ is lower bounded by the probability of choosing cycles in the basis independently, which is given by
\begin{equation}\label{Eq:Cycle_lower_bound_1D}
    \left<|\mathcal{C}|\right> \geq  N_{c}(Q)p^{4}
\end{equation}
where we used Eq.~\eqref{Eq:gen_cycle_structure_1D} along with the fact that the probability to choose each reaction is $p$.
Comparing Eq.~\eqref{Eq:Cycle_lower_bound_1D} with simulations in Fig.~\ref{fig:cycle_1D}b), we see that the lower bound is only tight for small $Q$ and small $p$. 
We note that, in the extensive limit with the rescaling of Eq.~\eqref{Eq:prob_rescaled}, the lower bound in Eq.~\eqref{Eq:Cycle_lower_bound_1D} goes to zero, as anticipated in Eq.~\eqref{Eq:cycle_hyp_gen}.

\section{Limiting form of degree distribution}\label{Sec:sp_proof}

We want to compute the limiting form of Eq.~\eqref{Eq:deg_from_num} in the limit of $|\mathcal{Z}|^{\text{max}} \to \infty$ and with the rescaled variable $p = \gamma/|\mathcal{Z}|^{\text{max}}$. 
Using Eq.~\eqref{Eq:Mean_deg}, 
\begin{align}
    \mu(k) &=  \sum_{\sigma} {I(\sigma) \choose k} p^k (1-p)^{I(\sigma)-k} \,, \nonumber \\
    &= \gamma^{k}\sum_{\sigma}  {I(\sigma) \choose k}\left(\frac{1}{|\mathcal{Z}|^{\text{max}}}\right)^{k} \left(1 - \frac{\gamma}{|\mathcal{Z}|^{\text{max}}}\right)^{I(\sigma) - k}  \,, \notag\\
    &\simeq \frac{\gamma^{k}}{k!} \sum_\sigma \left(\frac{I(\sigma)}{|\mathcal{Z}|^{\text{max}}}\right)^{k}e^{-\gamma I(\sigma)/|\mathcal{Z}|^{\text{max}}} \,,
    \label{Eq:avg_sp_zero1}
\end{align}
where $I(\sigma)$ is the number of reactions a species participates in (see Eq.~\eqref{Eq:tot_rxn_I}) and where we used the Poisson approximation of the Binomial. Using Eqs.~\eqref{Eq:assoc_rxn_def},\eqref{eq:diss} and \eqref{eq:NS_max}, we can write
\begin{align}
    \frac{I(\sigma)}{|\mathcal{Z}|^{\text{max}}} &= 
    \frac{D(\sigma)}{|\mathcal{Z}|^{\text{max}}} + \frac{A(\sigma)}{|\mathcal{Z}|^{\text{max}}} \,, \label{Eq:tot_rxn_scaling1}\\
    &= \frac{1}{2}\prod_{a= 1}^{N_{A}} \frac{n_{a} + 1}{Q_{a} + 1} + \prod_{a = 1}^{N_{A}} \left(1 - \frac{n_{a}}{Q_{a} + 1}\right) + \frac{\delta(\sigma)}{|\mathcal{Z}|^{\text{max}}}\,, \nonumber
\end{align}
where in the quantity $\delta(\sigma)$ we have combined $\mathcal{O}(1)$ terms from Eqs.~\eqref{Eq:assoc_rxn_def} and \eqref{eq:diss}. Defining the fractions $x_{a} = n_{a}/(Q_a + 1) \approx n_a/Q_{a}$ and substituting Eq.~\eqref{Eq:tot_rxn_scaling1} in \eqref{Eq:avg_sp_zero1}, we find,
\begin{align}
\frac{\mu(k)}{|\mathcal Z|^{\max}}
&\simeq
\frac{\gamma^{k}}{k!\,|\mathcal Z|^{\max}}
\sum_{x_{1}=0}^{1}\!\!\cdots\!\!
\sum_{x_{N_A}=0}^{1}
\Bigl(\tfrac12 \prod_{i=1}^{N_A}x_i
      + \prod_{j=1}^{N_A}(1-x_j)\Bigr)^{k}
\nonumber\\
&\hspace{2em}\times
\exp\!\Bigl[
-\gamma\Bigl(\tfrac12 \prod_{i=1}^{N_A}x_i
            + \prod_{j=1}^{N_A}(1-x_j)\Bigr)
\Bigr]
\nonumber\\[6pt]
&\simeq
\frac{\gamma^{k}}{k!}
\int_{0}^{1}\!dx_1\!\cdots\!\int_{0}^{1}\!dx_{N_A}\,
\Bigl(\tfrac12 \prod_{i=1}^{N_A}x_i
      + \prod_{j=1}^{N_A}(1-x_j)\Bigr)^{k}
\nonumber\\
&\hspace{2em}\times
\exp\!\Bigl[
-\gamma\Bigl(\tfrac12 \prod_{i=1}^{N_A}x_i
            + \prod_{j=1}^{N_A}(1-x_j)\Bigr)
\Bigr]\,,
\label{Eq:scale_sp0_final}
\end{align}
where we converted the discrete sum in the first equation into an integral.

\section{Simulations for emergent cycles}\label{Sec:data_all}
The simulations in Sec.~\ref{Sec:conserv_laws} comprised of two steps. First, we generated random CRNs for prescribed values of $\gamma$ and ${Q_{1}, Q_{2}, \dots, Q_{N_{A}}}$. For each random CRN, we then randomly selected a subset of species in $\mathcal{Z}$ and asked whether chemostatting them produced an emergent cycle by evaluating the difference $\dim(\ker(\nabla^{X})) - \dim(\ker(\nabla))$. To perform this rank calculation efficiently for large $\nabla$, we used the package SuiteSparseQR\cite{davisSparseQRpaper} through its Python wrapper~\cite{sparseqr_python}. Below, we describe the specifics and the analysis of the resulting data for the cases $N_{A}=1$ and $N_{A}=2$.

\subsection{One atom}\label{Subsec:data_emer1D}
We outline here the procedure used to estimate the threshold $\gamma_c$ for the single-atom case. For each value of $Q$, samples of $n_i$ (as defined in Sec. \ref{Sec:conserv_laws}) were simulated for $\gamma \in [3.0, 4.5]$ with a step size of $0.01$. The $Q$ values used were $(100, 250, 500, 750, 1000, 1250, 1500, 1750, 2000)$, with the number of samples per $\gamma$ set to $(25000, 25000, 20000)$ for the three smallest $Q$ values, and $10000$ samples for all larger $Q$. The corresponding probabilities $\mathbb{P}_\epsilon(\gamma, Q)$ were computed as the sample mean of $n_i$ and are plotted in Fig.~\ref{fig:emer_cyc1}, along with the standard error of the mean for selected $Q$ values.

In principle, $\gamma_c$ can be estimated by extrapolating the crossing points of the $\mathbb{P}_\epsilon(\gamma, Q)$ curves for successive values of $Q$ (e.g., between $Q = 100$ and $250$, $250$ and $500$, and so on), followed by extrapolation to $Q \to \infty$. However, in our case, the crossing points were all tightly clustered around $\gamma \approx 3.62$, and applying standard extrapolation techniques did not yield reliable results. We speculate that $Q$ dependent corrections are already small for $Q \geq 100$.

To obtain a more accurate estimate, we instead used the peaks of the variance $\tilde{\chi}(\gamma, Q)$. For each $Q$, we focused on the range $\gamma \in (3.5, 3.74)$ and interpolated the data using a cubic spline on a refined grid with step size $0.0001$. The location of the peak was then identified for each $Q$, and these peak positions were extrapolated using the BST algorithm to get an estimate of $\gamma_{c}$~\cite{west2015}.

To estimate uncertainty, we repeated the entire procedure with 1000 bootstrap resamples of the underlying data. This was followed by computing $\tilde{\chi}(\gamma, Q)$ for the resampled data and the procedure outlined above was followed. The resulting distribution had a mean of $3.6242$ and a standard deviation of $0.0198$. Since the original data was defined on a grid with spacing $0.01$, we report the threshold as $\gamma_c = 3.62 \pm 0.02$.

\subsection{Two atom}\label{Subsec:data_emer2D}

We now describe the procedure used to extract the critical exponents for the two-atom case ($N_A = 2$) with $Q_1 = Q_2 = Q$. For each $Q$, samples of $n_i$ (as defined in Sec.~\ref{Sec:conserv_laws}) were simulated for $\gamma \in [5.5, 7.5]$ using a step size of $0.01$. The $Q$ values used were $(10, 15, 20, 24, \dots, 70)$, with 32,000 samples per $\gamma$ for all $Q$ except the three largest, for which 16,000 samples were generated. The corresponding probabilities $\mathbb{P}_\epsilon(\gamma, Q, Q)$ were computed as the sample mean of $n_i$ and are plotted in Fig.~\ref{fig:emer_cyc2}, along with the standard error of the mean for selected values of $Q$.
To compute the numerical derivative, we first smoothed the data using the gaussianfilter1d routine from scipy.ndimage with a smoothing parameter $\sigma = 4.0$. The derivative was then calculated using a forward Euler scheme. We show the magnitude of the difference between $\mathbb{P}_{\epsilon}(\gamma,Q, Q)$ for $Q = 25, 35, 45$ and the resulting data after smoothing in Fig.~\ref{fig:smooth_diff}

\begin{figure}
    \centering
    \includegraphics[scale=.5]{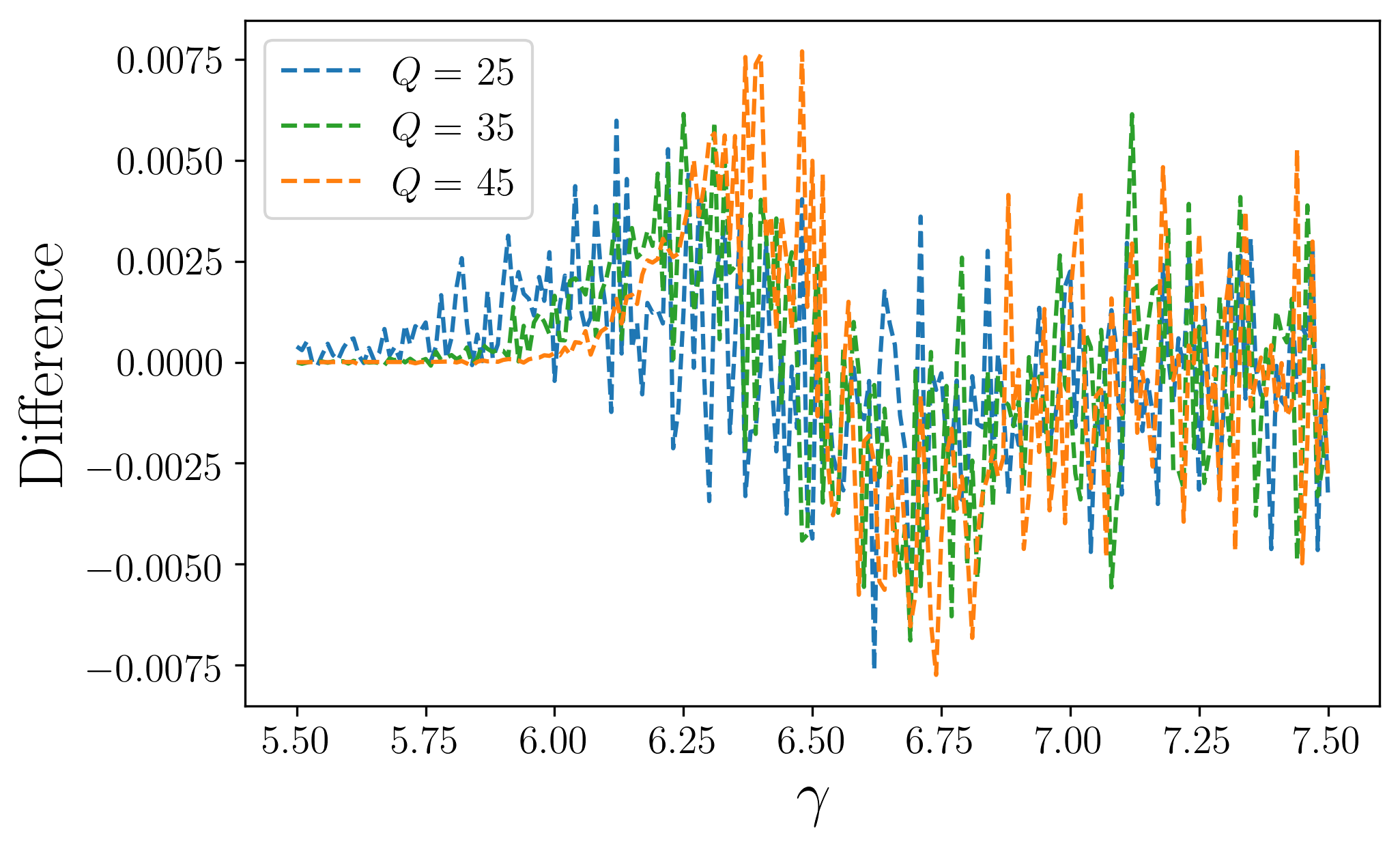}
    \caption{Data from Fig.~\ref{fig:emer_cyc2} smoothened by a Gaussian Filter and the difference between the smoothened and original data plotted as a function of $\gamma$. The parameters are $N_{A} = 2$ and $Q_{1} = Q_{2} = Q$ and $\sigma = 4$ for the filtering.}
    \label{fig:smooth_diff}
\end{figure}

For the scaling analysis, we followed the approach of Refs.~\onlinecite{houdayerhartmann2004, melchert2009} and defined a collapse quality function $S$, where a good collapse corresponds to $S \approx 1$. To determine the critical exponents, we minimized $S$ using the Nelder-Mead algorithm from scipy.optimize, starting from the initial guess $\gamma_c = 6.76$, $a = 2.24$, and $b = 0.09$, chosen by eye.

A crucial point here is to estimate the limits of the rescaled variable, denoted $x = (\gamma - \gamma_{c})|\mathcal{Z}^{\text{max}}|^{1/a}$ in which the collapse is valid. If we denote the upper limit (resp. lower limit) of this interval as $x_{\text{max}}$ (resp. $x_{\text{min}}$), the collapse is perfectly valid in the limit of $x_{\text{max}},x_{\text{min}} \to 0 $. However, choosing a narrow interval may lead to overfitting and unphysical values of the exponents (e.g., $a < 0$). Typically, in such cases both $x_{\text{min}}$ and $x_{\text{max}}$ lie on the same side of zero. We scanned over values of $x_{\text{max}}$ from 1.5 to 15 in steps of 0.5. For each candidate window, we minimized $S$ with respect to the five parameters $(\gamma_c, a, b, x_{\text{min}}, x_{\text{max}})$. Fits yielding unphysical exponents were discarded, and the best-fit parameters were selected from the remaining cases.

This procedure yielded the estimates $\gamma_c = 6.7415$, $a = 2.2191$, and $b = 0.0949$, with corresponding limits $x_{\text{min}}$, $x_{\text{max}}$, and a minimum collapse score of $S_{\text{min}} = 1.454$. To estimate uncertainties, we varied the exponents until $S = S_{\text{min}} + 1$, resulting in: $\delta\gamma_c = 0.0057$, $\delta a = 0.1315$, and $\delta b = 0.0052$. Given the original grid spacing of $0.01$, we report the final estimates as $\gamma_{c} = 6.74 \pm 0.01$, $a = 2.219 \pm 0.131$ and $b= 0.095 \pm 0.005$. This rounding changes the collapse score only slightly to $S = 1.56$, which still indicates a good fit.

The data in Fig.~\ref{fig:emer_cyc2} is rescaled according to Eq.~\eqref{Eq:data_collapse} using the above exponents and plotted in Fig.~\ref{fig:emer_cyc2_collapse}. The black dashed lines denote the best-fit region, which includes $x = 0$. Fig.~\ref{fig:emer_cyc2_der_pl1} shows the peak height of the smoothed numerical derivative $\partial \mathbb{P}_\epsilon(\gamma_r, Q, Q)/\partial \gamma$ plotted against $|\mathcal{Z}^{\max}|$. A linear fit on a log-log scale (excluding the first three data points) yields $r^2 > 0.99$, and the extracted exponent is consistent with the scaling exponents reported above.  The error in the exponent was calculated by bootstrap resampling and repeating the smoothing and differentiation steps. Finally, the location of the peak for each $|\mathcal{Z}|^{\text{max}}$ approaches $\gamma_{c}$ as shown in Fig.~\ref{fig:emer_cyc2_der_pl2}. Once again, a linear fit on a log-log scale (excluding the first three data points) yields $r^2 > 0.99$ and the extracted exponent is consistent with the scaling analysis. 

We note two caveats here: First, while $r^2 > 0.99$ indicates a good linear fit, it does not in itself confirm a power-law relationship. Second, the derivative peaks and their locations were extracted after applying smoothing. However, given that the resulting exponents are consistent with the scaling analysis, we believe any bias introduced by this step is negligible.

\begin{figure}
    \centering
    \includegraphics[scale=.5]{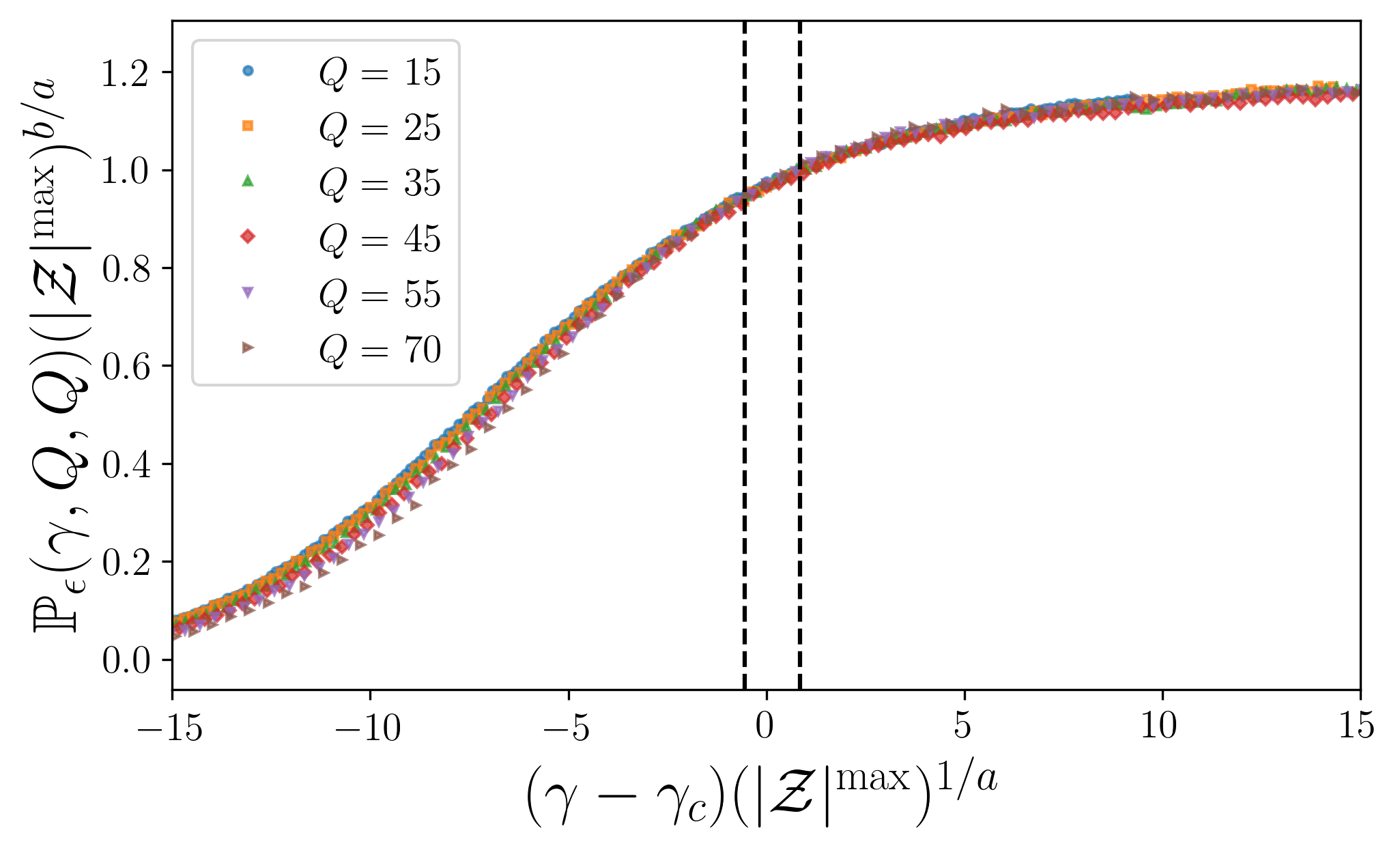}
    \caption{Data from Fig.~\ref{fig:emer_cyc2} rescaled as in Eq.~\eqref{Eq:data_collapse} and plotted as a function of $(\gamma-\gamma_{c})(|\mathcal{Z}|^{\text{max}})^{\frac{1}{a}}$. The parameters are $N_{A} = 2$ and $Q_{1} = Q_{2} = Q$. The dotted black lines indicate the regime of the best fit.}
    \label{fig:emer_cyc2_collapse}
\end{figure}

\begin{figure}
    \centering
    \includegraphics[scale=.5]{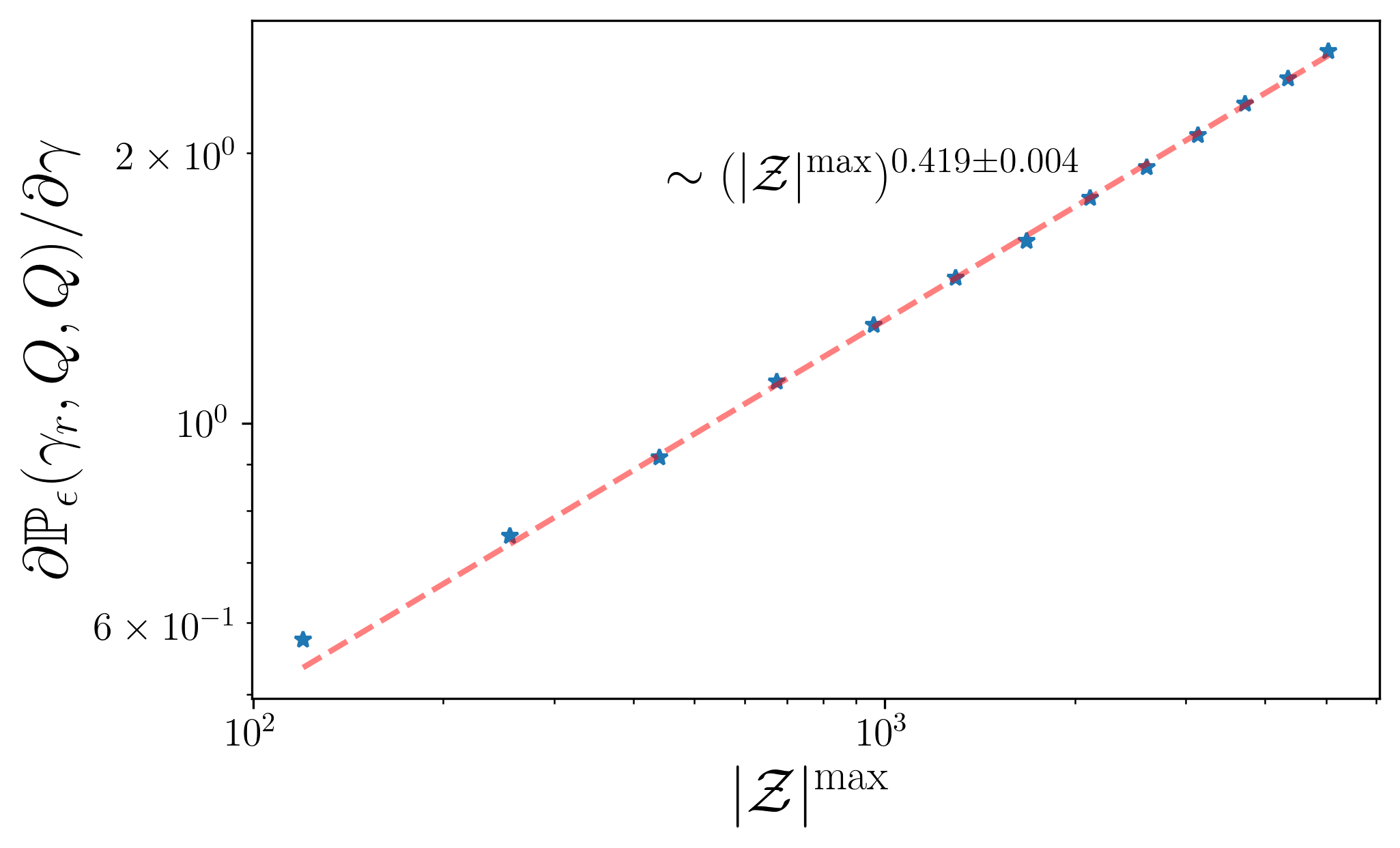}
    \caption{Peak of the derivative of the probability of having an emergent cycle wrt $\gamma$ plotted as a function of $|\mathcal{Z}|^\text{max}$ in log-log scale. The parameters are $N_{A} = 2$ and $Q_{1} = Q_{2} = Q$. The red line is a linear fit.}
    \label{fig:emer_cyc2_der_pl1}
\end{figure}

\begin{figure}
    \centering
    \includegraphics[scale=.5]{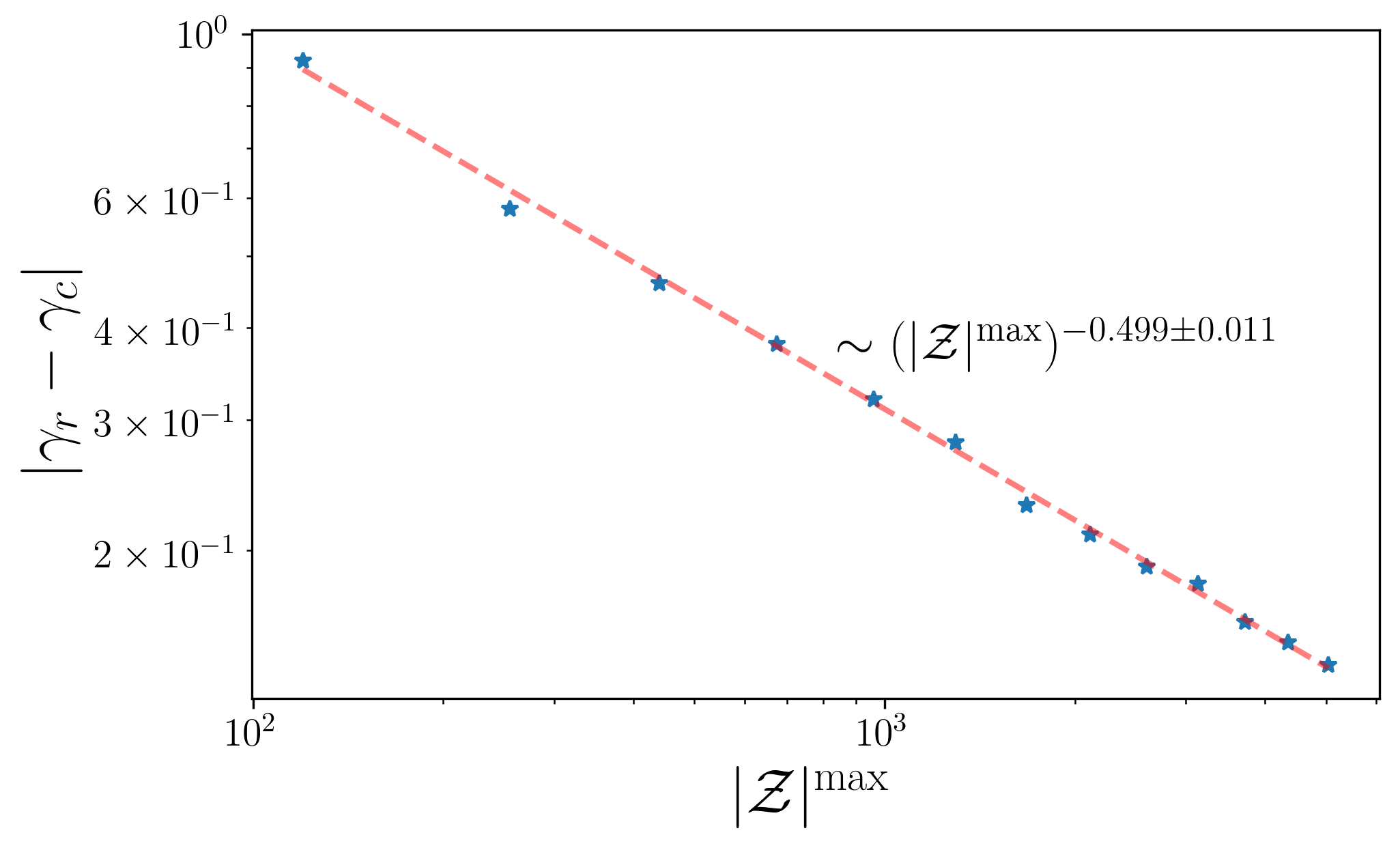}
    \caption{Magnitude of the difference between the location of the peak of the derivative of the probability of having an emergent cycle wrt $\gamma$ from $\gamma_{c}$ plotted as a function of $|\mathcal{Z}|^\text{max}$ in log-log scale. The parameters are $N_{A} = 2$ and $Q_{1} = Q_{2} = Q$. The red line is a linear fit.}
    \label{fig:emer_cyc2_der_pl2}
\end{figure}

\section{Proofs}\label{Sec:proofs_final}
\subsection{Conservation laws at $p=1$}\label{ap:proof_basis} 
Given an arbitrary conservation law $\boldsymbol{\ell}^{\lambda} = \left(....,\ell^{\lambda}_{\sigma},....\right)^{\intercal}$, its components satisfy a set of equalities of the form,
\begin{equation}\label{Eq:conserv_law_constraint}
    \ell^{\lambda}_{r(\sigma)} + \ell^{\lambda}_{r'(\sigma)} = \ell^{\lambda}_{p(\sigma)}\,,
\end{equation}
where we used the definition $\boldsymbol{\ell}^{\lambda}\cdot\nabla = 0$ along with Eq.~\eqref{Eq:rxn_redefined}.
A solution to this set of equations is given by the set of vectors \{$\boldsymbol{\ell}^{a}$\} with entries
\begin{equation}\label{eq:claw_soln}
    \ell^{a}_{\sigma} = n_{a}\,,
\end{equation}
where $\sigma = \left(\dots, n_{a},\dots\right)$ and the index $a$ goes from $1$ to $N_{A}$. 
Here, each vector $\boldsymbol{\ell}^{a}$ counts the number of atoms of type $a$ in $\sigma$. By construction, each vector $\boldsymbol{\ell}^{a}$ is a conservation law: (see  Eqs.~\eqref{eq:claw_soln}, \eqref{Eq:rxn_chem} and \eqref{Eq:conserv_law_constraint}). 
As each of the vectors $\boldsymbol{\ell}^{a}$ are independent, the number of conservation laws of $\nabla$ when $p \to 1$ are lower bounded by $N_{A}$. 

Now, consider a generic conservation law $\boldsymbol{\ell}^{1}$ and a set of species $\sigma_{a}$, which are just singlet atoms, i.e, $\sigma_{a} = (0,0,0....1,0,0..)^{\intercal}$. 
Since $p \to 1$, we can expect that any realization of the random CRN will include all reactions of the form Eq~\eqref{Eq:rxn_chem}.
Starting with this set of species, we can construct a chain of reactions, taking us to any arbitrary species $\sigma = \left(n_{1},n_{2}\dots n_{N_{A}}\right)$. 
Iterating Eq.~\eqref{Eq:conserv_law_constraint} along this chain of reactions, the entries of $\boldsymbol{\ell}^{1}$ satisfy, 
\begin{equation}
    \ell^{1}_{\sigma} = \sum_{a = 1}^{N_{A}} n_{a} {\ell}^{1}_{\sigma_{a}} = \sum_{a = 1}^{N_{A}} \ell^{a}_{\sigma} {\ell}^{1}_{\sigma_{a}} \,,
\end{equation}
which implies that arbitrary conservation laws can be expressed as linear combinations of the set $\{\boldsymbol{\ell}^{a}\}$,
showing that at probabilities close to one, the number of conservation laws should be exactly $N_{A}$.

\subsection{Applying our ansatz to the ER transition} \label{Sec:ER_appendix}

We use the approach of Subs.~\ref{Sec:rank_arguments} to show here that our approach can only approximate the percolation threshold in the ER case .
We are interested in the number of connected components scaled by the number of vertices (denoted  $\mathfrak{c}$) as a function of the rescaled probability $\gamma$. We first show the exact result~\cite{engel2004} and then our ansatz for comparison.

\subsubsection{Exact result}
The order parameter for the ER graph, denoted here with $\mathcal{S}$, is the (average) fraction of nodes in the largest component of the random graph and it is known to be the (stable) root of the equation~\cite{newman2018, engel2004}:
\begin{equation}\label{Eq:ER_op}
  1 - \mathcal{S}(\gamma) - e^{-\gamma \mathcal{S}(\gamma)} = 0\,.
\end{equation}

From Ref.~\onlinecite{engel2004}, the scaled average number of components satisfies the equation,
\begin{equation}\label{Eq:ER_comp1}
  \langle \mathfrak{c} \rangle = (1 - \mathcal{S}(\gamma))\left(1 - \frac{(1-\mathcal{S}(\gamma))\gamma}{2}\right)\,.
\end{equation}

However, the quantity $\langle \mathfrak{c} \rangle$ also includes nodes that are not connected to any other component in the graph. While this is a common practice in complex networks, we exclude all such nodes in our analysis.  The probability of a node having degree zero at all is $e^{-\gamma}$. Thus, we get a modified number of components denoted $\langle \bar{\mathfrak{c}} \rangle$ which takes the form,
\begin{equation}\label{Eq:ER_compfull}
  \langle \bar{\mathfrak{c}} \rangle = (1 - \mathcal{S}(\gamma))\left(1 - \frac{(1-\mathcal{S}(\gamma))\gamma}{2}\right) - e^{-\gamma} \,.
\end{equation}
Substituting Eq.~\eqref{Eq:ER_op} in Eq.~\eqref{Eq:ER_compfull}, close to the critical probability $\gamma_{c} = 1$, we have the expression
\begin{equation}\label{Eq:ER_compf3}
 \langle \bar{\mathfrak{c}} \rangle  = \begin{cases}
       1 - \frac{\gamma}{2}-e^{-\gamma} ~~~\gamma \leq 1\,, \\
       (3-2\gamma)\left(1-\gamma(\frac{3}{2}-\gamma)\right) - e^{-\gamma}~~~\gamma > 1\,.
  \end{cases}
\end{equation}
Fig.~\ref{fig:ER_trans2} shows a simulation of $\langle \bar{c} \rangle$ as a function of $\gamma$.
Crucially, both the simulation in Fig.~\ref{fig:ER_trans2} and the theory in Eq.~\eqref{Eq:ER_compf3} show no discontinuities or kinks at $\gamma = 1$. In fact, the same is true for the first derivative of $\langle \bar{\mathfrak{c}} \rangle$ and there is only a kink in the second derivative $\langle \bar{\mathfrak{c}} \rangle$. 

\begin{figure}
    \centering
    \includegraphics[scale=.5]{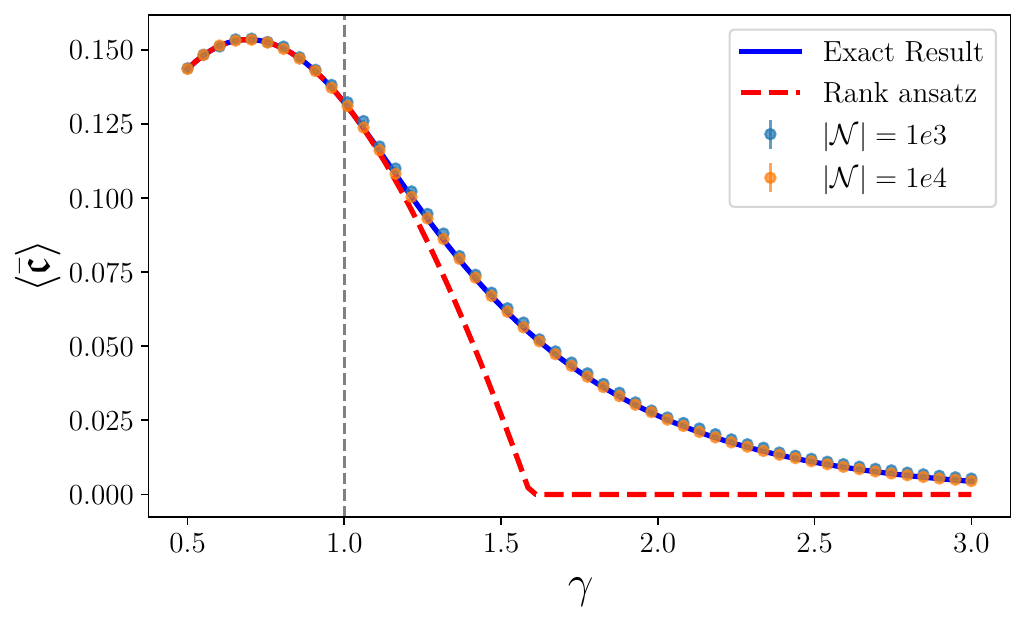}
    \caption{Average number of components (scaled) in the ER random graph for $|\mathcal{N}| = 1000$ and $|\mathcal{N}| = 10000$ plotted as a function of the scaled probability to select a reaction compared with the exact theory (blue solid line) from Eq.~\eqref{Eq:ER_compfull} and the rank ansatz in Eq.~\eqref{Eq:hypothesis_ER_Full} (dashed red line). For each value of $p$ we generated $1000$ graphs for averaging. The gray line is  $\gamma_{c} = 1$.}
    \label{fig:ER_trans2}
\end{figure}

\subsubsection{Using the ansatz}
We can write an ansatz with the rank of the incidence matrix $\partial$ similar to Eq.~\eqref{Eq:rank_bound} and the analysis in Sec.~\ref{Sec:rank_arguments}. For an ER graph, this ansatz would take the form,
\begin{equation}\label{Eq:rank_ER_hypothesis}
     \langle \text{rk}(\partial)\rangle \leq \begin{cases}
         \left<|\mathcal{E}|\right>  ~~ & p \leq p^{*} \\
         \langle|\mathcal{N}|\rangle - 1    ~~ &p> p^{*}\,.
     \end{cases} 
\end{equation}
Intuitively, the term $N_{A}$ in Eq.~\eqref{Eq:claw_hypothesisb} gets replaced by $1$ as in the fully connected graph, there is only one conservation law which is the vector $\left(1,1,1...\right)^{\intercal}$.
Transforming into the rescaled variables $p = \gamma/(|\mathcal{N}| - 1)$ and using  Eq.~\eqref{Eq:rank_ER_hypothesis} along with the rank nullity theorem, we find  that $\langle \bar{\mathfrak{c}} \rangle$ must satisfy,
\begin{equation}\label{Eq:hypothesis_ER_Full}
   \langle \bar{\mathfrak{c}} \rangle  \geq \begin{cases}
         1 - e^{-\gamma} - \frac{\gamma}{2} ~~ & \gamma  \leq \gamma^{*} \\
         0    ~~ & \gamma> \gamma^{*}\,,
     \end{cases} 
\end{equation}
where $\gamma^{*} \approx 1.493$. Note that while Eq.~\eqref{Eq:hypothesis_ER_Full} matches Eq.~\eqref{Eq:ER_compf3} exactly below the percolation transition, it is a lower bound above it. Particularly, the predicted transition point $\gamma^{*}$ has no physical meaning in the ER context. We compare both Eq.~\eqref{Eq:ER_compf3} and Eq.~\eqref{Eq:hypothesis_ER_Full} along with simulations in Fig.~\ref{fig:ER_trans2} to show the difference.

{
\subsection{Results related to roots}\label{Sec:ap_root_proof}

Consider a CRN with the stoichiometric matrix $\nabla$ and all reactions of the form of Eq.~\eqref{Eq:rxn_chem} and each reaction oriented as in Eq.~\eqref{Eq:rxn_redefined}. 
We first identify the root species by finding those species such that all the entries in the corresponding row of the stoichiometric matrix $\nabla$ are non-positive.
We then split the set of species $\mathcal{Z}$ as 
$\mathcal{Z} = \mathcal{Z}^{r} \cup \mathcal{Z}^{l}$  where the set of roots (resp. non-roots) is denoted $\mathcal{Z}^{r}$ (resp. $\mathcal{Z}^{l}$). 
Correspondingly, the stoichiometric matrix $\nabla$ splits into 
\begin{equation}\label{Eq:stoc_split}
    \nabla =
    \begin{pmatrix}
        \nabla^{l} \\ 
        \nabla^{r}\\
    \end{pmatrix}\,,
\end{equation}
where $\nabla^{r}$ (resp. $\nabla^{l}$) is the stoichiometric matrix with rows corresponding only to the root species (resp. non-root species).
The columns of $\nabla^{l}$ can take the following four forms (indicating only the nonzero elements):
\begin{equation}\label{Eq:stoc_l_form}
    \nabla^{l}_{\rho} = \begin{pmatrix}
        1\\
        .\\
        .\\
        .\\  
        .\\
        .\\
    \end{pmatrix}, \begin{pmatrix}
        .\\
        -1\\
        1\\
        .\\
        .\\
        .\\
    \end{pmatrix}, \begin{pmatrix}
        -2\\
        .\\
        .\\
        .\\
        1\\
        .\\
    \end{pmatrix},
    \begin{pmatrix}
        .\\
        .\\
        -1\\
        -1\\
        1\\
        .\\
    \end{pmatrix}\,,
\end{equation}
where the first two columns are reactions involving root species (see Eqs.~\ref{Eq:reac1_root} and \ref{Eq:reac2_root}) while the other two columns involve only non-root species.

\subsubsection{Result I}\label{Sec:root_ap1} 
We now argue that the left nullspace of $\nabla^{l}$ is trivial.

Consider any vector $\boldsymbol{\ell}$ such that $\boldsymbol{\ell}\cdot\nabla^{l} = 0$ where $\nabla^{l}$ has been defined in Eq.~\eqref{Eq:stoc_split}. 
Using the forms of the columns of $\nabla^{l}$ as shown in Eq.~\eqref{Eq:stoc_l_form}, we show that the entries of $\boldsymbol{\ell}$ are identically zero.
Consider a given species $\sigma$ which is not a root. 
Since it is not a root, it must be the product of atleast one reaction. 
Consider the corresponding column of $\nabla^{l}$.
If it is of the first form in Eq.~\eqref{Eq:stoc_l_form}, it is clear that $\ell_{\sigma} = 0$.
Otherwise, assume it is of the second form in Eq.~\eqref{Eq:stoc_l_form}.
Then, we get a linear relation between the entries $\ell_{\sigma} = \ell_{\sigma{'}}$, where the species $\sigma{'}$ is the non root reactant of the same reaction.
Starting with $(\sigma')$, we look for a reaction $\rho$ where $(\sigma')$ is a product (if no such reaction were to exist, $\sigma'$ would be a root) and repeat the process. 
Eventually, this process will terminate in a reaction of the form of column one (in each step, the atom numbers of the species reduce - eventually the process has to stop) implying $\ell_{\sigma} = 0$.
If the reaction is instead of the third or the fourth form in Eq.~\eqref{Eq:stoc_l_form}, then we get an equation of the form $\ell_{\sigma} = \ell_{\sigma_{r(\rho)}} + \ell_{\sigma_{r'(\rho)}}$. We repeat the process for both the species $\sigma_{r(\rho)}$ and $\sigma_{r'(\rho)}$ and arrive at the result.

\subsubsection{Result II}\label{Sec: root_ap2}
We show that the number of roots is always greater than or equal to the number of conservation laws.

Since the left nullspace of $\nabla^{l}$ is trivial, applying the rank-nullity theorem to its transpose, we find $\text{rk}(\nabla^{l}) =  |\mathcal{Z}|- |\mathcal{Z}^{r}|$.
Substituting the above in Eq.~\eqref{Eq:rank_nullity}, we find
\begin{equation}\label{Eq:root_resa}
    |\mathcal{Z}| + \text{dim}(\text{ker}(\nabla^{l})) = |\mathcal{R}| +  |\mathcal{Z}^{r}|\,.
\end{equation}
Substituting for $|\mathcal{Z}|$ from Eq.~\eqref{Eq:rank_nullity} into Eq.~\eqref{Eq:root_resa}, we get, 
\begin{equation}\label{Eq:root_resb}
    |\mathcal{L}| + \text{dim}(\text{ker}(\nabla^{l})) - |\mathcal{C}| =   |\mathcal{Z}^{r}|\,.
\end{equation}
Finally, using Eq.~\eqref{Eq:emer_cyc_defn} with Eq.~\eqref{Eq:root_resb}, we obtain the Eq.~\eqref{Eq:root_res}, which directly implies the result. 

\subsubsection{Result III}\label{Sec:root_ap3}
Here, we define the construction of independent roots.

Consider the matrix $\mathbb{L}$ whose rows are the conservation laws and denote its submatrix where the columns are restricted to the root species as $\mathbb{L}^{r}$. 
The rank of $\mathbb{L}^{r}$ must be equal to $|\mathcal{L}|$.

If not, there is a basis of the conservation laws, where at least one row of $\mathbb{L}^{r}$ has all entries zero. Call this conservation law $\boldsymbol{\ell}'$.
Since $\boldsymbol{\ell}^{'} \cdot \nabla = 0$, we have the relation
\begin{equation}
    \boldsymbol{\ell}^{'}_l \cdot \nabla^{l} + \boldsymbol{\ell}^{'}_{r} \cdot \nabla^{r} = 0\,,
\end{equation}
where we wrote the conservation law as $\boldsymbol{\ell}' = (\boldsymbol{\ell}^{'}_l, \boldsymbol{\ell}^{'}_r)^{\intercal}$.
Since $\boldsymbol{\ell}^{'}_r = 0$, $\boldsymbol{\ell}^{'}_l$ must satisfy $\boldsymbol{\ell}^{'}_l \cdot \nabla^{l} = 0$, which is a contradiction.
Since the matrix $\mathbb{L}^{r}$ has full rank, 
it must have at least one invertible square submatrix, denoted $\mathbb{L}^{\mathcal{I}}$. 
We call the subset of root species associated with this square invertible submatrix independent roots and denote the set of independent roots by $\mathcal{I} \subseteq \mathcal{Z}^{r}$.
The number of independent roots must therefore be equal to the number of conservation laws. 
Assume we now choose the set of independent roots $\mathcal{I}$ (resp. $\mathcal{Z}/ \mathcal{I}$) as the external species (resp. internal species) with the stoichiometric matrices $\nabla^{\mathcal{I}}$ (resp. $\nabla^{X}$).
Then, as in Sec.~\ref{Sec:root_ap1}, the left null space of $\nabla^{X}$ must be trivial.
If not, given a vector $\boldsymbol{\ell}_{0}$ such that $\boldsymbol{\ell}_{0} \cdot \nabla^{X} = 0$, we can then define a new  conservation law of the form:
\begin{equation}
    \boldsymbol{\ell}^{'} \cdot \nabla = \kbordermatrix{
    &\color{gray} X & \color{gray} \mathcal{I}\\
    & \boldsymbol{\ell}_{0} & \boldsymbol{0}} \cdot \begin{pmatrix}
        \nabla^{X} \\
        \nabla^{\mathcal{I}}  
    \end{pmatrix} = \boldsymbol{0}\,.
\end{equation}
Finally, using this conservation law $\boldsymbol{\ell}^{'}$, we see that a linear combination of the rows of $\mathbb{L}^{\mathcal{I}}$ should be zero - which contradicts the assumption that $\mathbb{L}^{\mathcal{I}}$ is invertible. 
Since the left nullspace of $\nabla^{X}$ is trivial, following the steps outlined in Sec.~\ref{Sec: root_ap2}, it is seen that $|\mathcal{\epsilon}|_{\mathcal{I}} = 0$ i.e, there are no emergent cycles. 
We note that there might be multiple possible choices of the set of independent roots, and the results do not depend on the choice. 

\subsubsection{Result IV}\label{Sec:root_ap4}
We now show how to construct any species from the independent roots via a net reaction.

Assume we wish to construct an arbitrary species $\sigma_{0}$ from the set of independent roots. 
Consider as the set of external species $Y' = \mathcal{I} \cup \{\sigma_{0}\}$ and the remaining species as the internal species $X'$ with stoichiometric matrices $\nabla^{Y'}$ and $\nabla^{X'}$.
But $\nabla^{X'} \subset \nabla^{X}$ where $X$ is the set of internal species if only the independent roots are considered the external species. 
In Sec.~\ref{Sec:root_ap3}, we showed that the left nullspace of $\nabla^{X}$ is trivial. 
Thus, the left nullspace of $\nabla^{X'}$ is also trivial (else vectors in the left nullspace of $\nabla^{X'}$ can be extended into vectors in the left nullspace of $\nabla^{X}$).
Thus, the rank of $\nabla^{X'}$ must be equal to $|\mathcal{Z}| - |\mathcal{I}| - 1$. Applying the rank-nullity theorem to $\nabla^{X'}$, we find,
\begin{equation}\label{Eq:root_resc}
    |\mathcal{Z}|  + \text{dim}(\text{ker}(\nabla^{X'})) = |\mathcal{R}| + |\mathcal{I}| + 1
\end{equation}
Substituting for $|\mathcal{Z}|$ from Eq.~\eqref{Eq:rank_nullity} into Eq.~\eqref{Eq:root_resc}, we get, 
\begin{equation}\label{Eq:root_resd}
    |\mathcal{L}| + \text{dim}(\text{ker}(\nabla^{l})) - |\mathcal{C}| = |\mathcal{I}| + 1  \implies |\epsilon|_{Y'} = 1\,,
\end{equation}
where we used Eqs.~\eqref{Eq:emer_cyc_defn} (also see Sec.~\ref{Sec:root_ap3}).
Thus, we have a unique emergent cycle $\boldsymbol{c}_{\sigma_{0}}$ associated with this set of external species. 
Finally, using Eq.~\eqref{Eq:net_reactions} , the net reaction constructing $\sigma_{0}$ is obtained from the vector $\nabla^{Y'} \cdot \boldsymbol{c}_{\sigma_{0}}$ upon suitable rescaling in the form
\begin{equation}\label{Eq:construc_rxn}
    \sigma_{0} \xrightleftharpoons[]{} \sum_{i \in \mathcal{I}} \nu_{\sigma_{0},i}\hat{\sigma}_{i}\,
\end{equation}
Unlike the coefficients in Eq.~\eqref{Eq:net_reactions}, all the coefficients in Eq.~\eqref{Eq:construc_rxn} are integers but need not be positive.
We note that the coefficients $\nu_{\sigma_{0},i}$ give us the representation of the species $\sigma_{0}$ when we regard the independent roots as atoms. 

\subsubsection{Result V} \label{Sec:root_ap5}

We now derive a basis set of conservation laws starting from the set of independent roots.

For every species $\sigma \in \mathcal{Z}$, following the prescription of Sec.~\ref{Sec:root_ap4}, we can write a representation as shown in Eq.~\eqref{Eq:construc_rxn}.
Associated with a given independent root $\hat{\sigma}_{i}$, we define the  vector $\boldsymbol{\ell}^{i}$ whose entry for the species $\sigma$ is the coefficient $\nu_{\sigma, i}$.
Given $\boldsymbol{\ell}^{i}$, the entries corresponding to the independent roots $\hat{\sigma}_{j}$ are given by the Kronecker symbol $\delta_{i,j}$. Thus, we have $|\mathcal{I}| = |\mathcal{L}|$ independent vectors. 
If a vector $\bar{\boldsymbol{\ell}}_{i}$ is a conservation law, given a reaction of form Eq.~\eqref{Eq:rxn_redefined}, the coefficients of Eq.~\eqref{Eq:construc_rxn} need to satisfy $\nu_{\sigma_{r(\rho)}, i} + \nu_{\sigma_{r'(\rho)}, i} = \nu_{\sigma_{p(\rho)}, i}$ for every reaction $\rho$.
To prove this, consider a particular reaction $\rho$ and consider as the external species $Y = \mathcal{I} \cup \{\sigma_{r(\rho)},\sigma_{r'(\rho)},\sigma_{p(\rho)}\}$ and the remaining species as internal species. This splits the stoichiometric matrix as
\begin{equation}
    \nabla = 
     \kbordermatrix{    
   \\
   \color{gray} X &\nabla^{X}\\
   \\
    \color{gray} \sigma_{r(\rho)} &\nabla^{\sigma_{r(\rho)}}\\
    \\
    \color{gray} \sigma_{r'(\rho)} &\nabla^{\sigma_{r'(\rho)}}\\
    \\
    \color{gray} \sigma_{p(\rho)} &\nabla^{\sigma_{p(\rho)}}\\  
    \\
    \color{gray} \mathcal{I} &\nabla^{\mathcal{I}}
   }
\end{equation}
Arguments along the lines in Sec.~\ref{Sec: root_ap2} and \ref{Sec:root_ap4} show us that the number of emergent cycles in this case must be exactly three. 
Consider the three appropriately scaled vectors $\boldsymbol{c}_{\sigma_{r(\rho)}}, \boldsymbol{c}_{\sigma_{r'(\rho)}}, \boldsymbol{c}_{\sigma_{p(\rho)}}$ defined as in Sec.~\ref{Sec:root_ap4}.
Each of these vectors are emergent cycles and construct the substrates and products of the reaction $\rho$ from the independent roots. Finally, consider the vector $\boldsymbol{c}$ which is zero for all reactions and is one for the reaction $\rho$. This is also an emergent cycle as
\begin{equation}
    \nabla \cdot \boldsymbol{c} = \kbordermatrix{    
   \\
   \color{gray} X & \boldsymbol{0}\\
   \\
    \color{gray} \sigma_{r(\rho)} &-1\\
    \\
    \color{gray} \sigma_{r'(\rho)} &-1\\
    \\
    \color{gray} \sigma_{p(\rho)} &1\\  
    \\
    \color{gray} \mathcal{I} &\boldsymbol{0}
   }\,.
\end{equation}
Since $\boldsymbol{c}$ is an emergent cycle, we can decompose it as $\boldsymbol{c} = k_{1} \boldsymbol{c}_{\sigma_{r(\rho)}} + k_{2} \boldsymbol{c}_{\sigma_{r'(\rho)}} + k_{3} \boldsymbol{c}_{\sigma_{r'(\rho)}}$. Then, we must have, 
\begin{align}\label{Eq:cyc_decomp}
 \nabla^{Y}\cdot\boldsymbol{c} & =   \kbordermatrix{   
   \\
    \color{gray} \sigma_{r(\rho)} &-1\\
    \\
    \color{gray} \sigma_{r'(\rho)} &-1\\
    \\
    \color{gray} \sigma_{p(\rho)} &1\\  
    \\
    \color{gray} \mathcal{I} &\boldsymbol{0}
   }\,, \notag \\
   & = k_{1} \nabla^{Y}\cdot\boldsymbol{c}_{\sigma_{r(\rho)}} + k_{2} \nabla^{Y}\cdot\boldsymbol{c}_{\sigma_{r'(\rho)}} + k_{3} \nabla^{Y}\cdot\boldsymbol{c}_{\sigma_{r'(\rho)}}\,. 
\end{align}
But from Eq.~\eqref{Eq:construc_rxn}, the form of the vectors $\nabla^{Y}\cdot\boldsymbol{c}_{\sigma_{r(\rho)}}$ is given by
\begin{equation}\label{Eq:cyc_decomp2}
    \nabla^{Y}\cdot\boldsymbol{c}_{\sigma_{r(\rho)}} = \kbordermatrix{   
   \\
    \color{gray} \sigma_{r(\rho)} &1\\
    \\
    \color{gray} \sigma_{r'(\rho)} & 0\\
    \\
    \color{gray} \sigma_{p(\rho)} & 0\\  
    \\
     & -\nu_{\sigma_{r(\rho), 1}}\\   
    \color{gray} \mathcal{I} & . \\
     & .\\
     &-\nu_{\sigma_{r(\rho)}, i}\\
   }\,.
\end{equation}
Substituting Eq.~\eqref{Eq:cyc_decomp2} in Eq.~\eqref{Eq:cyc_decomp}, we find that $k_{1} = -1, k_{2} = -1, k_{3} = 1$, which directly also implies $\nabla^{Y}\cdot\boldsymbol{c}_{\sigma_{r(\rho)}} +\nabla^{Y}\cdot\boldsymbol{c}_{\sigma_{r'(\rho)}} = \nabla^{Y}\cdot\boldsymbol{c}_{\sigma_{p(\rho)}} $ as needed.     

\subsubsection{Example}\label{Sec:root_apeg}
Here, we show an example to illustrate the results of the subsection.

\textit{Example a}: Consider the following CRN, generated randomly with $N_{A} = 1$ and $Q = 10$:
\begin{align}\label{Eq:CRN_root_eg1}
    (1) + (4) \xrightleftharpoons[-1]{+1}  (5) \,,\notag \\
    (3) + (4) \xrightleftharpoons[-2]{+2}  (7)\,, \notag\\
    (1) + (7) \xrightleftharpoons[-3]{+3}  (8)\, ,\notag\\
    (1) + (8) \xrightleftharpoons[-4]{+4}  (9)\, ,\notag\\
    (1) + (9) \xrightleftharpoons[-5]{+5}  (10)\, ,\notag\\
    (5) + (5) \xrightleftharpoons[-6]{+6}  (10)\,,
\end{align}  
with eight species and six reactions. 
The stoichiometric matrix of the CRN above is given by
\begin{equation}\label{Eq:stoc_m_egap1}
  \nabla = \kbordermatrix{
    & {\color{gray}\rho_{1}} & {\color{gray}\rho_{2}}  & {\color{gray}\rho_{3}}  & {\color{gray}\rho_{4}}&  {\color{gray}\rho_{5}} &  {\color{gray}\rho_{6}}\\
{\color{gray} (1)} & -1 & 0 & -1 & -1 & -1 & 0 \\
{\color{gray} (4)} & -1 & -1 & 0 & 0 & 0 & 0 \\
{\color{gray} (5)} & 1 & 0 & 0 & 0 & 0 & -2 \\
{\color{gray} (3)} & 0 & -1 & 0 & 0 & 0 & 0 \\
{\color{gray} (7)} & 0 & 1 & -1 & 0 & 0 & 0 \\
{\color{gray} (8)} &0 & 0 & 1 & -1 & 0 & 0 \\
{\color{gray} (9)} & 0 & 0 & 0 & 1 & -1 & 0 \\
{\color{gray} (10)} & 0 & 0 & 0 & 0 & 1 & 1 \\
}\,.
\end{equation}
Note that the rank of the stoichiometric matrix is six implying that there are two conservation laws and zero cycles.
Furthermore, it is seen from the stoichiometric matrix that there are three root species $(1), (3)$ and $(4)$ respectively. 
Taking the root species as the $Y$ species, from Eq.~\eqref{Eq:stoc_m_egap1}, we get the substochiometric matrix $\nabla^{l}$ (see Sec.~\ref{Sec:root_ap1}) to be, 

\begin{equation}\label{Eq:stoc_x_egap1}
  \nabla^{l} = \kbordermatrix{
    & {\color{gray}\rho_{1}} & {\color{gray}\rho_{2}}  & {\color{gray}\rho_{3}}  & {\color{gray}\rho_{4}}&  {\color{gray}\rho_{5}} &  {\color{gray}\rho_{6}}\\
{\color{gray} (5)} & 1 & 0 & 0 & 0 & 0 & -2 \\
{\color{gray} (7)} & 0 & 1 & -1 & 0 & 0 & 0 \\
{\color{gray} (8)} &0 & 0 & 1 & -1 & 0 & 0 \\
{\color{gray} (9)} & 0 & 0 & 0 & 1 & -1 & 0 \\
{\color{gray} (10)} & 0 & 0 & 0 & 0 & 1 & 1 \\
}\,.
\end{equation}
Firstly, we note that the left nullspace is trivial as expected from Sec.~\ref{Sec:root_ap1}.
Secondly, we find the emergent cycle $\boldsymbol{c}_{r} = \left(-2, 1, 1, 1, 1, -1 \right)^{\intercal}$.
This immediately verifies Eq.~\eqref{Eq:root_res} as there are three roots and one emergent cycle, i.e, two independent roots. 
The net reaction reaction connecting the roots is given by
\begin{equation}
    \nabla^{Y}\cdot\boldsymbol{c}_{r} = \kbordermatrix{    
   \\
   \color{gray} (1) & -1\\
   \\
    \color{gray} (4) & 1\\
    \\
     \color{gray} (3) & -1\\
   }\,,
\end{equation}
corresponding to the net reaction $(4) \xrightleftharpoons[]{} (1) + (3)$.
Considering the independent roots as $\mathcal{I} = \{ (1), (3)\}$, we now start constructing the representation of different species as outlined in Sec.~\ref{Sec:root_ap4}.
Assume we want to find the net reaction constructing the species $(7)$. 
Taking the $Y$ species as $\{(1), (3), (7)\}$, we find the substoichiometric matrix $\nabla^{X}$ from Eq.~\eqref{Eq:stoc_m_egap1} to be of the form,

\begin{equation}\label{Eq:stoc_x2_egap1}
  \nabla^{X} = \kbordermatrix{
    & {\color{gray}\rho_{1}} & {\color{gray}\rho_{2}}  & {\color{gray}\rho_{3}}  & {\color{gray}\rho_{4}}&  {\color{gray}\rho_{5}} &  {\color{gray}\rho_{6}}\\
{\color{gray} (4)} & -1 & -1 & 0 & 0 & 0 & 0 \\
{\color{gray} (5)} & 1 & 0 & 0 & 0 & 0 & -2 \\
{\color{gray} (8)} &0 & 0 & 1 & -1 & 0 & 0 \\
{\color{gray} (9)} & 0 & 0 & 0 & 1 & -1 & 0 \\
{\color{gray} (10)} & 0 & 0 & 0 & 0 & 1 & 1 \\
}\,,
\end{equation}
with the emergent cycle $\boldsymbol{c}' = \left(2, -2, -1, -1, -1, 1 \right)^{\intercal}$. The corresponding net reaction is seen to be,
\begin{equation}
    \nabla^{Y}\cdot\boldsymbol{c}' = \kbordermatrix{    
   \\
   \color{gray} (1) & 1\\
   \\
    \color{gray} (3) & 2\\
    \\
     \color{gray} (7) & -1\\
   }\,,
\end{equation}
corresponding to the net reaction $(7) \xrightleftharpoons[]{} (1) + 2 (3)$. 
We emphasize again that if we had represented the species $\{(1), (3), (4), (5), (7)\dots\}$ by the labels $\{A, B, C, D, E\dots\}$, the above net reaction corresponds to stating that the species $E$ can be written as $AB_{2}$ which is nontrivial.
Finally, following Sec.~\ref{Sec:root_ap5}, we derive the basis of conservation laws:
\begin{align}
    \boldsymbol{\ell}^{1} = \kbordermatrix{
    & \color{gray} (1) & \color{gray} (4) & \color{gray} (5) & \color{gray} (3) & \color{gray} (7)& \color{gray} (8)& \color{gray} (9)& \color{gray} (10) \\
    & 1 & 1 & 2 & 0 & 1 & 2 & 3 & 4 
    }\,,\\
     \boldsymbol{\ell}^{2} = \kbordermatrix{
    & \color{gray} (1) & \color{gray} (4) & \color{gray} (5) & \color{gray} (3) & \color{gray} (7)& \color{gray} (8)& \color{gray} (9)& \color{gray} (10) \\
    & 0 & 1 & 1 & 1 & 2 & 2 & 2 & 2 
    }\,.
\end{align}
corresponding to the conservation of the moieties $(1)$ and $(3)$.    

\section{Simulations for Reachability}\label{Subsec:data_reach1D}

\begin{figure}
    \centering
    \includegraphics[scale=.50]{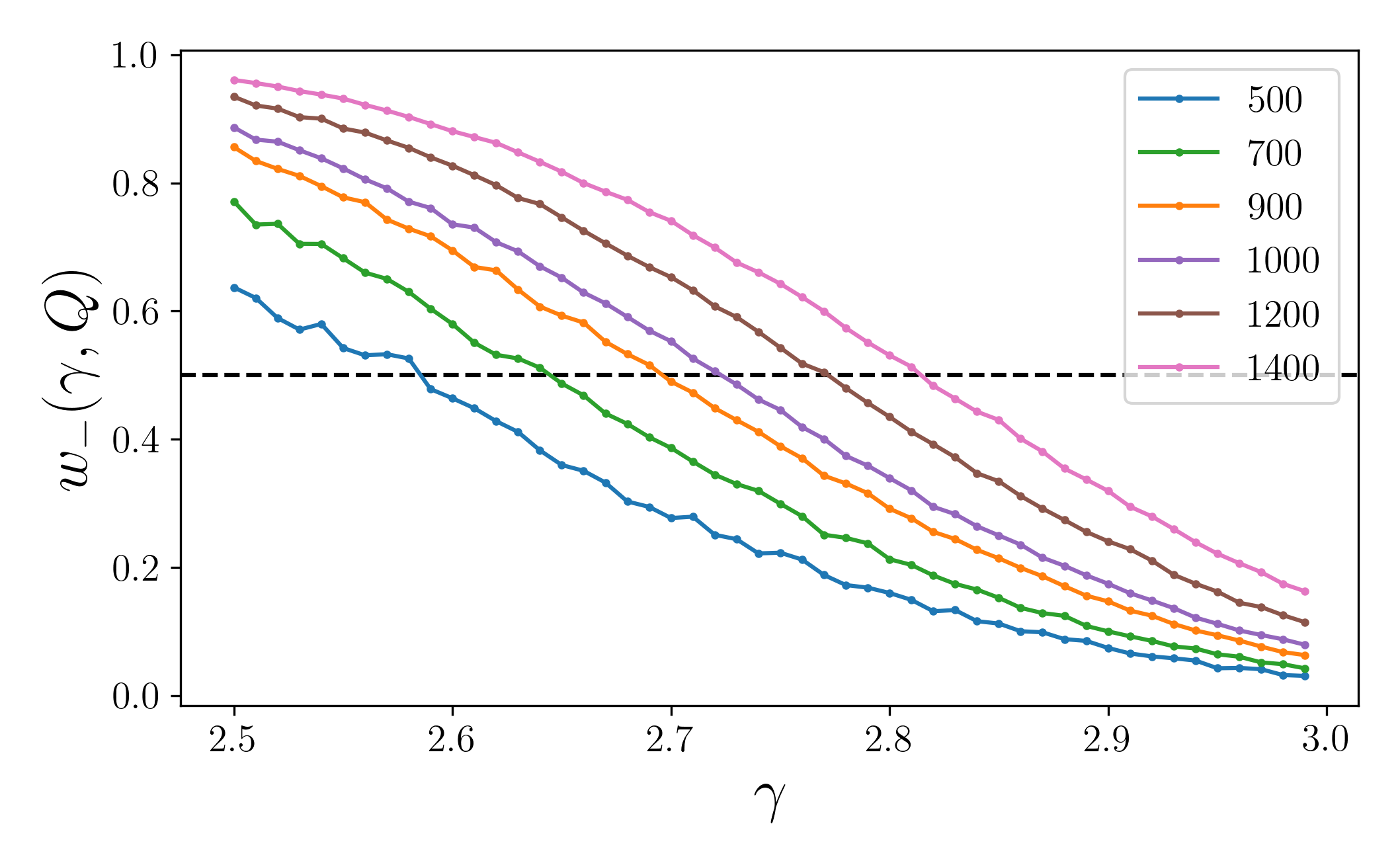}
    \caption{Probability weight $w_{-}(\gamma,Q)$ of the left peak of the order parameter histogram plotted against $\gamma$ for different $Q$. The parameters are $N_{A} = 1$ and $Q$. The horizontal dashed line is $0.5$ which represents the equal weight point.}
    \label{fig:reach_weights}
\end{figure}

We first outline the method to determine the forward reachable set of a species given a stoichiometric matrix \(\nabla\).  Let \(S\) be a set of species, initially containing only the species \(\sigma\) whose reachable set we wish to compute. We then loop over all reactions: if all substrates of a reaction \(\rho\) are in \(S\), we add all the products of \(\rho\) to \(S\) and then do the same for the reverse reaction. This process is iterated until \(S\) no longer changes. The final set \(S\) is defined as the forward reachable set \(\mathcal{F}(\sigma)\).  This procedure is repeated for each species \(\sigma\), and the largest reachable set is recorded. The resulting order parameter is the rescaled size of the largest reachable set, \(|f|\).  
Note that this procedure depends only on the stoichiometric matrix and is independent of the number of atoms.

\subsection{One atom}
We generated ensembles for \(Q = 100, 200, \dots, 1400\), but restricted analysis to \(Q \geq 500\).  
For each value of \(Q\), we generated CRNs for \(\gamma \in [2.0, 4.0]\) in steps of \(0.01\).  
Each \(\gamma\)–\(Q\) pair had 39,000 samples for \(Q \leq 1200\), and 25,000 samples for \(Q = 1300, 1400\).  The sample mean of \(|f|\) is plotted with standard errors in Fig.~\ref{fig:reach_op_1}, for selected values of \(Q\).

\begin{figure}
    \centering
    \includegraphics[scale=.50]{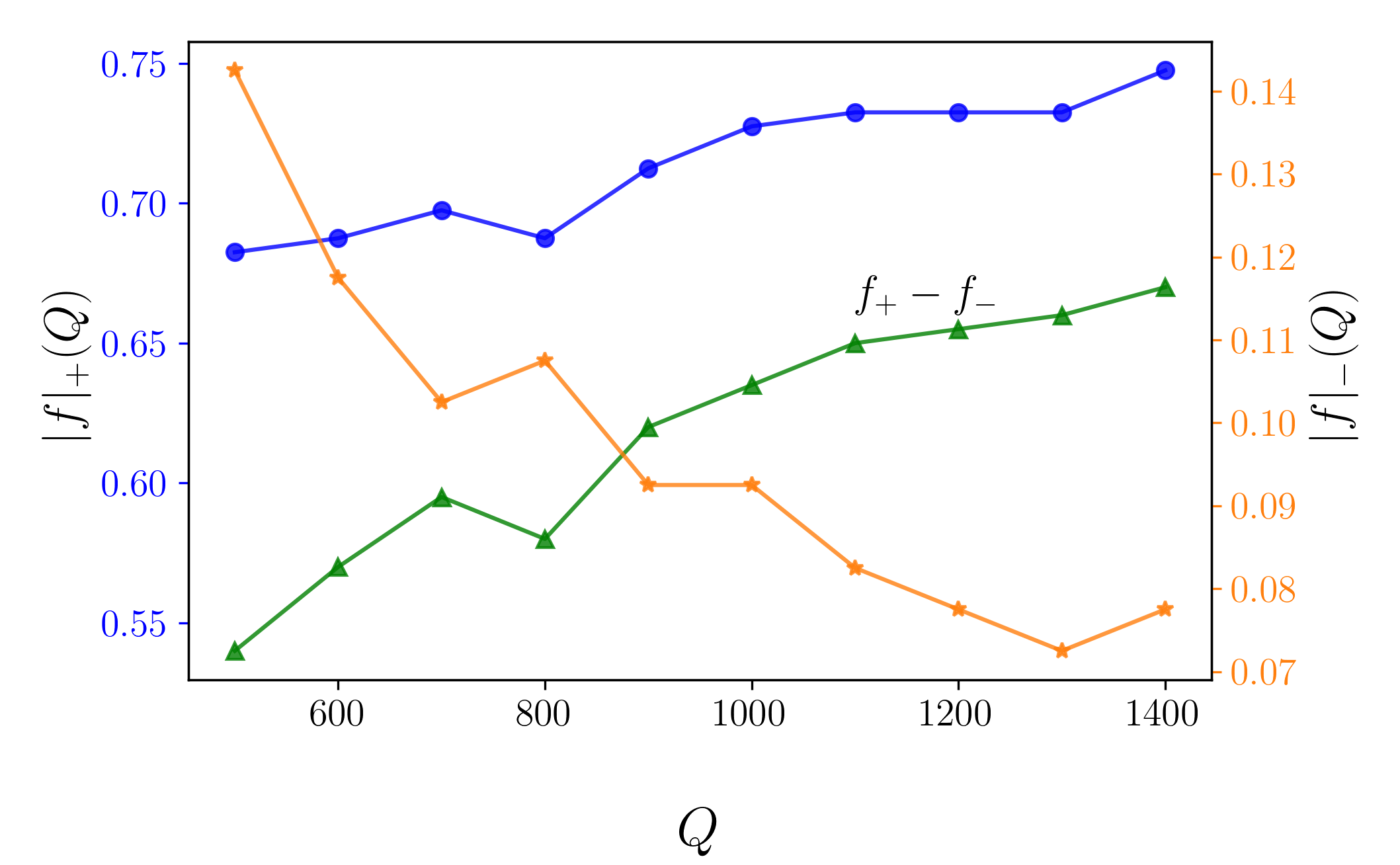}
    \caption{Location of the left (resp. right) peak of the order parameter histogram, denoted $|f|_{-}(Q)$ (yellow stars)(resp. $|f|_{+}(Q)$ (blue dots)) plotted against $Q$. Note that the two y-axes start at different values. The peaks are evaluated at $\gamma \approx \gamma_{1/2}(Q)$. The parameters are $N_{A} = 1$ and $Q$. The difference is plotted as green triangles.}
    \label{fig:reach_heights}
\end{figure}

To estimate \(\gamma_c\), we analyzed histograms of \(|f|\) for \(\gamma \in [2.5, 3.0]\). For each value of $\gamma$, the histogram was made using $201$ bins between $0$ and $1$ (via histogram method in numpy). We located the histogram peaks (maximum counts) and then found the histogram valley (minimum counts) between the peaks. The probability weight assigned to the left peak was the area from the leftmost bin, i.e., zero until the histogram valley. We plot the left probability weight, denoted $w_{-}(\gamma,Q)$ as a function of $\gamma$ in Fig.~\ref{fig:reach_weights} for different $Q$. We then interpolated between the two closest \(\gamma\) values where \(w_-\) crosses 0.5 to obtain \(\gamma_{1/2}(Q)\), the equal-weight probability.

\begin{figure}
    \centering
    \includegraphics[scale=.5]{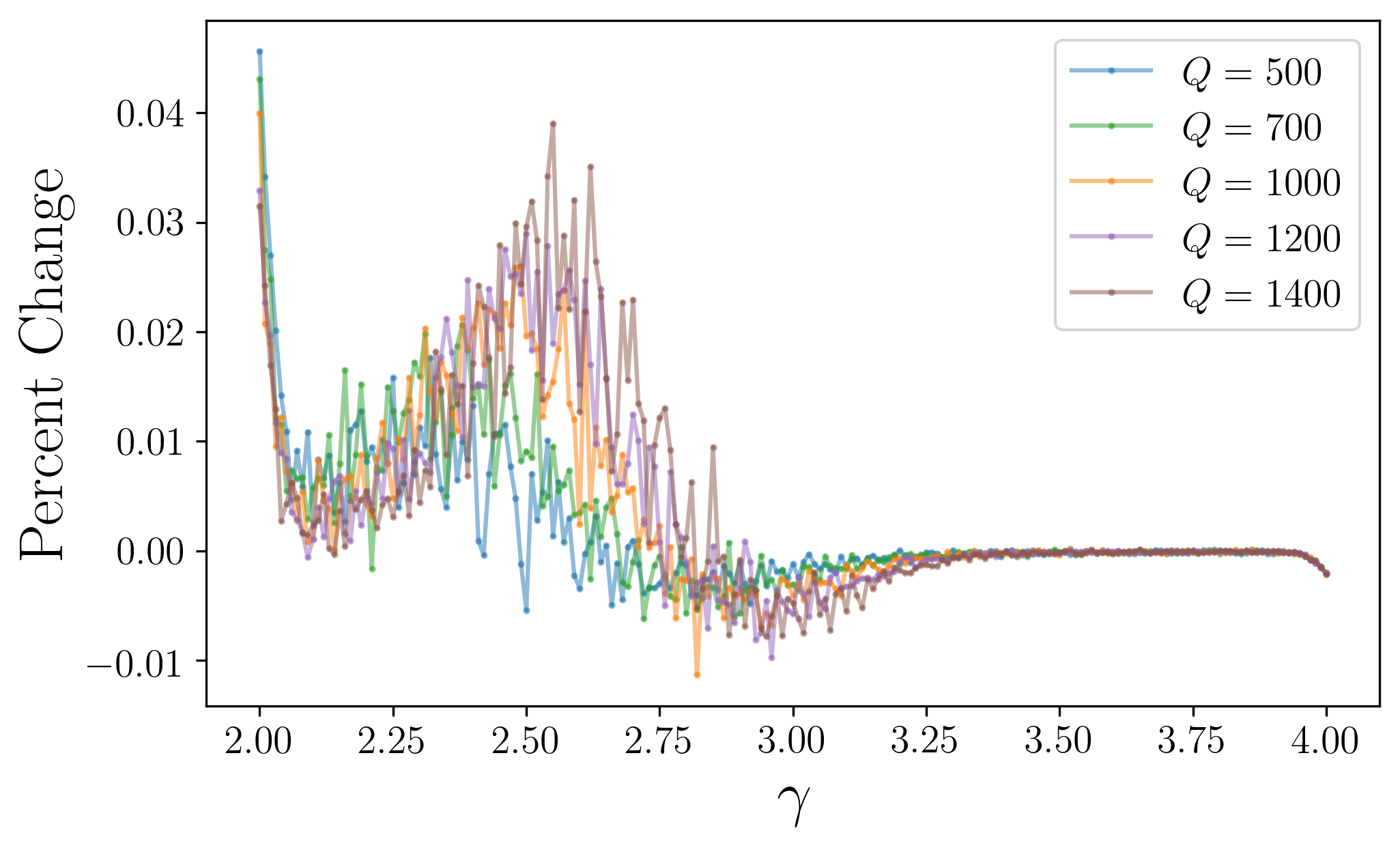}
    \caption{Data from Fig.~\ref{fig:reach_op_1} smoothened by a Gaussian Filter and the ratio of the difference between the smoothened and original data and the original data (percentage change) plotted as a function of $\gamma$. The parameters are $N_{A} = 1$ and $\sigma = 4$ for the filtering.}
    \label{fig:smooth_diff_reach}
\end{figure}

Uncertainty in \(\gamma_{1/2}(Q)\) arises from both histogram binning and the discretization of \(\gamma\).  %
To estimate binning uncertainty, we repeated the procedure for $181, 191, 201, 211,221$ bins and computed the standard deviation of results.  
To estimate \(\gamma_c\), we linearly fit \(\gamma_{1/2}(Q)\) (averaged over bin counts) versus \(1/Q\).  
The fit yields \(\gamma_c = 2.921 \pm 0.008\) with \(r^2 \approx 0.94\). Taking into account the \(\gamma\) discretization, we report \(\gamma_c = 2.92 \pm 0.01\).

\begin{figure}
    \centering
    \includegraphics[scale=.5]{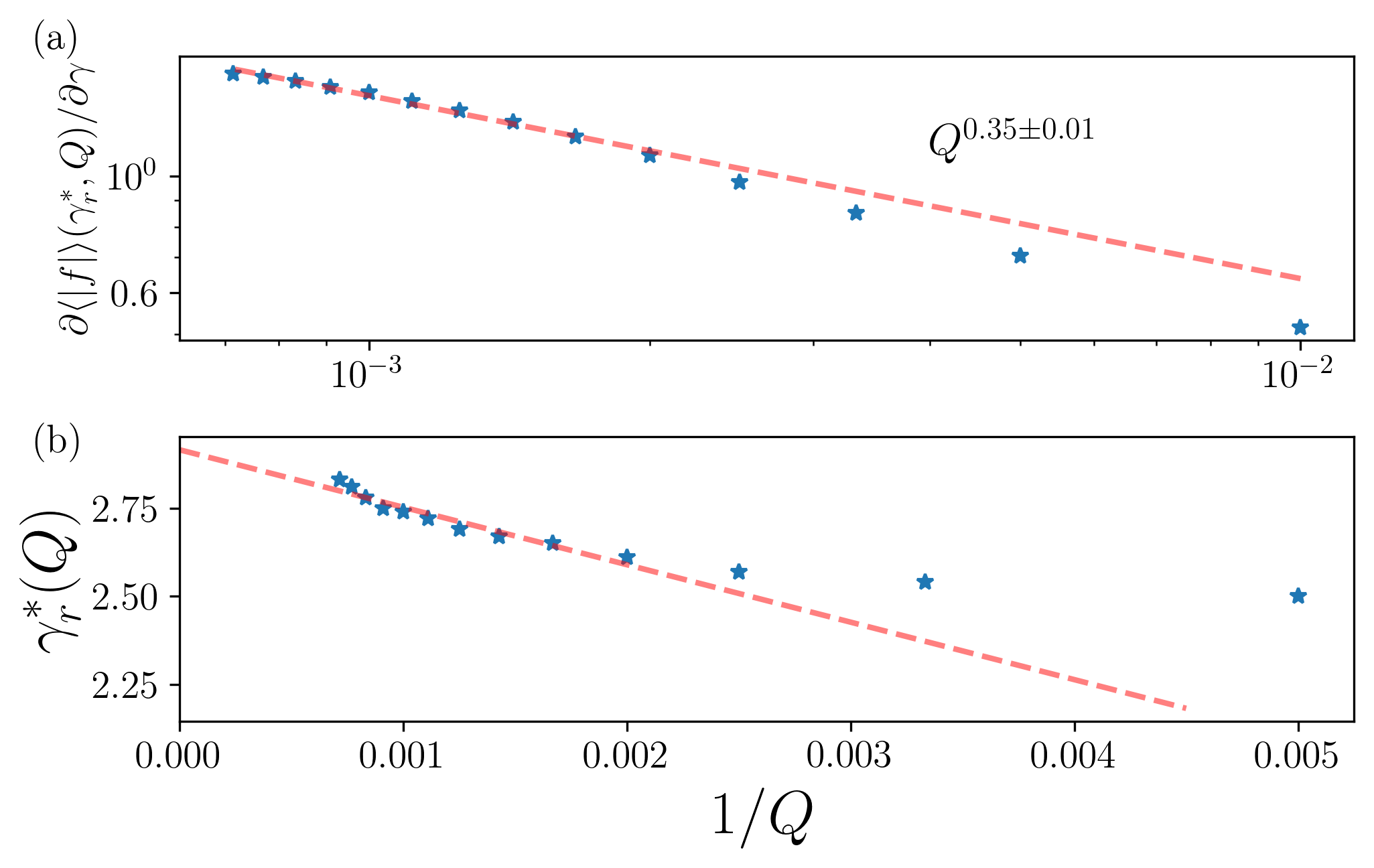}
    \caption{a) Peak of the derivative of the average reachability wrt $\gamma$ (data from Fig.~\ref{fig:reach_op_1}) plotted as a function of $1/Q$ in log-log scale. The redline is a linear fit in log-log scale. b) Location of the peak of the derivative of the average reachability, denoted $\gamma^{*}_{r}(Q)$ wrt $\gamma$ (data from Fig.~\ref{fig:reach_op_1} ) plotted as a function of $1/Q$. The redline is a linear fit. For both graphs, the parameters are $N_{A} = 1$.}
    \label{fig:reach_der2}
\end{figure}

\begin{figure}
    \centering
    \includegraphics[scale=.5]{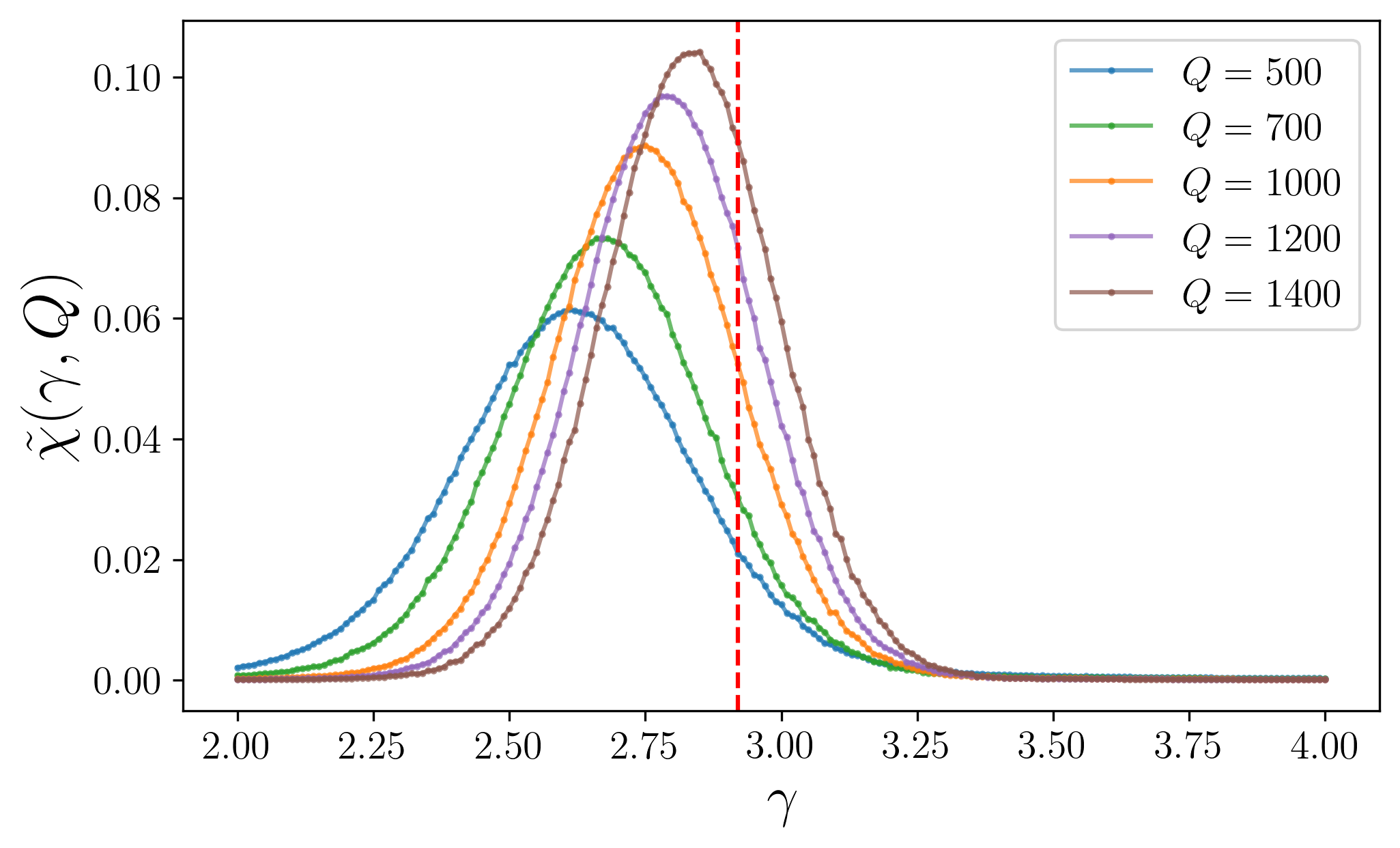}
    \caption{Fluctuations of the average reachability in Fig.~\ref{fig:reach_op_1} labelled $\tilde{\chi}(\gamma, Q)$ plotted as a function of $\gamma$. The parameters are $N_{A} = 1$. The redline is the estimated location of the percolation threshold.}
    \label{fig:sus_reach}
\end{figure}

For each \(Q\), we selected the \(\gamma\) closest to \(\gamma_{1/2}(Q)\) and extracted the locations of the two histogram peaks, denoted \(|f|_\pm(Q)\).  
The peak locations and their separation are shown in Fig.~\ref{fig:reach_heights}.  
We observe that \(|f|_+(Q)\) increases and appears to saturate, while \(|f|_-(Q)\) decreases, confirming separation between the two regimes.

\begin{figure}
    \centering
    \includegraphics[scale=.5]{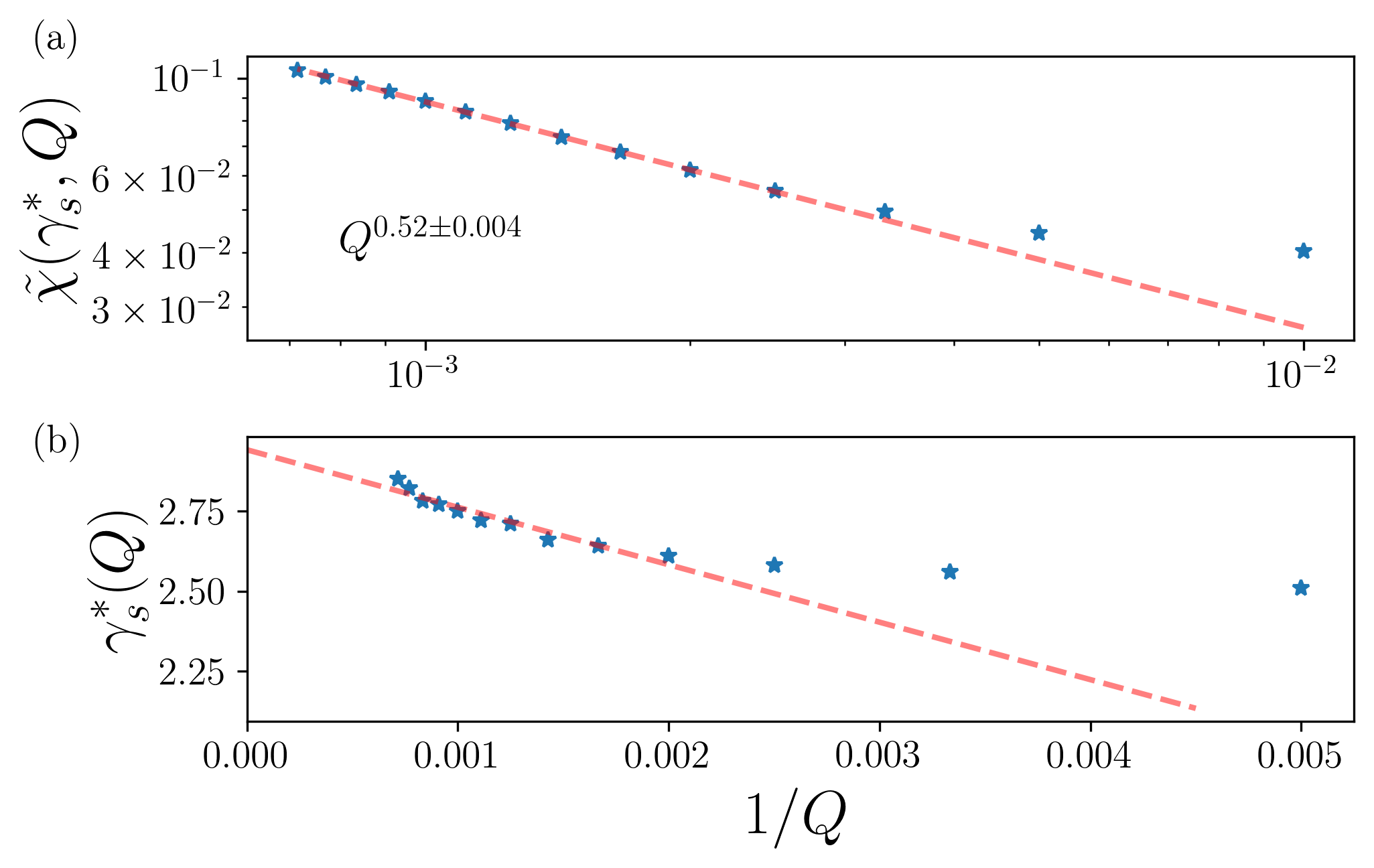}
    \caption{a) Peak of the fluctuations of the reachability (data from Fig.~\ref{fig:sus_reach}) plotted as a function of $1/Q$ in log-log scale. The redline is a linear fit in log-log scale. b) Location of the peak of $\tilde{\chi}(\gamma, Q)$, denoted $\gamma^{*}_{s}(Q)$ (data from Fig.~\ref{fig:sus_reach} ) plotted as a function of $1/Q$. The redline is a linear fit. For both graphs, the parameters are $N_{A} = 1$.}
    \label{fig:reach_sus2}
\end{figure}

To confirm the discontinuous nature of the transition, we computed the derivative \(\partial\langle |f| \rangle/\partial \gamma\) using the data in Fig.~\ref{fig:reach_op_1}. We first smoothed the data using the gaussianfilter1d routine from scipy.ndimage with a smoothing parameter $\sigma = 4.0$.
The percent difference due to smoothing is shown in Fig.~\ref{fig:smooth_diff_reach}, and does not exceed four percent.
The peak height of the derivative, as a function of \(Q\), is shown in Fig.~\ref{fig:reach_der2}a and fits well to a power law with exponent 0.35.
Fig.~\ref{fig:reach_der2}b shows the extrapolation of the location of the derivative peak (denoted $\gamma^{*}_{r}(Q)$) to \(Q \to \infty\), using a linear fit in \(1/Q\).  
The intercept is \(2.92 \pm 0.02\), consistent with the estimate from histogram analysis (\(r^2 \approx 0.93\)).
We define the fluctuations of the order parameter as \(\tilde{\chi}(\gamma, Q) = \langle |f|^2 \rangle - \langle |f| \rangle^2\).  
The behavior of \(\tilde{\chi}(\gamma, Q)\) is shown in Fig.~\ref{fig:sus_reach}.  
As \(Q\) increases, the peak in \(\tilde{\chi}\) becomes sharper and shifts toward \(\gamma_c\).  
The peak height scales with \(Q\) as a power law with exponent \(0.52\), as shown in Fig.~\ref{fig:reach_sus2}a.  
Fig.~\ref{fig:reach_sus2}b shows that the location of the fluctuation peak (denoted $\gamma^{*}_{s}(Q)$) also extrapolates to \(\gamma_c = 2.94 \pm 0.02\), consistent with the other estimates. The extrapolation is a linear fit in $1/Q$.
\end{document}